%% file: main.tex
\documentclass{article}

\usepackage{arxiv}

\usepackage{setspace}
\usepackage[utf8]{inputenc} 
\usepackage[T1]{fontenc}    
\usepackage{hyperref}       
\usepackage{url}            
\usepackage{booktabs}       
\usepackage{amsfonts}       
\usepackage{nicefrac}       
\usepackage{microtype}      
\usepackage{lipsum}
\usepackage{graphicx}
\graphicspath{ {./images/} }

\usepackage{bm}
\usepackage{amsmath, amssymb} 
\usepackage{graphicx}        
\usepackage{pgfplots}        
\pgfplotsset{compat=newest}    
\usepackage{caption}         
\usepackage{subcaption}      
\usepgfplotslibrary{groupplots}

\usepackage{tabularx}
\usepackage{tikz}
\usepackage{xcolor}
\usepackage{placeins}
\usepackage{mathtools}

\usepackage{siunitx}
\usepackage{tikz-3dplot}
\usepackage{graphicx}
\usetikzlibrary{calc,backgrounds}
\usepackage{pgfplotstable}
\usepgfplotslibrary{groupplots}
\usetikzlibrary{intersections}
\usetikzlibrary{positioning}
\usepgfplotslibrary{fillbetween}

\definecolor{rwth1}{RGB}{0,84,159}      
\definecolor{rwth2}{RGB}{142,186,229}   
\definecolor{rwth3}{RGB}{0,97,101}      
\definecolor{rwth4}{RGB}{0,152,161}     
\definecolor{rwth5}{RGB}{87,171,39}     
\definecolor{rwth6}{RGB}{189,205,0}     
\definecolor{rwth7}{RGB}{255,237,0}     
\definecolor{rwth8}{RGB}{246,168,0}     
\definecolor{rwth9}{RGB}{227,0,102}     
\definecolor{rwth10}{RGB}{204,7,30}     
\definecolor{rwth11}{RGB}{161,16,53}    
\definecolor{rwth12}{RGB}{97,33,88}     
\definecolor{rwth13}{RGB}{122,111,172}  

\definecolor{rwthb1}{HTML}{e8f1fa}      
\definecolor{rwthb2}{HTML}{c7ddf2}      
\definecolor{rwthb3}{HTML}{8ebae5}      
\definecolor{rwthb4}{HTML}{407fb7}      
\definecolor{rwthb5}{HTML}{00549f}      

\definecolor{rwtho1}{HTML}{fff7ea}      
\definecolor{rwtho2}{HTML}{feeac9}      
\definecolor{rwtho3}{HTML}{fdd48f}      
\definecolor{rwtho4}{HTML}{fabe50}      
\definecolor{rwtho5}{HTML}{f6a800}      

\definecolor{rwthg1}{RGB}{255,245,80}  
\definecolor{rwthg2}{RGB}{204,190,0}  
\definecolor{rwthg3}{RGB}{153,140,0}    

\definecolor{lightblue}{RGB}{173,216,230}

\title{Autoencoder-based non-intrusive model order reduction in continuum mechanics}

\author{
Jannick Kehls \\
  Institute of Applied Mechanics\\
  RWTH Aachen University\\
  Aachen, 52074, Germany \\
  \texttt{jannick.kehls@ifam.rwth-aachen.de} \\
  \And
Ellen Kuhl \\
  Department of Mechanical Engineering\\
  Stanford University\\
  Stanford, CA 94305, United States \\
  \texttt{ekuhl@stanford.edu} \\
  \And
Tim Brepols \\
  Institute of Applied Mechanics\\
  RWTH Aachen University\\
  Aachen, 52074, Germany \\
  \texttt{tim.brepols@rwth-aachen.de} \\
  \And
Kevin Linka \\
  Institute for Continuum and Material Mechanics\\
  Hamburg University of Technology\\
  Hamburg, 21073, Germany \\
  \texttt{kevin.linka@tuhh.de} \\
  \And
Hagen Holthusen \\
  Institute of Applied Mechanics\\
  University of Erlangen-Nuremberg\\
  Erlangen, 91058, Germany \\
  \texttt{hagen.holthusen@fau.de}
}
\begin{document}
\maketitle
\begin{abstract}
  We propose a non-intrusive, Autoencoder-based framework for reduced-order modeling in continuum mechanics.
  Our method integrates three stages: (i) an unsupervised Autoencoder compresses high-dimensional finite element solutions into a compact latent space,
  (ii) a supervised regression network maps problem parameters to latent codes, and (iii) an end-to-end surrogate reconstructs full-field solutions directly from input parameters.

  To overcome limitations of existing approaches, we propose two key extensions: a force-augmented variant that jointly predicts displacement fields and reaction forces at Neumann boundaries, and a multi-field architecture that enables coupled field predictions, such as in thermo-mechanical systems.
  The framework is validated on nonlinear benchmark problems involving heterogeneous composites, anisotropic elasticity with geometric variation, and thermo-mechanical coupling.
  Across all cases, it achieves accurate reconstructions of high-fidelity solutions while remaining fully non-intrusive.

  These results highlight the potential of combining deep learning with dimensionality reduction to build efficient and extensible surrogate models.
  Our publicly available implementation provides a foundation for integrating data-driven model order reduction into uncertainty quantification, optimization, and digital twin applications.
\end{abstract}


\input{sections/section1}
\input{sections/section2}
\input{sections/section3}
\input{sections/section4}
\input{sections/section5}

\FloatBarrier

\appendix

\input{sections/appendixA}
\input{sections/appendixB}

\bibliographystyle{unsrt}
\bibliography{literature}  


\end{document}

%% file: sections/section1.tex
\section{Introduction}
\label{sec1:1}

Today's continuum mechanics problems often entail repeatedly solving extremely large equation systems generated by high-fidelity, high-dimensional finite element (FE) models. Despite the ever-increasing power of modern computers, such computations can quickly become very time-consuming -- or even prohibitive -- in many practical applications. The growing demand for real-time predictions and many-query scenarios involving varying input parameters in areas such as design optimization, uncertainty quantification, control, and digital twin applications has amplified the challenge, spurring significant interest in advanced model order reduction (MOR) techniques. Generally speaking, these methods aim to drastically reduce the number of degrees of freedom in high-dimensional discretized systems, thereby accelerating computations while maintaining a sufficiently high level of accuracy.

Traditional projection-based MOR procedures, such as proper orthogonal decomposition (POD) \cite{Chatterjee2000} and other reduced-basis methods, have demonstrated considerable success in compressing high-dimensional simulation data into low-dimensional subspaces. For instance, POD has been successfully applied to problems related to dynamical systems \cite{HinzeVolkwein2005, KerschenGolinvalEtAl2005, LuJinEtAl2021}, turbulent flow \cite{BerkoozHolmesEtAl1993, SmithMoehlisEtAl2005, HijaziStabileEtAl2020}, and solid mechanics \cite{ChenKareem2005, RadermacherReese2013, RitzertMacekEtAl2024, RitzertKehlsEtAl2025}, among others. When combined with modern hyper-reduction techniques, such as the a priori hyper-reduction method \cite{Ryckelynck2005, Ryckelynck2009, MiledRyckelynckEtAl2013}, the discrete empirical interpolation method \cite{ChaturantabutSorensen2010, RadermacherReese2016}, the energy-conserving sampling and weighting method \cite{AnKimEtAl2008, FarhatAveryEtAl2014, FarhatChapmanEtAl2015, TrainottiMarinkoEtAl2024}, or the empirical cubature method \cite{HernandezCaicedoEtAl2017, BravoAresdePargaEtAl2023, HernandezBravoEtAl2024}, even highly nonlinear problems can be tackled within a reasonable amount of time.

Nevertheless, these approaches are typically intrusive in the sense that they require sophisticated modifications to the underlying simulation code. Furthermore, constructing a linear subspace as in classical POD to approximate a potentially highly nonlinear solution manifold may not in all cases capture the behavior of a system effectively. Finally, incorporating abstract data in the modeling framework or additional input parameters to steer the solution is usually either not possible or not straightforward with these methods. Consequently, there is a growing interest in non-intrusive, data-driven strategies that bypass the need for a direct access to the system's governing equations while enabling the construction of fast surrogate models that can, additionally, rely on arbitrary input data.

\subsection{Non-intrusive MOR techniques based on dimensionality reduction and a subsequent mapping of the latent space}

The field of non-intrusive model order reduction is extremely broad, with developments evolving rapidly in many different directions. This overview does not aim to give a fully comprehensive overview of all currently ongoing developments, but rather focuses on approaches that align very closely with the ideas presented in this paper.

A fundamental idea behind many non-intrusive, data-driven MOR approaches -- including the one proposed in this paper -- is to first transform a system's high-dimensional state into a compact, low-dimensional latent space representation. Subsequently, a mapping is constructed that predicts or approximates the latent space states directly, based on certain input parameters determining the system's behavior. This mapping can be obtained, for example, through regression or interpolation.

A frequently used physics-based method for the dimension reduction in this context is POD. In \cite{XiaoFangEtAl2015}, a non-intrusive reduced-order modeling approach for the Navier–Stokes equations is proposed that combines POD with radial basis function interpolation to efficiently approximate complex fluid flow dynamics. The interpolation is employed to rapidly predict the POD coefficients at the current time step based on previously computed ones. Afterwards, the high-dimensional solution can be reconstructed by projecting the reduced coefficients back into the original state space. In \cite{HesthavenUbbiali2018}, the authors present a MOR approach for parametrized nonlinear partial differential equations that combines POD with artificial neural networks, which learn a mapping between the problem's input parameters and the reduced-space coefficients via regression. An extension of this approach to time-dependent nonlinear problems is developed in \cite{WangHesthavenEtAl2019}. Conceptually similar approaches are presented in \cite{SwischukMaininiEtAl2019} and \cite{KneiflGrunertEtAl2021}, where the authors explore a range of machine learning techniques -- such as neural networks, multivariate polynomial regression, $k$-nearest neighbors, decision trees, and Gaussian process regression -- to learn the specific mapping from the input parameters to the POD coefficients, respectively. The advantages and disadvantages of each method are thoroughly examined. As one of the few studies in the current literature, \cite{KneiflGrunertEtAl2021} focuses on structural dynamics problems, demonstrating significant simulation speedups while maintaining sufficient accuracy in crash simulations of a lightweight racing kart. Furthermore, in \cite{SalvadorDedeEtAl2021} kernel POD is used to generate a reduced basis with far fewer modes compared to traditional POD. Also here, a neural network is trained to learn the mapping from the input parameters to the reduced coefficients. As a novel point, the network's architecture is adaptively constructed such that its complexity scales with the reduced basis dimension, generally resulting in a smaller network, being both easier and faster to train.

Despite their success, POD-based methods naturally face inherent limitations. The extraction of a fixed, linear subspace may not adequately capture the complex, nonlinear behavior present in many continuum mechanical problems. If the solution manifold exhibits significant curvature, the linear basis provided by POD may fail to represent important dynamic features, leading to reduced accuracy.

In light of these shortcomings, the research community is increasingly turning its attention toward nonlinear alternatives. Among others, approaches based on Autoencoders are particularly promising, as the latter can be flexibly adapted to the nonlinearity of the problem under consideration, possess a scalable architecture, and are comparatively easy to implement and use due to many freely available and well-documented software libraries. Autoencoders are specialized neural networks that learn to reconstruct their input at the output layer by passing the data through a typically low-dimensional `bottleneck'. This latent space represents the core dynamics of the system \cite{FeffermanMitterEtAl2016}. Therefore, even a relatively simple regression model is assumed to be able to provide accurate approximations within this reduced space.
They can be seen as a nonlinear generalization of POD, capable of learning more complex, nonlinearly embedded low-dimensional representations of data. In fact, as is well-known, linear Autoencoders with a single hidden layer, linear activation functions, and mean-squared error loss are mathematically equivalent to POD \cite{BourlardKamp1988, BaldiHornik1989, Plaut2018}.

In \cite{GonzalezBalajewicz2018}, a deep convolutional Autoencoder is trained to achieve the dimensionality reduction of a problem and, afterwards, a long short-term memory (LSTM) network is used to evolve the system's states in the latent space. A comparable methodology is developed in \cite{HernandezBadiasEtAl2021}, with the distinctive feature that the Autoencoder learns by itself the suitable size of the reduced space, requiring no prior knowledge from the user; a special neural network is trained afterwards to predict the system's latent space states in a thermodynamically consistent manner. Numerical examples from fluid dynamics and solid mechanics demonstrate that these approaches outperform traditional POD–Galerkin-based model order reduction methods in terms of prediction quality and stability. However, the mentioned methods do not explicitly account for any parameter dependence in the partial differential equations describing the underlying problem. In contrast, conceptually very similar strategies are presented in which the regression procedures to predict the latent space states directly can be categorized into either neural networks \cite{FrescaDedeEtAl2021, SimpsonEtAl2021, ShindeItierEtAl2023, PichiMoyaEtAl2024} or Gaussian process regression \cite{HalderFidkowskiEtAl2022, KneiflRosinEtAl2023, DeshpandeRappelEtAl2025}. A parametric dependence of the underlying problem is explicitly considered in the respective mappings. Particularly, in \cite{ShindeItierEtAl2023} a non-intrusive approach is presented to rapidly predict fracture patterns in two-dimensional structural problems solely based on (non-proportional) loading paths as input. The training data for the corresponding convolutional Autoencoder stem from preliminary simulations using a geometrically linear cohesive phase-field model for brittle fracture \cite{WuNguyen2018}. Another very interesting application case is shown in \cite{KneiflRosinEtAl2023} where a real-time capable surrogate model of a musculoskeletal system is developed. In the latter work, (kernel) POD and (variational) Autoencoders are applied, respectively, to achieve the dimensionality reduction of a model, based on snapshot data which is obtained from continuum mechanical simulations based on a geometrically nonlinear anisotropic material model. Gaussian process regression is finally employed to successfully learn the relation between muscle activation and the system's latent space states. Interestingly enough, while it is reported that all methods used to achieve the dimensionality reduction are able to predict the displacements in the musculoskeletal system in a satisfactory manner, the `standard' non-variational Autoencoder leads overall to the best results. 

\subsection{Research gap and hypothesis of the study}

\paragraph{Research gap.}
Despite substantial progress in projection-based and data-driven model order reduction, important challenges remain. 
Classical approaches struggle with strongly nonlinear systems, require intrusive modifications, and rarely extend to multiphysical settings. 
Autoencoder-based methods provide efficient compression of displacement fields, yet the reliable prediction of Neumann boundary reactions has hardly been addressed. 
Moreover, existing strategies typically focus on single fields, lacking generality for coupled thermo-mechanical or other multiphysical problems. 
Thus, a framework that combines efficient dimensionality reduction with consistent prediction of both multiphysical fields and boundary forces is still missing.

\paragraph{Hypothesis.}
We hypothesize that a neural-network-based framework, combining an Autoencoder with parameter-to-latent regression, 
can efficiently represent high-dimensional solution manifolds across nonlinear and multiphysical problems. 
By extending the architecture with a force-augmented variant, the framework further enables the consistent prediction of Neumann boundary reactions. 
Together, these capabilities are expected to provide a robust trade-off between computational efficiency and physical accuracy, 
establishing a general-purpose tool for surrogate modeling in continuum mechanics with applications in uncertainty quantification, optimization, and digital twin technologies.

\subsection{Outline}

The paper is organized as follows. In Section \ref{sec2:1}, we present our Autoencoder-based framework for non-intrusive model order reduction. We begin by describing the design and training of an unsupervised Autoencoder that compresses high-dimensional finite element snapshots into a low-dimensional latent space and reconstructs them via its decoder. Subsequently, we introduce a regression network that maps a few key input parameters directly onto the latent space, and we demonstrate how coupling its output with the decoder yields an end-to-end surrogate model capable of predicting full finite element solutions. In Section \ref{sec3:1}, we illustrate the effectiveness of our approach through several benchmark problems. One example involves a heterogeneous unit cell problem where both the displacement field and reaction forces are predicted under parametrized loading and stiffness contrasts between the cell’s constituents. We then consider a fibre-reinforced plate with an elliptic hole, in which both the geometry of the hole and the fiber angle are parameterized, followed by a transient thermo-mechanical multi-field problem where the heat conductivity and the plate's thickness are varied. Section \ref{sec4:1} discusses our results by critically assessing the surrogate model’s accuracy, computational efficiency, and limitations. Finally, Section \ref{sec5:1} concludes with suggestions for potential improvements and directions for future research.

%% file: sections/section2.tex
\section{Network architecture}
\label{sec2:1}
\paragraph{Non-intrusive machine learning approach.}
To enable efficient prediction of continuum mechanical states for unseen parameters, we adapt the three-stage, non-intrusive model order reduction framework by \cite{SimpsonEtAl2021}.
The approach begins with \textit{Unsupervised Autoencoder Discovery}, where a low-dimensional latent space is learned solely from solution snapshots, without any parameter information. 
This latent space captures the essential features of the system’s behavior. 
In the second phase, \textit{Supervised Latent Space Prediction}, we train a regression network to map input parameters to their corresponding locations in the latent space, using matched snapshot–parameter pairs. 
Finally, the \textit{End-to-End Surrogate Model} combines this parameter-to-latent network with the trained decoder, enabling rapid and accurate reconstruction of solution states for new, unseen parameter inputs.
To account for both the force field strongly associated with the solution snapshots as well as multiphysical interactions, an \textit{End-to-End Force-augmented Model} and a \textit{Multi-field extension} are presented.
For interested readers, we present a simplified analogy of the proposed approach to classical, intrusive model order reduction in Appendix~\ref{app:analogy}.

\paragraph{Unsupervised Autoencoder Discovery.} The encoder network $\mathcal{E}$ maps the solution space $\bm{\phi}$ onto a latent representation $\bm{z}$
\begin{equation}
    \bm{z} = \mathcal{E}(\bm{\phi}; \bm{w}_\mathcal{E}),
    \label{eq:Encoder}
\end{equation}
while the decoder $\mathcal{D}$ reconstructs the input from the latent space
\begin{equation}
    \hat{\bm{\phi}} = \mathcal{D}(\bm{z}; \bm{w}_\mathcal{D}) = \mathcal{D}(\mathcal{E}(\bm{\phi}; \bm{w}_\mathcal{E}); \bm{w}_\mathcal{D}),
    \label{eq:Decoder}
\end{equation}
where $\bm{w}_\mathcal{E}$ and $\bm{w}_\mathcal{D}$ denote the weights of the encoder and decoder, respectively.
The Autoencoder architecture — comprising encoder, decoder, and latent space — is illustrated in Figure~\ref{fig:autoencoder}.
The input $\bm{\phi} \subset \bm{\psi}$ includes only the \emph{active} degrees of freedom, with $\bm{\psi}$ representing the \textit{entire} solution field of the system, including those constrained by Dirichlet boundary conditions.

Training is performed on a finite set of snapshots $\{\bm{\phi}^s\}_{s=1}^{n_S}$, which sample the solution space. 
The training is unsupervised, i.e., the Autoencoder receives no information about the associated parameters $\bm{\theta}^s$. 
It is expected to autonomously learn the underlying structure and encode it in the latent space.

The loss function is defined as the mean squared reconstruction error, regularized by $\mathcal{R}$
\begin{equation}
    \mathcal{L}_\mathcal{A} = \frac{1}{n_S\ n_N\ n_D} \sum_{s=1}^{n_S} \sum_{n=1}^{n_N} \sum_{m=1}^{n_D} \left( \hat{\phi}_{n,m}^s - \phi_{n,m}^s \right)^2 + \mathcal{R}(\bm{w}_\mathcal{E}, \bm{w}_\mathcal{D}),
    \label{eq:loss_AE}
\end{equation}
where $n_S$ is the number of snapshots, $n_N$ the number of active nodes, and $n_D$ the number of active degrees of freedom per node.
Hence, the training is performed according to
\begin{equation}
    (\bm{w}_\mathcal{E}^*,\bm{w}_\mathcal{D}^*) = \arg\min_{\bm{w}_\mathcal{E},\bm{w}_\mathcal{D}} \; \mathcal{L}_\mathcal{A}
\end{equation}
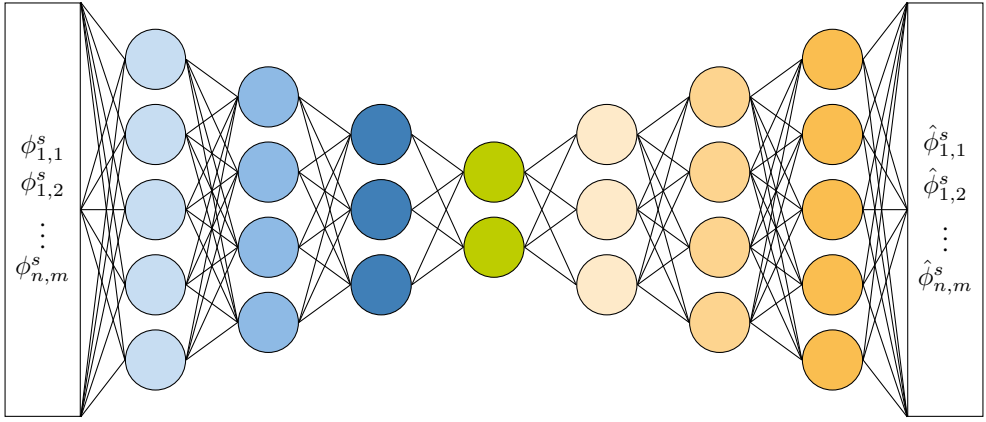
\begin{figure}
    \centering
    \input{figures/net_autoencoder}
    \caption{\textbf{Unsupervised Autoencoder Discovery.} Unlabeled snapshots $(\bullet)^s$ from the input space $\bm{\phi}$ are encoded into a latent representation and subsequently reconstructed by the decoder. The encoder’s hidden layers are shown in blue, the latent space neurons in green, and the decoder’s hidden layers in orange. The architectures of the encoder and decoder are not necessarily the same.}
    \label{fig:autoencoder}
\end{figure}

\begin{figure}
    \centering
    \input{figures/net_FFNtoLS}
    \caption{\textbf{Supervised Latent Space Prediction.} The regression network (yellow) is trained to infer the mapping from input parameters $\bm{\theta}$ to the solution field $\bm{\phi}$ via a latent representation. To this end, the encoder (dashed, frozen during training) provides the latent space for each snapshot.}
    \label{fig:FFN}
\end{figure}
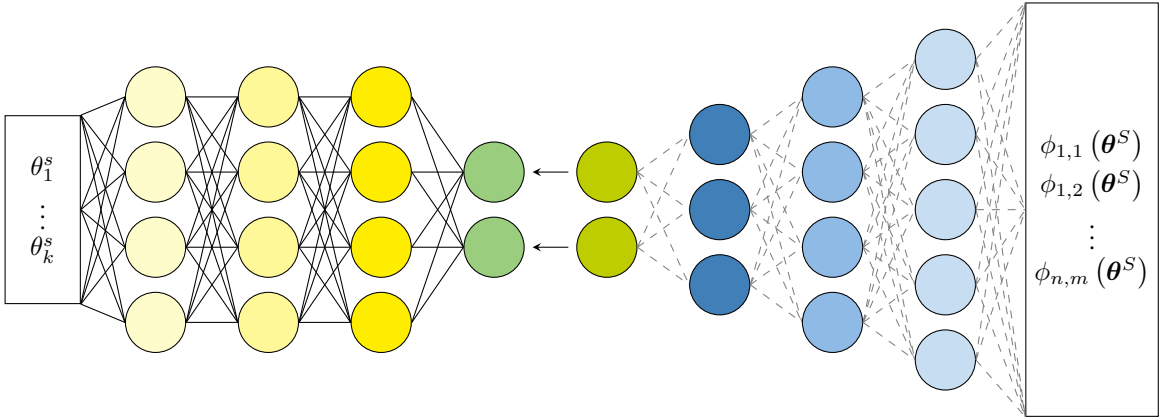

\paragraph{Supervised Latent Space Prediction.} Next, we aim to learn the relation between the set of parameters $\{\bm{\theta}^s\}_{s=1}^{n_S}$ and the corresponding snapshots of the solution field $\{\bm{\phi}^s\}_{s=1}^{n_S}$.
Therefore, a regression network $\mathcal{P}$ is introduced, which generally approximates the latent space representation
\begin{equation}
    \hat{\bm{z}} = \mathcal{P}(\bm{\theta};\bm{w}_\mathcal{P}),
\end{equation}
with the network's weights denoted by $\bm{w}_\mathcal{P}$.
For the time being, the regression model is chosen as a standard feedforward neural network, however, more sophisticated choices are possible.
Its architecture, including the previously trained encoder network, which is considered to be frozen, is schematically illustrated in Figure~\ref{fig:FFN}

In contrast to the training of the Autoencoder, we provide the regression network with information between the solution space $\bm{\phi}$ and the parameters $\bm{\theta}$, i.e., the training is supervised.
In fact, the snapshots of the solution can be considered intrinsically dependent on the parameters, i.e., $\bm{\phi}^S=\bar{\bm{\phi}}^S(\bm{\theta}^S)$.

The minimization procedure during training is expressed in terms of the mean squared error of the latent representation (cf. Equation~\eqref{eq:Encoder})
\begin{equation}
    \bm{w}_\mathcal{P}^* = \arg\min_{\bm{w}_\mathcal{P}} \; \frac{1}{n_S\ n_K} \sum_{s=1}^{n_S} \sum_{k=1}^{n_K} \left( \hat{z}_k^s - z_k^s \right)^2 + \mathcal{C}(\bm{w}_\mathcal{P}),
    \label{eq:LossLS}
\end{equation} 
which is regularized by $\mathcal{C}$.
The number of neurons in the latent space representation is given by $n_K$.
Noteworthy, the latent representation, $\bm{z}$, in Equation~\eqref{eq:LossLS} can be precomputed as the weights of the encoder are frozen during the training of the regression network.

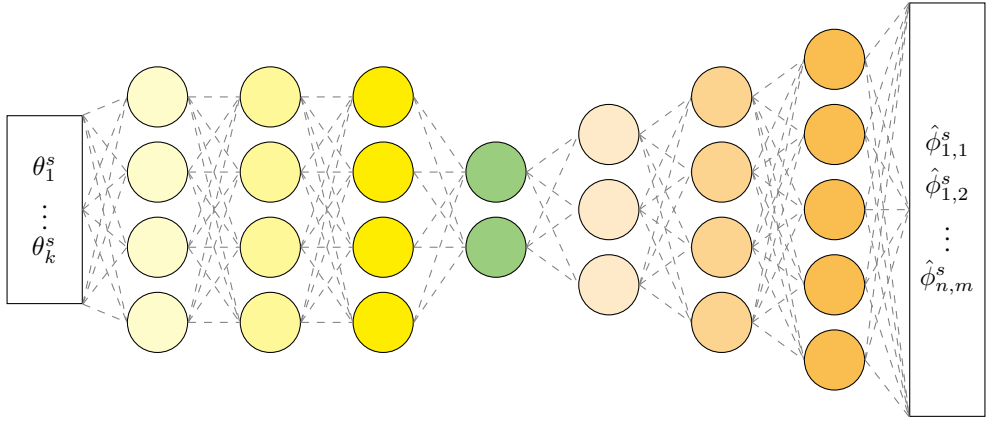
\begin{figure}
    \centering
    \input{figures/net_EndtoEnd}
    \caption{\textbf{End-to-End Surrogate Model.} The latent representation generated by the encoder is replaced by the output of the trained regression network, enabling a direct and efficient pipeline for predicting the solution field from input parameters.}
    \label{fig:EndtoEnd}
\end{figure}

\paragraph{End-to-End Surrogate Model.} Lastly, we aim to construct a surrogate model enabling us to efficiently predict the solution field within the parametric space spanned by the snapshots. 
Therefore, we integrate the regression network $\mathcal{P}$ into the workflow of the decoder network $\mathcal{D}$ to obtain an End-to-End surrogate model.
Its architecture is shown in Figure~\ref{fig:EndtoEnd}, which can be mathematically described as
\begin{equation}
    \hat{\bm{\phi}} = \mathcal{D}\left( \mathcal{P}(\bm{\theta};\bm{w}_\mathcal{P}); \bm{w}_\mathcal{D} \right).
\end{equation}
To which is extent the prediction matches unseen states within the parametric space likely depends on the individual training procedures and is to be investigated. 

\paragraph{End-to-End Force-augmented Model.} To date, we have developed an end-to-end surrogate model that maps input parameters directly to the solution field. In practical applications, however, it is also essential to estimate the force vector $\bm{f}$ associated with \textit{non-active} degrees of freedom, i.e., those constrained by Dirichlet boundary conditions.

To this end, we augment the latent space of the previous model by incorporating the force information at these non-active degrees. 
Recognizing the distinct structural characteristics of the solution field and the force vector, we introduce separate encoder-decoder pairs: $\prescript{}{\phi}{}\mathcal{E}, \prescript{}{\phi}{}\mathcal{D}$ for the solution field, and $\prescript{}{f}{}\mathcal{E}, \prescript{}{f}{}\mathcal{D}$ for the force terms. A shared latent representation is formed by summing the outputs of both encoders
\begin{equation}
    \bm{z} = \prescript{}{\phi}{}\mathcal{E}(\bm{\phi}; \prescript{}{\phi}{}\bm{w}_\mathcal{E}) + \prescript{}{f}{}\mathcal{E}(\bm{f}; \prescript{}{f}{}\bm{w}_\mathcal{E}),
    \label{eq:EncoderReac}
\end{equation}
where $\prescript{}{i}{}\bm{w}_\mathcal{E}$ denote the encoder parameters. The shared latent variable $\bm{z}$ is then decoded to reconstruct both fields:
\begin{align}
    \hat{\bm{\phi}} &= \prescript{}{\phi}{}\mathcal{D}(\bm{z}; \prescript{}{\phi}{}\bm{w}_\mathcal{D}), \\
    \hat{\bm{f}} &= \prescript{}{f}{}\mathcal{D}(\bm{z}; \prescript{}{f}{}\bm{w}_\mathcal{D}),
\end{align}
where $\bm{w}_\mathcal{D}$ are the decoder parameters. 
This architectural choice is motivated by the underlying assumption that both the solution field and the corresponding reaction forces originate from the same governing physical laws and boundary conditions; thus, they encode complementary but fundamentally linked information that can be jointly represented in a unified latent space.

\begin{figure}
    \centering
    \input{figures/net_DecoupledForce}
    \caption{\textbf{End-to-End Force-augmented Model.} Top: The solution space $\bm{\phi}$ and the corresponding force terms $\bm{f}$ are provided to individual encoders. 
    Subsequently, the encoders learn a shared latent space.
    Thereafter, individual decoder networks reconstruct the solution field and force terms.
    The \textit{non-active} nodes and degrees of freedom for the force terms are denotes by $\alpha$ and $\beta$, respectively
    Bottom: The trained End-to-End Force-augmented surrogate model with a single regression network predicting the shared latent space.}
    \label{fig:ReacForce}
\end{figure}
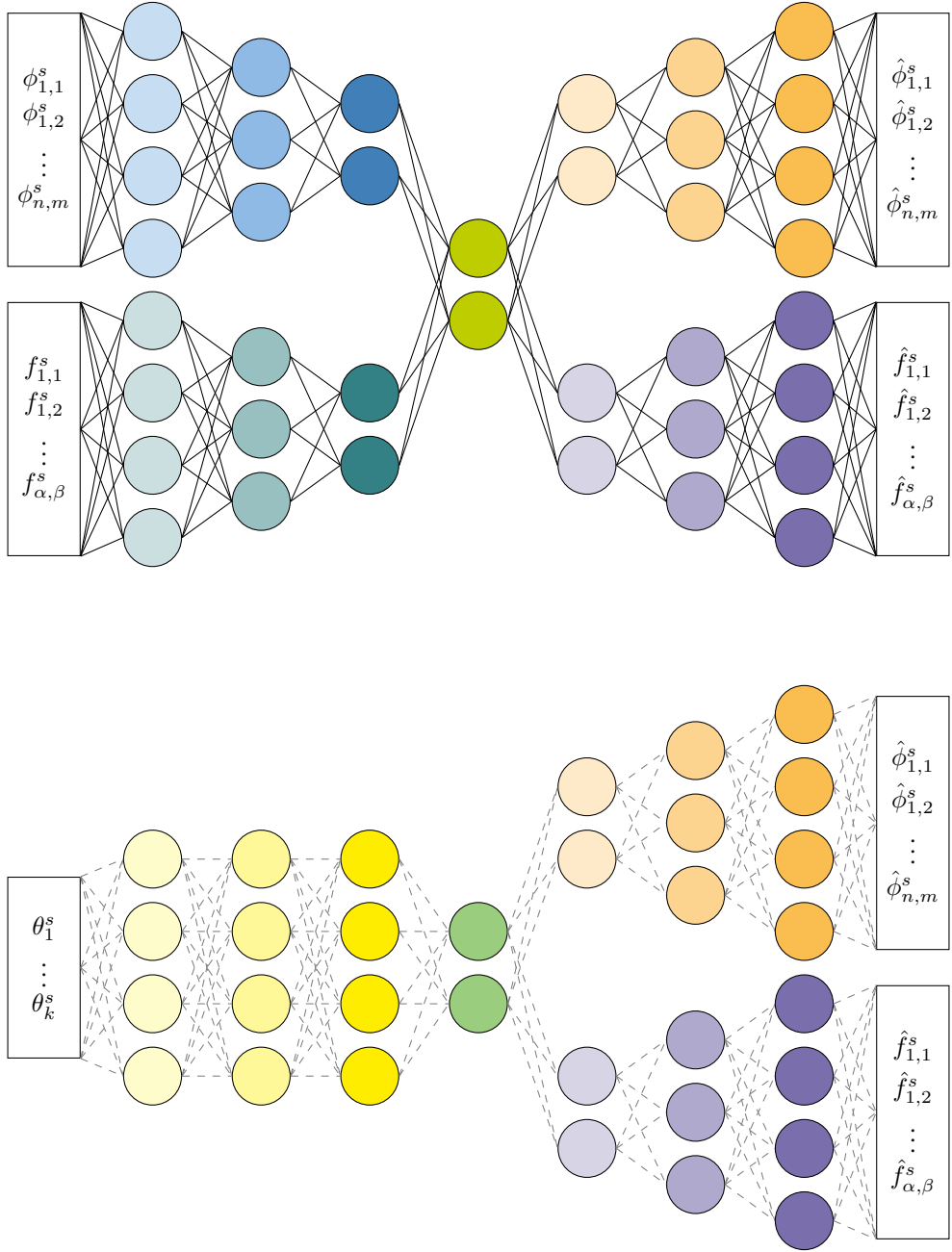

To balance the contributions of both outputs during training, we minimize a variance-normalized ($\mathrm{Var}$) mean squared error ($\mathrm{MSE}$) loss function, cf. Equation~\eqref{eq:loss_AE}
\begin{equation}
\mathcal{L}_\mathcal{A} = \frac{\mathrm{MSE}(\hat{\bm{\phi}}, \bm{\phi})}{\mathrm{Var}(\bm{\phi})} + \frac{\mathrm{MSE}(\hat{\bm{f}}, \bm{f})}{\mathrm{Var}(\bm{f})} + \mathcal{R}(\{\bm{w}\}),
\label{eq:loss_Force_enhanced}
\end{equation}
where $\mathcal{R}$ denotes a regularization term and $\{\bm{w}\}$ the collection of all model parameters. Subsequently, the shared latent representation is mapped from input parameters via a regression network $\mathcal{P}$, as described earlier and illustrated in Figure~\ref{fig:ReacForce}.

For comparison, we also investigated a simplified variant predicting force vectors from the solution latent space alone (cf. \ref{app:force}). However, as shown in Section~\ref{sec3:1}, this approach failed to achieve satisfactory accuracy.

\paragraph{Multi-field extension.} Thus far, we have outlined a general framework for dimensionality reduction, independent of the specific physical interpretation of the system state $\bm{\phi}$.  
However, within the context of \emph{classical intrusive} model order reduction (see Appendix~\ref{app:analogy}), it has been demonstrated that decomposing the state variable $\bm{\phi} = \prescript{1}{}{\bm{\phi}} \cup \hdots \cup \prescript{N}{}{\bm{\phi}}$ into $N$ physically meaningful subfields — each approximated individually via its own reduced basis $\prescript{1}{}{\bm{\Xi}},\hdots,\prescript{N}{}{\bm{\Xi}}$ — consistently outperforms the global reduction approach employing a single monolithic basis $\bm{\Xi}$~\cite{ZhangEtAl2025,KehlsRitzertEtAl2025}.  
In what follows, we extend this concept to the non-intrusive Autoencoder-based setting.

For clarity, we restrict the explanation to two representative physical fields, e.g., displacement and temperature.  
In~\cite{ZhangEtAl2025}, Equation~\eqref{eq:MOR_reduce} is formulated independently for each field, yielding
\begin{equation}
    \prescript{1}{}{\bm{\phi}} \approx \prescript{1}{}{\bm{\Xi}}\, \prescript{1}{}{\bm{a}}, \quad \prescript{2}{}{\bm{\phi}} \approx \prescript{2}{}{\bm{\Xi}}\, \prescript{2}{}{\bm{a}},
\end{equation}
and thus the reduced-order system — consisting of the reduced stiffness matrix and right-hand side vector (cf. Equation~\eqref{eq:MOR_reduced_system}) — takes the form
\begin{equation}
    \tilde{\bm{K}} = \begin{pmatrix}
        \prescript{1}{}{\bm{\Xi}} & \bm{0} \\
        \bm{0} & \prescript{2}{}{\bm{\Xi}}
    \end{pmatrix}^T \begin{pmatrix}
        \prescript{11}{}{\bm{K}}    &   \prescript{12}{}{\bm{K}} \\
        \prescript{21}{}{\bm{K}}    &   \prescript{22}{}{\bm{K}}
    \end{pmatrix}
    \begin{pmatrix}
        \prescript{1}{}{\bm{\Xi}} & \bm{0} \\
        \bm{0} & \prescript{2}{}{\bm{\Xi}}
    \end{pmatrix}, \quad 
    \tilde{\bm{r}} = \begin{pmatrix}
        \prescript{1}{}{\bm{\Xi}} & \bm{0} \\
        \bm{0} & \prescript{2}{}{\bm{\Xi}}
    \end{pmatrix}^T \begin{pmatrix}
        \prescript{1}{}{\bm{r}} \\
        \prescript{2}{}{\bm{r}}
    \end{pmatrix},
    \label{eq:MOR_multi}
\end{equation}
where $\prescript{ij}{}{\bm{K}}$ and $\prescript{i}{}{\bm{r}}$ denote the submatrices of the global stiffness matrix and right-hand side vector, including the contributions from the individual fields as well as their couplings.

\begin{figure}
    \centering
    \input{figures/net_multiphysics}
    \caption{\textbf{Multi-field extension.} Top: The input space $\bm{\phi}$ is provided to individual encoders in a semi-labeled fashion. Subsequently, each encoder learns its own latent space representation, which are then concatenated to an unified latent representation. A fully-connected decoder network reconstructs the input space from this unification. Bottom: The trained End-to-End surrogate model consists of individual regression networks while sharing the same input parameters.}
    \label{fig:AE_multi}
\end{figure}
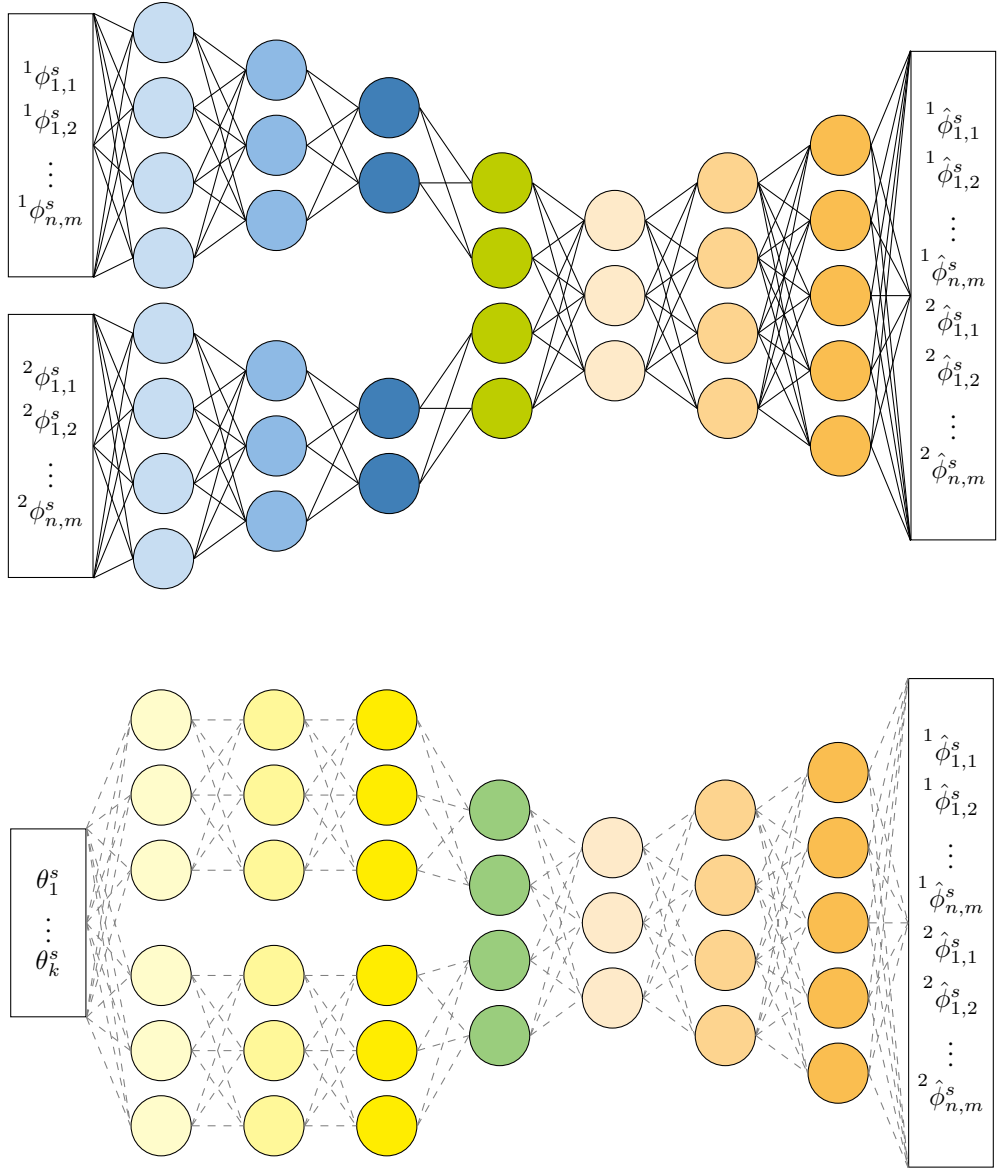

In essence, the snapshot data is decomposed field-wise, while the reduced-order model still accounts for field coupling through the structure of the reduced system~\eqref{eq:MOR_multi}.  
This concept can be transferred to the non-intrusive Autoencoder framework by introducing $N$ separate encoder networks — one for each physical field (see Figure~\ref{fig:AE_multi}).  
Each encoder maps its corresponding input field to an individual latent representation, analogous to the field-specific reduced bases $\prescript{i}{}{\bm{\Xi}}$.  
In contrast, a single, fully connected decoder is employed to jointly reconstruct the complete system state from the concatenated latent vectors.  
This architectural choice is motivated by the fact that, despite the decomposition of the encoder, the reduced system remains strongly coupled and is solved as a whole, as reflected in Equation~\eqref{eq:MOR_multi}.

The resulting Autoencoder architecture can be expressed as
\begin{equation}
    \hat{\bm{\phi}} = \mathcal{D}\left( \prescript{1}{}{\mathcal{E}}\left(\prescript{1}{}{\bm{\phi}}; \prescript{1}{}{\bm{w}_\mathcal{E}}\right), \prescript{2}{}{\mathcal{E}}\left(\prescript{2}{}{\bm{\phi}}; \prescript{2}{}{\bm{w}_\mathcal{E}}\right); \bm{w}_\mathcal{D} \right),
\end{equation}
while the end-to-end surrogate model is given by
\begin{equation}
    \hat{\bm{\phi}} = \mathcal{D}\left( \prescript{1}{}{\mathcal{P}}\left(\bm{\theta}; \prescript{1}{}{\bm{w}_\mathcal{P}}\right), \prescript{2}{}{\mathcal{P}}\left(\bm{\theta}; \prescript{2}{}{\bm{w}_\mathcal{P}}\right); \bm{w}_\mathcal{D} \right).
\end{equation}
It is important to note that while the full-field variable $\bm{\phi}$ must be partitioned into its physical components, the parameter vector $\bm{\theta}$ serves as a shared input across all regression networks.
To prevent the Autoencoder from being overly influenced by fields with larger absolute errors — due to significant differences in their respective error magnitudes — the mean squared error ($\mathrm{MSE}$) for each field is normalized by the corresponding data variance ($\mathrm{Var}$), cf. Equations~\eqref{eq:loss_AE} and \eqref{eq:loss_Force_enhanced}, viz.
\begin{equation}
\mathcal{L}_\mathcal{A} = \frac{\mathrm{MSE}\left(\prescript{1}{}{}\hat{\bm{\phi}},\prescript{1}{}{\bm{\phi}}\right)}{\mathrm{Var}\left(\prescript{1}{}{\bm{\phi}}\right)} + \frac{\mathrm{MSE}\left(\prescript{2}{}{}\hat{\bm{\phi}},\prescript{2}{}{\bm{\phi}}\right)}{\mathrm{Var}\left(\prescript{2}{}{\bm{\phi}}\right)} + \mathcal{R}(\{\bm{w}\}).
\end{equation}
The loss function for the supervised training of the regression networks remains unchanged.

An important advantage of the proposed decomposed architecture is its flexibility in tailoring the complexity of the encoder and regression networks to the characteristics of each physical field.  
In particular, fields governed by relatively simple or weakly nonlinear relations can be represented using shallower or narrower networks, thereby reducing model complexity and training cost.  
Such adaptivity is not readily achievable in monolithic, non-decomposed approaches, where the entire multi-field problem is treated uniformly, potentially leading to over-parameterization or inefficient learning for simpler subfields.

Alternative architectural choices could be conceived, such as a fully connected encoder with a field-wise (partially decoupled) decoder, or even entirely separate Autoencoder branches for each field.  
However, in direct analogy to the intrusive approach, the proposed configuration offers a balanced trade-off between field-specific expressiveness and coupled-system consistency.  
A comprehensive investigation of such alternative designs is beyond the scope of the present work.

%% file: figures/net_autoencoder.tex
\begin{center}
\begin{tikzpicture}[
    >=stealth,
    neuronEncOne/.style={circle, draw, fill=rwthb2, minimum size=8mm},
    neuronEncTwo/.style={circle, draw, fill=rwthb3, minimum size=8mm},
    neuronEncThree/.style={circle, draw, fill=rwthb4, minimum size=8mm},
    neuronLatent/.style={circle, draw, fill=rwth6, minimum size=8mm},
    neuronDecOne/.style={circle, draw, fill=rwtho2, minimum size=8mm},
    neuronDecTwo/.style={circle, draw, fill=rwtho3, minimum size=8mm},
    neuronDecThree/.style={circle, draw, fill=rwtho4, minimum size=8mm},
    neuron/.style={circle, draw, minimum size=8mm}
]

\def\hsep{1.5cm}

\def\nEncOne{5}
\def\nEncTwo{4}
\def\nEncThree{3}
\def\nLatent{2}
\def\nDecOne{3}
\def\nDecTwo{4}
\def\nDecThree{5}

\newcommand{\drawlayer}[5]{ 
  \foreach \i in {1,...,#2} {
    \node[#5] (#1\i) at (#3,{-#2/2 + \i}) {};
  }
  \node[below=0.2cm of #1#2, yshift=-1mm] {#4};
}

\drawlayer{enc1}{\nEncOne}{\hsep}{}{neuronEncOne}
\drawlayer{enc2}{\nEncTwo}{2*\hsep}{}{neuronEncTwo}
\drawlayer{enc3}{\nEncThree}{3*\hsep}{}{neuronEncThree}

\drawlayer{latent}{\nLatent}{4*\hsep}{}{neuronLatent}

\drawlayer{dec1}{\nDecOne}{5*\hsep}{}{neuronDecOne}
\drawlayer{dec2}{\nDecTwo}{6*\hsep}{}{neuronDecTwo}
\drawlayer{dec3}{\nDecThree}{7*\hsep}{}{neuronDecThree}

\node[draw, fill=white, minimum width=1cm, minimum height=5.5cm, anchor=center] (inputBox) at (0,0.5) {\shortstack{$\phi_{1,1}^s$ \\ $\phi_{1,2}^s$ \\ $\vdots$ \\ $\phi_{n,m}^s$}};
\node[below=2mm of inputBox] {};

\node[draw, fill=white, minimum width=1cm, minimum height=5.5cm, anchor=center] (outputBox) at (8*\hsep,0.5) {\shortstack{$\hat{\phi}_{1,1}^s$ \\ $\hat{\phi}_{1,2}^s$ \\ $\vdots$ \\ $\hat{\phi}_{n,m}^s$}};
\node[below=2mm of outputBox] {};

\foreach \j in {1,...,\nEncOne} {
  \draw[-] (inputBox.south east) -- (enc1\j.west);
  \draw[-] (inputBox.east) -- (enc1\j.west);
  \draw[-] (inputBox.north east) -- (enc1\j.west);
}
\foreach \i in {1,...,\nEncOne} {
  \foreach \j in {1,...,\nEncTwo} {
    \draw[-] (enc1\i.east) -- (enc2\j.west);
  }
}
\foreach \i in {1,...,\nEncTwo} {
  \foreach \j in {1,...,\nEncThree} {
    \draw[-] (enc2\i.east) -- (enc3\j.west);
  }
}
\foreach \i in {1,...,\nEncThree} {
  \foreach \j in {1,...,\nLatent} {
    \draw[-] (enc3\i.east) -- (latent\j.west);
  }
}
\foreach \i in {1,...,\nLatent} {
  \foreach \j in {1,...,\nDecOne} {
    \draw[-] (latent\i.east) -- (dec1\j.west);
  }
}
\foreach \i in {1,...,\nDecOne} {
  \foreach \j in {1,...,\nDecTwo} {
    \draw[-] (dec1\i.east) -- (dec2\j.west);
  }
}
\foreach \i in {1,...,\nDecTwo} {
  \foreach \j in {1,...,\nDecThree} {
    \draw[-] (dec2\i.east) -- (dec3\j.west);
  }
}
\foreach \j in {1,...,\nDecThree} {
  \draw[-] (dec3\j.east) -- (outputBox.south west);
  \draw[-] (dec3\j.east) -- (outputBox.west);
  \draw[-] (dec3\j.east) -- (outputBox.north west);
}
\end{tikzpicture}
\end{center}

%% file: figures/net_FFNtoLS.tex
\begin{tikzpicture}[
    >=stealth,
    neuronEncOne/.style={circle, draw, fill=rwthb2, minimum size=8mm},
    neuronEncTwo/.style={circle, draw, fill=rwthb3, minimum size=8mm},
    neuronEncThree/.style={circle, draw, fill=rwthb4, minimum size=8mm},
    neuronLatent/.style={circle, draw, fill=rwth6, minimum size=8mm},
    neuronLatentFFN/.style={circle, draw, fill=rwth5!60, minimum size=8mm},
    neuronFFNOne/.style={circle, draw, fill=rwthg1!30, minimum size=8mm},
    neuronFFNTwo/.style={circle, draw, fill=rwth7!40, minimum size=8mm},
    neuronFFNThree/.style={circle, draw, fill=rwth7, minimum size=8mm},
    neuron/.style={circle, draw, minimum size=8mm}
]

\def\hsep{1.5cm}

\def\nEncOne{5}
\def\nEncTwo{4}
\def\nEncThree{3}
\def\nLatent{2}

\def\nFFNOne{4}
\def\nFFNTwo{4}
\def\nFFNThree{4}

\newcommand{\drawlayer}[5]{ 
  \foreach \i in {1,...,#2} {
    \node[#5] (#1\i) at (#3,{-#2/2 + \i}) {};
  }
  \node[below=0.2cm of #1#2, yshift=-1mm] {#4};
}

\drawlayer{latent}{\nLatent}{5*\hsep}{}{neuronLatent}

\drawlayer{FFN1}{\nFFNOne}{1*\hsep}{}{neuronFFNOne}
\drawlayer{FFN2}{\nFFNTwo}{2*\hsep}{}{neuronFFNTwo}
\drawlayer{FFN3}{\nFFNThree}{3*\hsep}{}{neuronFFNThree}
\drawlayer{LatentFFN}{\nLatent}{4*\hsep}{}{neuronLatentFFN}

\drawlayer{enc3}{\nEncThree}{6*\hsep}{}{neuronEncThree}
\drawlayer{enc2}{\nEncTwo}{7*\hsep}{}{neuronEncTwo}
\drawlayer{enc1}{\nEncOne}{8*\hsep}{}{neuronEncOne}

\node[draw, fill=white, minimum width=1cm, minimum height=5.5cm, anchor=center] (inputBox) at (9.3*\hsep,0.5) {\shortstack{$\phi_{1,1}\left(\bm{\theta}^S\right)$ \\ $\phi_{1,2}\left(\bm{\theta}^S\right)$ \\ $\vdots$ \\ $\phi_{n,m}\left(\bm{\theta}^S\right)$}};
\node[below=2mm of inputBox] {};

\node[draw, fill=white, minimum width=1cm, minimum height=2.5cm, anchor=center] (inputBoxFFN) at (0,0.5) {\shortstack{$\theta_{1}^s$ \\ $\vdots$ \\ $\theta_{k}^s$}};
\node[below=2mm of inputBoxFFN] {};

\foreach \i in {1,...,\nLatent} {
  \foreach \j in {1,...,\nEncThree} {
    \draw[-,dashed,gray] (latent\i.east) -- (enc3\j.west);
  }
}
\foreach \i in {1,...,\nEncThree} {
  \foreach \j in {1,...,\nEncTwo} {
    \draw[-,dashed,gray] (enc3\i.east) -- (enc2\j.west);
  }
}
\foreach \i in {1,...,\nEncTwo} {
  \foreach \j in {1,...,\nEncOne} {
    \draw[-,dashed,gray] (enc2\i.east) -- (enc1\j.west);
  }
}
\foreach \j in {1,...,\nEncOne} {
  \draw[-,dashed,gray] (enc1\j.east) -- (inputBox.south west);
  \draw[-,dashed,gray] (enc1\j.east) -- (inputBox.west);
  \draw[-,dashed,gray] (enc1\j.east) -- (inputBox.north west);
}
\foreach \j in {1,...,\nFFNOne} {
  \draw[-] (inputBoxFFN.north east) -- (FFN1\j.west);
  \draw[-] (inputBoxFFN.east) -- (FFN1\j.west);
  \draw[-] (inputBoxFFN.south east) -- (FFN1\j.west);
}
\foreach \i in {1,...,\nFFNOne} {
  \foreach \j in {1,...,\nFFNTwo} {
    \draw[-] (FFN1\i.east) -- (FFN2\j.west);
  }
}
\foreach \i in {1,...,\nFFNTwo} {
  \foreach \j in {1,...,\nFFNThree} {
    \draw[-] (FFN2\i.east) -- (FFN3\j.west);
  }
}
\foreach \i in {1,...,\nFFNThree} {
  \foreach \j in {1,...,\nLatent} {
    \draw[-] (FFN3\i.east) -- (LatentFFN\j.west);
  }
}
\foreach \j in {1,...,\nLatent} {
  \draw[->, shorten >= 3pt, shorten <= 3pt] (latent\j.west) -- (LatentFFN\j.east);
}
\end{tikzpicture}

%% file: figures/net_EndtoEnd.tex
\begin{center}
\begin{tikzpicture}[
    >=stealth,
    neuronDecOne/.style={circle, draw, fill=rwtho2, minimum size=8mm},
    neuronDecTwo/.style={circle, draw, fill=rwtho3, minimum size=8mm},
    neuronDecThree/.style={circle, draw, fill=rwtho4, minimum size=8mm},
    neuronLatentFFN/.style={circle, draw, fill=rwth5!60, minimum size=8mm},
    neuronFFNOne/.style={circle, draw, fill=rwthg1!30, minimum size=8mm},
    neuronFFNTwo/.style={circle, draw, fill=rwth7!40, minimum size=8mm},
    neuronFFNThree/.style={circle, draw, fill=rwth7, minimum size=8mm},
    neuron/.style={circle, draw, minimum size=8mm}
]

\def\hsep{1.5cm}

\def\nLatent{2}
\def\nDecOne{3}
\def\nDecTwo{4}
\def\nDecThree{5}

\def\nFFNOne{4}
\def\nFFNTwo{4}
\def\nFFNThree{4}

\newcommand{\drawlayer}[5]{ 
  \foreach \i in {1,...,#2} {
    \node[#5] (#1\i) at (#3,{-#2/2 + \i}) {};
  }
  \node[below=0.2cm of #1#2, yshift=-1mm] {#4};
}

\drawlayer{FFN1}{\nFFNOne}{1*\hsep}{}{neuronFFNOne}
\drawlayer{FFN2}{\nFFNTwo}{2*\hsep}{}{neuronFFNTwo}
\drawlayer{FFN3}{\nFFNThree}{3*\hsep}{}{neuronFFNThree}

\drawlayer{LatentFFN}{\nLatent}{4*\hsep}{}{neuronLatentFFN}

\drawlayer{dec1}{\nDecOne}{5*\hsep}{}{neuronDecOne}
\drawlayer{dec2}{\nDecTwo}{6*\hsep}{}{neuronDecTwo}
\drawlayer{dec3}{\nDecThree}{7*\hsep}{}{neuronDecThree}

\node[draw, fill=white, minimum width=1cm, minimum height=2.5cm, anchor=center] (inputBoxFFN) at (0,0.5) {\shortstack{$\theta_{1}^s$ \\ $\vdots$ \\ $\theta_{k}^s$}};
\node[below=2mm of inputBoxFFN] {};

\node[draw, fill=white, minimum width=1cm, minimum height=5.5cm, anchor=center] (outputBox) at (8*\hsep,0.5) {\shortstack{$\hat{\phi}_{1,1}^s$ \\ $\hat{\phi}_{1,2}^s$ \\ $\vdots$ \\ $\hat{\phi}_{n,m}^s$}};
\node[below=2mm of outputBox] {};

\foreach \j in {1,...,\nFFNOne} {
  \draw[-,dashed,gray] (inputBoxFFN.north east) -- (FFN1\j.west);
  \draw[-,dashed,gray] (inputBoxFFN.east) -- (FFN1\j.west);
  \draw[-,dashed,gray] (inputBoxFFN.south east) -- (FFN1\j.west);
}
\foreach \i in {1,...,\nFFNOne} {
  \foreach \j in {1,...,\nFFNTwo} {
    \draw[-,dashed,gray] (FFN1\i.east) -- (FFN2\j.west);
  }
}
\foreach \i in {1,...,\nFFNTwo} {
  \foreach \j in {1,...,\nFFNThree} {
    \draw[-,dashed,gray] (FFN2\i.east) -- (FFN3\j.west);
  }
}
\foreach \i in {1,...,\nFFNThree} {
  \foreach \j in {1,...,\nLatent} {
    \draw[-,dashed,gray] (FFN3\i.east) -- (LatentFFN\j.west);
  }
}
\foreach \i in {1,...,\nLatent} {
  \foreach \j in {1,...,\nDecOne} {
    \draw[-,dashed,gray] (LatentFFN\i.east) -- (dec1\j.west);
  }
}
\foreach \i in {1,...,\nDecOne} {
  \foreach \j in {1,...,\nDecTwo} {
    \draw[-,dashed,gray] (dec1\i.east) -- (dec2\j.west);
  }
}
\foreach \i in {1,...,\nDecTwo} {
  \foreach \j in {1,...,\nDecThree} {
    \draw[-,dashed,gray] (dec2\i.east) -- (dec3\j.west);
  }
}
\foreach \j in {1,...,\nDecThree} {
  \draw[-,dashed,gray] (dec3\j.east) -- (outputBox.south west);
  \draw[-,dashed,gray] (dec3\j.east) -- (outputBox.west);
  \draw[-,dashed,gray] (dec3\j.east) -- (outputBox.north west);
}
\end{tikzpicture}
\end{center}

%% file: figures/net_DecoupledForce.tex
\begin{center}
\begin{tikzpicture}[
    >=stealth,
    neuronEncOne/.style={circle, draw, fill=rwthb2, minimum size=8mm},
    neuronEncTwo/.style={circle, draw, fill=rwthb3, minimum size=8mm},
    neuronEncThree/.style={circle, draw, fill=rwthb4, minimum size=8mm},
    neuronEncROne/.style={circle, draw, fill=rwth3!20, minimum size=8mm},
    neuronEncRTwo/.style={circle, draw, fill=rwth3!40, minimum size=8mm},
    neuronEncRThree/.style={circle, draw, fill=rwth3!80, minimum size=8mm},    
    neuronReacOne/.style={circle, draw, fill=rwth13!30, minimum size=8mm},
    neuronReacTwo/.style={circle, draw, fill=rwth13!60, minimum size=8mm},
    neuronReacThree/.style={circle, draw, fill=rwth13, minimum size=8mm},
    neuronDecOne/.style={circle, draw, fill=rwtho2, minimum size=8mm},
    neuronDecTwo/.style={circle, draw, fill=rwtho3, minimum size=8mm},
    neuronDecThree/.style={circle, draw, fill=rwtho4, minimum size=8mm},
    neuronLatentFFN/.style={circle, draw, fill=rwth5!60, minimum size=8mm},
    neuronFFNOne/.style={circle, draw, fill=rwthg1!30, minimum size=8mm},
    neuronFFNTwo/.style={circle, draw, fill=rwth7!40, minimum size=8mm},
    neuronFFNThree/.style={circle, draw, fill=rwth7, minimum size=8mm},
    neuronLatent/.style={circle, draw, fill=rwth6, minimum size=8mm},
    neuron/.style={circle, draw, minimum size=8mm}
]

\def\hsep{1.5cm}

\def\nEncOne{4}
\def\nEncTwo{3}
\def\nEncThree{2}
\def\nLatent{2}
\def\nDecOne{2}
\def\nDecTwo{3}
\def\nDecThree{4}

\newcommand{\drawlayer}[6][0]{ 
  \foreach \i in {1,...,#3} {
    \node[#6] (#2\i) at (#4,{-#3/2 + \i + #1}) {};
  }
  \node[below=0.2cm of #2#3, yshift=-1mm, yshift=#1 cm] {#5};
}

\drawlayer[2]{enc1}{\nEncOne}{\hsep}{}{neuronEncOne}
\drawlayer[2]{enc2}{\nEncTwo}{2*\hsep}{}{neuronEncTwo}
\drawlayer[2]{enc3}{\nEncThree}{3*\hsep}{}{neuronEncThree}
\drawlayer[-2]{encR1}{\nEncOne}{\hsep}{}{neuronEncROne}
\drawlayer[-2]{encR2}{\nEncTwo}{2*\hsep}{}{neuronEncRTwo}
\drawlayer[-2]{encR3}{\nEncThree}{3*\hsep}{}{neuronEncRThree}

\drawlayer{latent}{\nLatent}{4*\hsep}{}{neuronLatent}

\drawlayer[2]{dec1}{\nDecOne}{5*\hsep}{}{neuronDecOne}
\drawlayer[2]{dec2}{\nDecTwo}{6*\hsep}{}{neuronDecTwo}
\drawlayer[2]{dec3}{\nDecThree}{7*\hsep}{}{neuronDecThree}
\drawlayer[-2]{reac1}{\nDecOne}{5*\hsep}{}{neuronReacOne}
\drawlayer[-2]{reac2}{\nDecTwo}{6*\hsep}{}{neuronReacTwo}
\drawlayer[-2]{reac3}{\nDecThree}{7*\hsep}{}{neuronReacThree}

\node[draw, fill=white, minimum width=1cm, minimum height=3.5cm, anchor=center] (inputBox) at (0,2.5) {\shortstack{$\phi_{1,1}^s$ \\ $\phi_{1,2}^s$ \\ $\vdots$ \\ $\phi_{n,m}^s$}};
\node[below=2mm of inputBox] {};
\node[draw, fill=white, minimum width=1cm, minimum height=3.5cm, anchor=center] (inputBoxReac) at (0,-1.5) {\shortstack{$f_{1,1}^s$ \\ $f_{1,2}^s$ \\ $\vdots$ \\ $f_{\alpha,\beta}^s$}};
\node[below=2mm of inputBoxReac] {};

\node[draw, fill=white, minimum width=1cm, minimum height=3.5cm, anchor=center] (outputBox) at (8*\hsep,2.5) {\shortstack{$\hat{\phi}_{1,1}^s$ \\ $\hat{\phi}_{1,2}^s$ \\ $\vdots$ \\ $\hat{\phi}_{n,m}^s$}};
\node[below=2mm of outputBox] {};
\node[draw, fill=white, minimum width=1cm, minimum height=3.5cm, anchor=center] (outputBoxReac) at (8*\hsep,-1.5) {\shortstack{$\hat{f}_{1,1}^s$ \\ $\hat{f}_{1,2}^s$ \\ $\vdots$ \\ $\hat{f}_{\alpha,\beta}^s$}};
\node[below=2mm of outputBoxReac] {};

\foreach \j in {1,...,\nEncOne} {
  \draw[-] (inputBox.south east) -- (enc1\j.west);
  \draw[-] (inputBox.east) -- (enc1\j.west);
  \draw[-] (inputBox.north east) -- (enc1\j.west);
}
\foreach \i in {1,...,\nEncOne} {
  \foreach \j in {1,...,\nEncTwo} {
    \draw[-] (enc1\i.east) -- (enc2\j.west);
  }
}
\foreach \i in {1,...,\nEncTwo} {
  \foreach \j in {1,...,\nEncThree} {
    \draw[-] (enc2\i.east) -- (enc3\j.west);
  }
}
\foreach \i in {1,...,\nEncThree} {
  \foreach \j in {1,...,\nLatent} {
    \draw[-] (enc3\i.east) -- (latent\j.west);
  }
}
\foreach \j in {1,...,\nEncOne} {
  \draw[-] (inputBoxReac.south east) -- (encR1\j.west);
  \draw[-] (inputBoxReac.east) -- (encR1\j.west);
  \draw[-] (inputBoxReac.north east) -- (encR1\j.west);
}
\foreach \i in {1,...,\nEncOne} {
  \foreach \j in {1,...,\nEncTwo} {
    \draw[-] (encR1\i.east) -- (encR2\j.west);
  }
}
\foreach \i in {1,...,\nEncTwo} {
  \foreach \j in {1,...,\nEncThree} {
    \draw[-] (encR2\i.east) -- (encR3\j.west);
  }
}
\foreach \i in {1,...,\nEncThree} {
  \foreach \j in {1,...,\nLatent} {
    \draw[-] (encR3\i.east) -- (latent\j.west);
  }
}
\foreach \i in {1,...,\nLatent} {
  \foreach \j in {1,...,\nDecOne} {
    \draw[-] (latent\i.east) -- (dec1\j.west);
  }
}
\foreach \i in {1,...,\nDecOne} {
  \foreach \j in {1,...,\nDecTwo} {
    \draw[-] (dec1\i.east) -- (dec2\j.west);
  }
}
\foreach \i in {1,...,\nDecTwo} {
  \foreach \j in {1,...,\nDecThree} {
    \draw[-] (dec2\i.east) -- (dec3\j.west);
  }
}
\foreach \j in {1,...,\nDecThree} {
  \draw[-] (dec3\j.east) -- (outputBox.south west);
  \draw[-] (dec3\j.east) -- (outputBox.west);
  \draw[-] (dec3\j.east) -- (outputBox.north west);
}
\foreach \i in {1,...,\nLatent} {
  \foreach \j in {1,...,\nDecOne} {
    \draw[-] (latent\i.east) -- (reac1\j.west);
  }
}
\foreach \i in {1,...,\nDecOne} {
  \foreach \j in {1,...,\nDecTwo} {
    \draw[-] (reac1\i.east) -- (reac2\j.west);
  }
}
\foreach \i in {1,...,\nDecTwo} {
  \foreach \j in {1,...,\nDecThree} {
    \draw[-] (reac2\i.east) -- (reac3\j.west);
  }
}
\foreach \j in {1,...,\nDecThree} {
  \draw[-] (reac3\j.east) -- (outputBoxReac.south west);
  \draw[-] (reac3\j.east) -- (outputBoxReac.west);
  \draw[-] (reac3\j.east) -- (outputBoxReac.north west);
}

\end{tikzpicture}
\end{center}

\begin{center}
\begin{tikzpicture}[
    >=stealth,
    neuronEncOne/.style={circle, draw, fill=rwth3!20, minimum size=8mm},
    neuronEncTwo/.style={circle, draw, fill=rwth3!40, minimum size=8mm},
    neuronEncThree/.style={circle, draw, fill=rwth3!80, minimum size=8mm},
    neuronReacOne/.style={circle, draw, fill=rwth13!30, minimum size=8mm},
    neuronReacTwo/.style={circle, draw, fill=rwth13!60, minimum size=8mm},
    neuronReacThree/.style={circle, draw, fill=rwth13, minimum size=8mm},
    neuronDecOne/.style={circle, draw, fill=rwtho2, minimum size=8mm},
    neuronDecTwo/.style={circle, draw, fill=rwtho3, minimum size=8mm},
    neuronDecThree/.style={circle, draw, fill=rwtho4, minimum size=8mm},
    neuronLatentFFN/.style={circle, draw, fill=rwth5!60, minimum size=8mm},
    neuronFFNOne/.style={circle, draw, fill=rwthg1!30, minimum size=8mm},
    neuronFFNTwo/.style={circle, draw, fill=rwth7!40, minimum size=8mm},
    neuronFFNThree/.style={circle, draw, fill=rwth7, minimum size=8mm},
    neuronLatent/.style={circle, draw, fill=rwth6, minimum size=8mm},
    neuron/.style={circle, draw, minimum size=8mm}
]

\def\hsep{1.5cm}

\def\nLatent{2}
\def\nDecOne{2}
\def\nDecTwo{3}
\def\nDecThree{4}

\def\nFFNOne{4}
\def\nFFNTwo{4}
\def\nFFNThree{4}

\newcommand{\drawlayer}[6][0]{ 
  \foreach \i in {1,...,#3} {
    \node[#6] (#2\i) at (#4,{-#3/2 + \i + #1}) {};
  }
  \node[below=0.2cm of #2#3, yshift=-1mm, yshift=#1 cm] {#5};
}

\drawlayer[0]{FFN1}{\nFFNOne}{1*\hsep}{}{neuronFFNOne}
\drawlayer[0]{FFN2}{\nFFNTwo}{2*\hsep}{}{neuronFFNTwo}
\drawlayer[0]{FFN3}{\nFFNThree}{3*\hsep}{}{neuronFFNThree}

\drawlayer{latent}{\nLatent}{4*\hsep}{}{neuronLatentFFN}

\drawlayer[2]{dec1}{\nDecOne}{5*\hsep}{}{neuronDecOne}
\drawlayer[2]{dec2}{\nDecTwo}{6*\hsep}{}{neuronDecTwo}
\drawlayer[2]{dec3}{\nDecThree}{7*\hsep}{}{neuronDecThree}
\drawlayer[-2]{reac1}{\nDecOne}{5*\hsep}{}{neuronReacOne}
\drawlayer[-2]{reac2}{\nDecTwo}{6*\hsep}{}{neuronReacTwo}
\drawlayer[-2]{reac3}{\nDecThree}{7*\hsep}{}{neuronReacThree}

\node[draw, fill=white, minimum width=1cm, minimum height=2.5cm, anchor=center] (inputBoxFFN) at (0,0.5) {\shortstack{$\theta_{1}^s$ \\ $\vdots$ \\ $\theta_{k}^s$}};
\node[below=2mm of inputBoxFFN] {};

\node[draw, fill=white, minimum width=1cm, minimum height=3.5cm, anchor=center] (outputBox) at (8*\hsep,2.5) {\shortstack{$\hat{\phi}_{1,1}^s$ \\ $\hat{\phi}_{1,2}^s$ \\ $\vdots$ \\ $\hat{\phi}_{n,m}^s$}};
\node[below=2mm of outputBox] {};
\node[draw, fill=white, minimum width=1cm, minimum height=3.5cm, anchor=center] (outputBoxReac) at (8*\hsep,-1.5) {\shortstack{$\hat{f}_{1,1}^s$ \\ $\hat{f}_{1,2}^s$ \\ $\vdots$ \\ $\hat{f}_{\alpha,\beta}^s$}};
\node[below=2mm of outputBox] {};

\foreach \j in {1,...,\nFFNOne} {
  \draw[-, dashed, gray] (inputBoxFFN.south east) -- (FFN1\j.west);
  \draw[-, dashed, gray] (inputBoxFFN.east) -- (FFN1\j.west);
  \draw[-, dashed, gray] (inputBoxFFN.north east) -- (FFN1\j.west);
}
\foreach \i in {1,...,\nFFNOne} {
  \foreach \j in {1,...,\nFFNTwo} {
    \draw[-, dashed, gray] (FFN1\i.east) -- (FFN2\j.west);
  }
}
\foreach \i in {1,...,\nFFNTwo} {
  \foreach \j in {1,...,\nFFNThree} {
    \draw[-, dashed, gray] (FFN2\i.east) -- (FFN3\j.west);
  }
}
\foreach \i in {1,...,\nFFNThree} {
  \foreach \j in {1,...,\nLatent} {
    \draw[-, dashed, gray] (FFN3\i.east) -- (latent\j.west);
  }
}
\foreach \i in {1,...,\nLatent} {
  \foreach \j in {1,...,\nDecOne} {
    \draw[-, dashed, gray] (latent\i.east) -- (dec1\j.west);
  }
}
\foreach \i in {1,...,\nDecOne} {
  \foreach \j in {1,...,\nDecTwo} {
    \draw[-, dashed, gray] (dec1\i.east) -- (dec2\j.west);
  }
}
\foreach \i in {1,...,\nDecTwo} {
  \foreach \j in {1,...,\nDecThree} {
    \draw[-, dashed, gray] (dec2\i.east) -- (dec3\j.west);
  }
}
\foreach \j in {1,...,\nDecThree} {
  \draw[-, dashed, gray] (dec3\j.east) -- (outputBox.south west);
  \draw[-, dashed, gray] (dec3\j.east) -- (outputBox.west);
  \draw[-, dashed, gray] (dec3\j.east) -- (outputBox.north west);
}
\foreach \i in {1,...,\nLatent} {
  \foreach \j in {1,...,\nDecOne} {
    \draw[-, dashed, gray] (latent\i.east) -- (reac1\j.west);
  }
}
\foreach \i in {1,...,\nDecOne} {
  \foreach \j in {1,...,\nDecTwo} {
    \draw[-, dashed, gray] (reac1\i.east) -- (reac2\j.west);
  }
}
\foreach \i in {1,...,\nDecTwo} {
  \foreach \j in {1,...,\nDecThree} {
    \draw[-, dashed, gray] (reac2\i.east) -- (reac3\j.west);
  }
}
\foreach \j in {1,...,\nDecThree} {
  \draw[-, dashed, gray] (reac3\j.east) -- (outputBoxReac.south west);
  \draw[-, dashed, gray] (reac3\j.east) -- (outputBoxReac.west);
  \draw[-, dashed, gray] (reac3\j.east) -- (outputBoxReac.north west);
}

\end{tikzpicture}
\end{center}

%% file: figures/net_multiphysics.tex
\begin{center}
\begin{tikzpicture}[
    >=stealth,
    neuronEncOne/.style={circle, draw, fill=rwthb2, minimum size=8mm},
    neuronEncTwo/.style={circle, draw, fill=rwthb3, minimum size=8mm},
    neuronEncThree/.style={circle, draw, fill=rwthb4, minimum size=8mm},
    neuronLatent/.style={circle, draw, fill=rwth6, minimum size=8mm},
    neuronDecOne/.style={circle, draw, fill=rwtho2, minimum size=8mm},
    neuronDecTwo/.style={circle, draw, fill=rwtho3, minimum size=8mm},
    neuronDecThree/.style={circle, draw, fill=rwtho4, minimum size=8mm},
    neuron/.style={circle, draw, minimum size=8mm}
]

\def\hsep{1.5cm}

\def\nEncOne{4}
\def\nEncTwo{3}
\def\nEncThree{2}
\def\nLatent{2}
\def\nDecOne{3}
\def\nDecTwo{4}
\def\nDecThree{5}

\newcommand{\drawlayer}[6][0]{ 
  \foreach \i in {1,...,#3} {
    \node[#6] (#2\i) at (#4,{-#3/2 + \i + #1}) {};
  }
  \node[below=0.2cm of #2#3, yshift=-1mm, yshift=#1 cm] {#5};
}

\drawlayer[2]{enc11}{\nEncOne}{\hsep}{}{neuronEncOne}
\drawlayer[2]{enc12}{\nEncTwo}{2*\hsep}{}{neuronEncTwo}
\drawlayer[2]{enc13}{\nEncThree}{3*\hsep}{}{neuronEncThree}
\drawlayer[-2]{enc21}{\nEncOne}{\hsep}{}{neuronEncOne}
\drawlayer[-2]{enc22}{\nEncTwo}{2*\hsep}{}{neuronEncTwo}
\drawlayer[-2]{enc23}{\nEncThree}{3*\hsep}{}{neuronEncThree}

\drawlayer[1]{latent1}{\nLatent}{4*\hsep}{}{neuronLatent}
\drawlayer[-1]{latent2}{\nLatent}{4*\hsep}{}{neuronLatent}

\drawlayer{dec1}{\nDecOne}{5*\hsep}{}{neuronDecOne}
\drawlayer{dec2}{\nDecTwo}{6*\hsep}{}{neuronDecTwo}
\drawlayer{dec3}{\nDecThree}{7*\hsep}{}{neuronDecThree}

\node[draw, fill=white, minimum width=1cm, minimum height=3.5cm, anchor=center] (inputBox1) at (0,2.5) {\shortstack{$\prescript{1}{}{\phi_{1,1}^s}$ \\ $\prescript{1}{}{\phi_{1,2}^s}$ \\ $\vdots$ \\ $\prescript{1}{}{\phi_{n,m}^s}$}};
\node[below=2mm of inputBox] {};
\node[draw, fill=white, minimum width=1cm, minimum height=3.5cm, anchor=center] (inputBox2) at (0,-1.5) {\shortstack{$\prescript{2}{}{\phi_{1,1}^s}$ \\ $\prescript{2}{}{\phi_{1,2}^s}$ \\ $\vdots$ \\ $\prescript{2}{}{\phi_{n,m}^s}$}};
\node[below=2mm of inputBox] {};

\node[draw, fill=white, minimum width=1cm, minimum height=6.5cm, anchor=center] (outputBox) at (8*\hsep,0.5) {\shortstack{$\prescript{1}{}{\hat{\phi}}_{1,1}^s$ \\ $\prescript{1}{}{\hat{\phi}_{1,2}^s}$ \\ $\vdots$ \\ $\prescript{1}{}{\hat{\phi}_{n,m}^s}$ \\ $\prescript{2}{}{\hat{\phi}_{1,1}^s}$ \\ $\prescript{2}{}{\hat{\phi}_{1,2}^s}$ \\ $\vdots$ \\ $\prescript{2}{}{\hat{\phi}_{n,m}^s}$}};
\node[below=2mm of outputBox] {};

\foreach \j in {1,...,\nEncOne} {
  \draw[-] (inputBox1.south east) -- (enc11\j.west);
  \draw[-] (inputBox1.east) -- (enc11\j.west);
  \draw[-] (inputBox1.north east) -- (enc11\j.west);
}
\foreach \i in {1,...,\nEncOne} {
  \foreach \j in {1,...,\nEncTwo} {
    \draw[-] (enc11\i.east) -- (enc12\j.west);
  }
}
\foreach \i in {1,...,\nEncTwo} {
  \foreach \j in {1,...,\nEncThree} {
    \draw[-] (enc12\i.east) -- (enc13\j.west);
  }
}
\foreach \i in {1,...,\nEncThree} {
  \foreach \j in {1,...,\nLatent} {
    \draw[-] (enc13\i.east) -- (latent1\j.west);
  }
}
\foreach \i in {1,...,\nLatent} {
  \foreach \j in {1,...,\nDecOne} {
    \draw[-] (latent1\i.east) -- (dec1\j.west);
  }
}
\foreach \j in {1,...,\nEncOne} {
  \draw[-] (inputBox2.south east) -- (enc21\j.west);
  \draw[-] (inputBox2.east) -- (enc21\j.west);
  \draw[-] (inputBox2.north east) -- (enc21\j.west);
}
\foreach \i in {1,...,\nEncOne} {
  \foreach \j in {1,...,\nEncTwo} {
    \draw[-] (enc21\i.east) -- (enc22\j.west);
  }
}
\foreach \i in {1,...,\nEncTwo} {
  \foreach \j in {1,...,\nEncThree} {
    \draw[-] (enc22\i.east) -- (enc23\j.west);
  }
}
\foreach \i in {1,...,\nEncThree} {
  \foreach \j in {1,...,\nLatent} {
    \draw[-] (enc23\i.east) -- (latent2\j.west);
  }
}
\foreach \i in {1,...,\nLatent} {
  \foreach \j in {1,...,\nDecOne} {
    \draw[-] (latent2\i.east) -- (dec1\j.west);
  }
}
\foreach \i in {1,...,\nDecOne} {
  \foreach \j in {1,...,\nDecTwo} {
    \draw[-] (dec1\i.east) -- (dec2\j.west);
  }
}
\foreach \i in {1,...,\nDecTwo} {
  \foreach \j in {1,...,\nDecThree} {
    \draw[-] (dec2\i.east) -- (dec3\j.west);
  }
}
\foreach \j in {1,...,\nDecThree} {
  \draw[-] (dec3\j.east) -- (outputBox.south west);
  \draw[-] (dec3\j.east) -- (outputBox.west);
  \draw[-] (dec3\j.east) -- (outputBox.north west);
}
\end{tikzpicture}
\end{center}
\begin{center}
\begin{tikzpicture}[
    >=stealth,
    neuronDecOne/.style={circle, draw, fill=rwtho2, minimum size=8mm},
    neuronDecTwo/.style={circle, draw, fill=rwtho3, minimum size=8mm},
    neuronDecThree/.style={circle, draw, fill=rwtho4, minimum size=8mm},
    neuronLatentFFN/.style={circle, draw, fill=rwth5!60, minimum size=8mm},
    neuronFFNOne/.style={circle, draw, fill=rwthg1!30, minimum size=8mm},
    neuronFFNTwo/.style={circle, draw, fill=rwth7!40, minimum size=8mm},
    neuronFFNThree/.style={circle, draw, fill=rwth7, minimum size=8mm},
    neuron/.style={circle, draw, minimum size=8mm}
]

\def\hsep{1.5cm}

\def\nLatent{2}
\def\nDecOne{3}
\def\nDecTwo{4}
\def\nDecThree{5}

\def\nFFNOne{3}
\def\nFFNTwo{3}
\def\nFFNThree{3}

\newcommand{\drawlayer}[6][0]{ 
  \foreach \i in {1,...,#3} {
    \node[#6] (#2\i) at (#4,{-#3/2 + \i + #1}) {};
  }
  \node[below=0.2cm of #2#3, yshift=-1mm, yshift=#1 cm] {#5};
}

\drawlayer[1.7]{FFN11}{\nFFNOne}{1*\hsep}{}{neuronFFNOne}
\drawlayer[1.7]{FFN12}{\nFFNTwo}{2*\hsep}{}{neuronFFNTwo}
\drawlayer[1.7]{FFN13}{\nFFNThree}{3*\hsep}{}{neuronFFNThree}

\drawlayer[-1.7]{FFN21}{\nFFNOne}{1*\hsep}{}{neuronFFNOne}
\drawlayer[-1.7]{FFN22}{\nFFNTwo}{2*\hsep}{}{neuronFFNTwo}
\drawlayer[-1.7]{FFN23}{\nFFNThree}{3*\hsep}{}{neuronFFNThree}

\drawlayer[1]{LatentFFN1}{\nLatent}{4*\hsep}{}{neuronLatentFFN}

\drawlayer[-1]{LatentFFN2}{\nLatent}{4*\hsep}{}{neuronLatentFFN}

\drawlayer{dec1}{\nDecOne}{5*\hsep}{}{neuronDecOne}
\drawlayer{dec2}{\nDecTwo}{6*\hsep}{}{neuronDecTwo}
\drawlayer{dec3}{\nDecThree}{7*\hsep}{}{neuronDecThree}

\node[draw, fill=white, minimum width=1cm, minimum height=2.5cm, anchor=center] (inputBoxFFN) at (0,0.5) {\shortstack{$\theta_{1}^s$ \\ $\vdots$ \\ $\theta_{k}^s$}};
\node[below=2mm of inputBoxFFN] {};

\node[draw, fill=white, minimum width=1cm, minimum height=6.5cm, anchor=center] (outputBox) at (8*\hsep,0.5) {\shortstack{$\prescript{1}{}{\hat{\phi}}_{1,1}^s$ \\ $\prescript{1}{}{\hat{\phi}_{1,2}^s}$ \\ $\vdots$ \\ $\prescript{1}{}{\hat{\phi}_{n,m}^s}$ \\ $\prescript{2}{}{\hat{\phi}_{1,1}^s}$ \\ $\prescript{2}{}{\hat{\phi}_{1,2}^s}$ \\ $\vdots$ \\ $\prescript{2}{}{\hat{\phi}_{n,m}^s}$}};

\foreach \j in {1,...,\nFFNOne} {
  \draw[-,dashed,gray] (inputBoxFFN.north east) -- (FFN11\j.west);
  \draw[-,dashed,gray] (inputBoxFFN.east) -- (FFN11\j.west);
  \draw[-,dashed,gray] (inputBoxFFN.south east) -- (FFN11\j.west);
  \draw[-,dashed,gray] (inputBoxFFN.north east) -- (FFN21\j.west);
  \draw[-,dashed,gray] (inputBoxFFN.east) -- (FFN21\j.west);
  \draw[-,dashed,gray] (inputBoxFFN.south east) -- (FFN21\j.west);
}
\foreach \i in {1,...,\nFFNOne} {
  \foreach \j in {1,...,\nFFNTwo} {
    \draw[-,dashed,gray] (FFN11\i.east) -- (FFN12\j.west);
  }
}
\foreach \i in {1,...,\nFFNTwo} {
  \foreach \j in {1,...,\nFFNThree} {
    \draw[-,dashed,gray] (FFN12\i.east) -- (FFN13\j.west);
  }
}
\foreach \i in {1,...,\nFFNThree} {
  \foreach \j in {1,...,\nLatent} {
    \draw[-,dashed,gray] (FFN13\i.east) -- (LatentFFN1\j.west);
  }
}
\foreach \i in {1,...,\nLatent} {
  \foreach \j in {1,...,\nDecOne} {
    \draw[-,dashed,gray] (LatentFFN1\i.east) -- (dec1\j.west);
  }
}
\foreach \i in {1,...,\nFFNOne} {
  \foreach \j in {1,...,\nFFNTwo} {
    \draw[-,dashed,gray] (FFN21\i.east) -- (FFN22\j.west);
  }
}
\foreach \i in {1,...,\nFFNTwo} {
  \foreach \j in {1,...,\nFFNThree} {
    \draw[-,dashed,gray] (FFN22\i.east) -- (FFN23\j.west);
  }
}
\foreach \i in {1,...,\nFFNThree} {
  \foreach \j in {1,...,\nLatent} {
    \draw[-,dashed,gray] (FFN23\i.east) -- (LatentFFN2\j.west);
  }
}
\foreach \i in {1,...,\nLatent} {
  \foreach \j in {1,...,\nDecOne} {
    \draw[-,dashed,gray] (LatentFFN2\i.east) -- (dec1\j.west);
  }
}
\foreach \i in {1,...,\nDecOne} {
  \foreach \j in {1,...,\nDecTwo} {
    \draw[-,dashed,gray] (dec1\i.east) -- (dec2\j.west);
  }
}
\foreach \i in {1,...,\nDecTwo} {
  \foreach \j in {1,...,\nDecThree} {
    \draw[-,dashed,gray] (dec2\i.east) -- (dec3\j.west);
  }
}
\foreach \j in {1,...,\nDecThree} {
  \draw[-,dashed,gray] (dec3\j.east) -- (outputBox.south west);
  \draw[-,dashed,gray] (dec3\j.east) -- (outputBox.west);
  \draw[-,dashed,gray] (dec3\j.east) -- (outputBox.north west);
}
\end{tikzpicture}
\end{center}

%% file: sections/section3.tex
\section{Numerical results}
\label{sec3:1}
In the following, we investigate the combined use of Autoencoders and latent space prediction networks, along with their specialized variants tailored for force prediction and multi-field problems.

We begin in Section~\ref{sec3:2} with a unit cell boundary value problem, in which we additionally investigate the prediction of reaction forces at Dirichlet boundaries. 
Subsequently, in Section~\ref{sec3:3}, we study a plate with an elliptical hole, where the geometry is nonlinearly parameterized, and a second Autoencoder architecture is introduced to predict the corresponding mesh morphing. 
Finally, we examine a thermo-mechanical multi-field problem in Section~\ref{sec3:4}.

Across all studies, we apply a consistent regularization strategy based on elastic net regularization, employing a penalty factor of $10^{-7}$ for both $L_1$ and $L_2$ terms. 
The learning rate is uniformly set to $10^{-3}$ throughout all experiments. 
These choices are applied to both the unsupervised Autoencoder and the supervised latent space prediction network. 
Additionally, we employ the \texttt{Adam} optimizer for training.

It is important to note that the choices for the regularization parameters, learning rate, number of hidden layers, and the corresponding number of neurons per layer are heuristic in nature. 
A comprehensive hyperparameter study is beyond the scope of this work.

The neural network models are implemented in \texttt{JAX}, utilizing the open-source \texttt{Flax} neural network library. 
The optimization routines are provided by the \texttt{Optax} library.
%
\subsection{Unit cell}
\label{sec3:2}
\begin{figure}
    \centering
    \input{figures/UC_sketch}
    \caption{Sketch of the three-dimensional unit cell with inclusions. 
    The dimensions of the unit cell are $3\times 3 \times 1~\mathrm{mm}^3$. 
    The diameter of one inclusion is $2~\mathrm{mm}$. 
    The horizontal and vertical displacements are parameterized at the top surface, while the remaining components of the displacement at the top and the entire displacements the bottom are considered to be clamped. In addition, the $z-$displacement component is set to zero at the rear and front surfaces (in $z-$direction).}
    \label{fig:UC}
\end{figure}
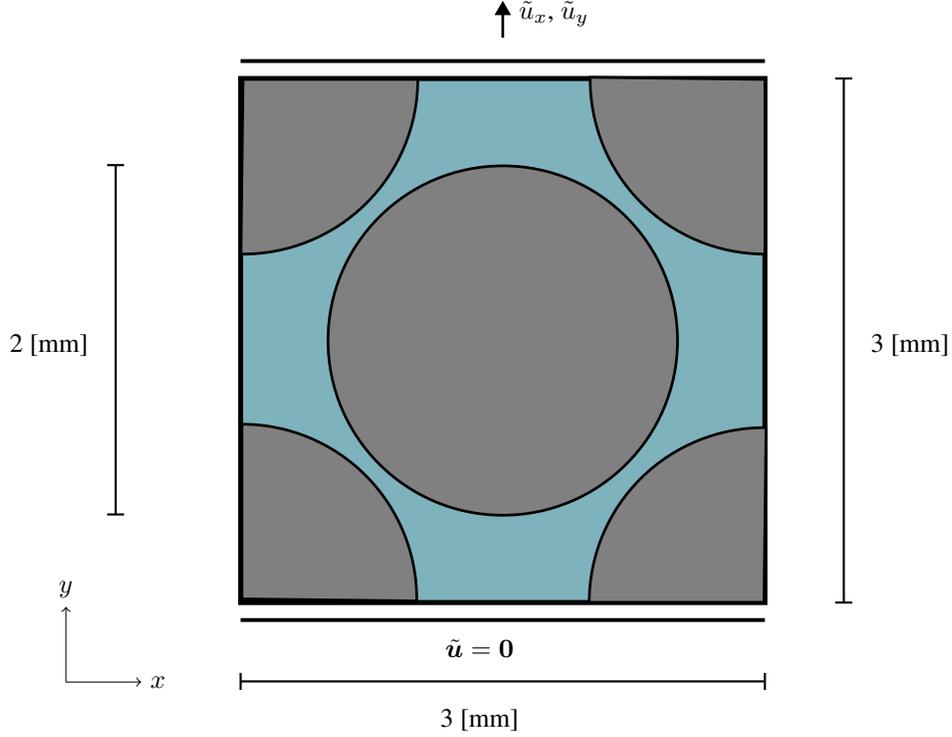

\paragraph{Boundary value problem.} To begin with, we investigate a unit cell of a heterogeneous composite material consisting of inclusions and a matrix material, as depicted in Figure~\ref{fig:UC}. 
The lower surface of the specimen is fully clamped, i.e., all displacement components are constrained to zero, while we parameterize the displacements in $x$- and $y$-direction at the top surface (see below), respectively. 
In addition, the displacement in $z$-direction at the front ($z=1~\mathrm{mm}$) as well as the rear surface ($z=0~\mathrm{mm}$) is constrained to be zero.

For the matrix material, in accordance with~\cite{HartmannNeff2003}, we choose a polyconvex Helmholtz free energy $\psi$ based on the Gent model~\cite{Gent1996}
\begin{equation}
    \psi = - \frac{\mu_M\,J_M}{2}\,\ln\left( 1 - \frac{I_1'-3}{J_M} \right)
           + \frac{\kappa_M}{4} \left( I_3 - 1 - \ln(I_3) \right),
\end{equation}
where the material parameters are given as $\mu_M = 100~\mathrm{MPa}$, $\kappa_M = 400~\mathrm{MPa}$, and $J_M = 10~[-]$. 
Furthermore, the invariants are defined as $I_1 = \mathrm{tr}\,\bm{C}$, $I_3 = \det\bm{C}$, and $I_1' = I_1 / I_3^{1/3}$, where $\bm{C}$ is the right Cauchy-Green deformation tensor.

Analogously, the material behavior of the inclusions is given by
\begin{equation}
    \psi = \frac{\mu_I}{2}\left( I_1' - 3 \right) 
         + \frac{\kappa_I}{4} \left( I_3 - 1 - \ln(I_3) \right),
\end{equation}
with the material parameters $\kappa_I = 400~\mathrm{MPa}$ and $\mu_I$ being parameterized (see below).

\paragraph{Parametrization and discretization.} In this first study, we explore the reduction capabilities in the case of parameterized loading as well as the stiffness ratio between matrix and inclusions. 
Therefore, we parameterize the nonlinear solution space by two natural coordinates $\bm{\theta} = (\xi, \eta) \in [0,1]^2$, i.e.,
\begin{gather}
    \tilde{u}_x = 2\,\cos(90^\circ - 90^\circ\,\xi), \quad 
    \tilde{u}_y = 2\,\sin(90^\circ - 90^\circ\,\xi), \\
    \mu_I = 150 + 750\,\eta^2~\mathrm{MPa},
\end{gather}
with the angle of the prescribed displacements is given in degree, see also Figure~\ref{fig:UC}.
For training the Autoencoder, we compute 25 snapshots by equally discretizing the natural coordinates as 
$\xi, \eta \in \left\{ 0, \frac{1}{4}, \frac{1}{2}, \frac{3}{4}, 1 \right\}$.

We employ a standard finite element formulation based on low-order trilinear hexahedral elements. 
The computational mesh comprises 7752 nodes and 5577 three-dimensional elements, leading to a total of 23256 degrees of freedom. 
Of these, 18580 degrees of freedom remain \textit{active}, i.e., they are not constrained by the imposed boundary conditions.

In this example, we evaluate the unit cell at $\xi=0.55$ and $\eta=0.35$.
%
\subsubsection{Prediction of solution field}
\label{sec3:2:1}
We begin by evaluating the performance of the proposed architecture in predicting the solution field of the unit cell, where the network is trained exclusively on solution data, i.e., the displacement field. 
The force-augmented extension, which incorporates additional force information during training, is analyzed in the following section.

\paragraph{Network topology.} The topology of the employed Autoencoder comprises an encoder and a decoder, each consisting of five hidden layers. 
The dimensionality of the latent space is fixed to four.

More specifically, the encoder architecture, including the activation functions, is given by
\begin{equation*}
    18580 \xrightarrow{\mathrm{GELU}} 1024 \xrightarrow{\mathrm{GELU}} 512 \xrightarrow{\mathrm{GELU}} 128 \xrightarrow{\mathrm{GELU}} 32 \xrightarrow{\mathrm{GELU}} 16 \xrightarrow{\phantom{\mathrm{GELU}}} 4,
\end{equation*}
where each arrow indicates a fully connected layer followed by the Gaussian Error Linear Unit (GELU) activation function~\cite{HendrycksGimpel2023}, except for the final mapping into the latent space of dimension four.

The decoder architecture is defined analogously as
\begin{equation*}
    4 \xrightarrow{\mathrm{SiLU}} 16 \xrightarrow{\mathrm{SiLU}} 16 \xrightarrow{\mathrm{SiLU}} 16 \xrightarrow{\mathrm{SiLU}} 128 \xrightarrow{\mathrm{GELU}} 1024 \xrightarrow{\phantom{\mathrm{SiLU}}} 18580,
\end{equation*}
where the Sigmoid Linear Unit (SiLU) activation function is employed after each fully connected layer, except the final hidden layer.

Furthermore, the topology of the network mapping from the natural coordinates to the solution space is chosen as
\begin{equation*}
    2 \xrightarrow{\mathrm{GELU}} 16 \xrightarrow{\mathrm{GELU}} 16 \xrightarrow{\mathrm{GELU}} 16 \xrightarrow{\mathrm{GELU}} 16 \xrightarrow{\phantom{\mathrm{GELU}}} 4,
\end{equation*}
comprising four hidden layers with GELU activations.\newline

\textbf{Results.} The training losses for both the unsupervised Autoencoder and the supervised latent space prediction network are shown on the left side of Figure~\ref{fig:UC_loss}. 
During training, both models converge to a loss on the order of $10^{-3}$, which we consider indicative of successful minimization.

Using the trained weights and biases, the model is evaluated at the parameter point $\xi = 0.55$ and $\eta = 0.35$, as depicted in Figure~\ref{fig:UC_prediction_disp_only}. 
To enhance the visibility of the stiffness contrast between matrix and inclusion, we visualize the gradient of the displacement field, corresponding to the strain field, along the vertical direction.

The predicted solution shows good overall agreement with the high-fidelity reference obtained from full-order finite element simulations, both qualitatively and quantitatively. 
Notably, the predicted locations of peak strain coincide with those of the reference solution.

Nevertheless, the predicted strain gradient field appears significantly less smooth compared to the reference. 
We attribute this discrepancy to the simplicity of the employed network architecture, which consists solely of classical fully connected layers without any convolutional or graph-based components. 
Furthermore, while the outer contours of the predicted solution generally align well with the reference, some deviations are noticeable near the top-right surface of the specimen. 
Despite these limitations, the results are deemed satisfactory given the architectural simplicity of the network.
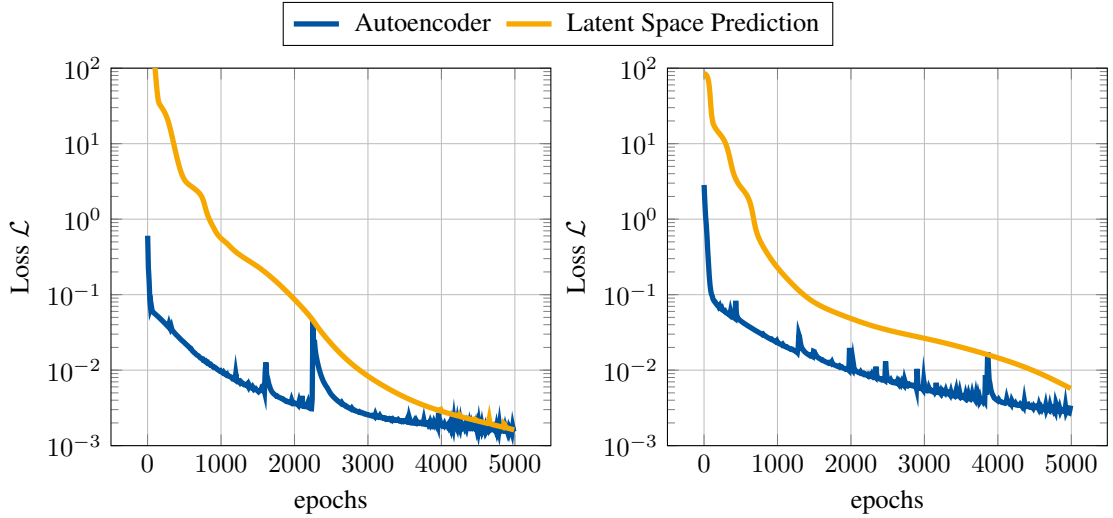
\begin{figure}
    \centering
    \input{figures/UC/UC_loss}
\caption{\textbf{Unit cell.} Training loss over the course of training for both neural networks, namely the Autoencoder and the Latent Space Prediction network, each trained for a maximum of 5000 epochs. 
Left: The end-to-end model is trained solely on solution field data, i.e., the displacement field; see Section~\ref{sec3:2:1}. 
Right: The end-to-end model is trained in a force-augmented manner by additionally incorporating the force terms at the Dirichlet boundaries during training; see Section~\ref{sec3:2:2}.}
    \label{fig:UC_loss}
\end{figure}
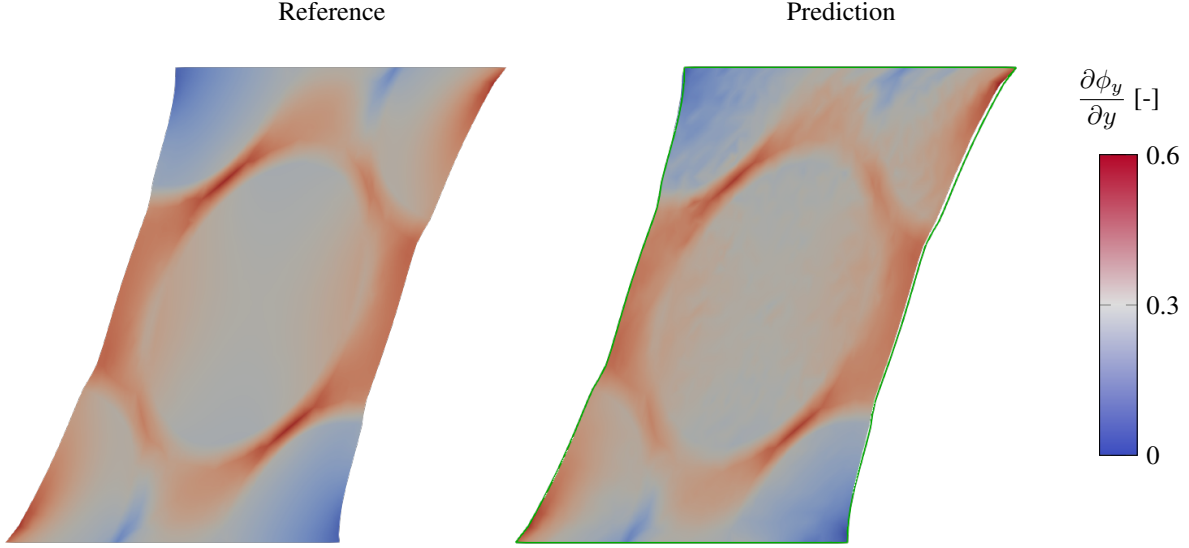
\begin{figure}
    \hspace*{-1cm}
    \centering
    \input{figures/UC/pure_solution}
\caption{Gradient of the $y$-component of the solution field $\bm{\phi}$ in $y$-direction, computed using ParaView's \texttt{Gradient filter}~\cite{AhrensGeveciEtAl2005}. 
The natural coordinates are evaluated at $\xi = 0.55$ and $\eta = 0.35$. 
On the left, the reference solution obtained from the full-order model is shown; the right shows the prediction of the trained end-to-end surrogate model. 
Green contours indicate the deformed reference configuration.}
\label{fig:UC_prediction_disp_only}
\end{figure}
%
\subsubsection{Prediction of force terms}
\label{sec3:2:2}
Next, we investigate the prediction of force terms at the Dirichlet boundaries. To this end, we employ the proposed force-augmented end-to-end model, which explicitly incorporates the force information at the Dirichlet boundaries during training.

For comparison, we also consider a force-reconstructed approach: 
following the unsupervised training of the Autoencoder on the solution field alone and the supervised training to predict the Latent space (cf. Section~\ref{sec3:2:1}), we train an additional feed-forward network in a supervised manner. 
This network takes the latent space representation as input and is tasked with predicting the corresponding force terms at the boundary.

Once trained, the force-reconstructed network enables the evaluation of the boundary forces for any natural coordinates within the parametric space $\bm{\theta}$.

\paragraph{Network topology.} For encoding and decoding the solution field, we retain the same network topology as introduced in the previous section. 
The architecture of the latent space prediction network also remains unchanged. 
However, a key distinction in the present setup is the use of a shared latent space representation for both displacement and force data.

To map the force terms onto this shared latent representation, we employ the following encoder architecture
\begin{equation*}
    4676 \xrightarrow{\mathrm{GELU}} 128 \xrightarrow{\mathrm{GELU}} 32 \xrightarrow{\mathrm{GELU}} 16 \xrightarrow{\phantom{\mathrm{GELU}}} 4,
\end{equation*}
and decode it back using the following decoder structure
\begin{equation*}
    4 \xrightarrow{\mathrm{SiLU}} 16 \xrightarrow{\mathrm{SiLU}} 32 \xrightarrow{\mathrm{SiLU}} 128 \xrightarrow{\phantom{\mathrm{SiLU}}} 4676.
\end{equation*}
It is important to highlight that, although force data is incorporated in this setting, the latent dimensionality is kept identical to the previous study. 
Consequently, the compression ratio is even higher, increasing the challenge of representing both displacement and force information jointly.

For the force-reconstruction network, which maps the latent space, which was solely trained on the displacement field, back to the force terms at the Dirichlet boundary, the architecture is given by
\begin{equation*}
\begin{split}
    4 \xrightarrow{} 8 &\xrightarrow{\mathrm{ReLU}} 16 \xrightarrow{\mathrm{ReLU}} 32 \xrightarrow{\mathrm{ReLU}} 64 \xrightarrow{\mathrm{ReLU}} 128 \xrightarrow{\mathrm{ReLU}} 256 \\
    &\xrightarrow{\mathrm{ReLU}} 512 \xrightarrow{\mathrm{ReLU}} 1024 \xrightarrow{\phantom{\mathrm{ReLU}}} 4676,
\end{split}
\end{equation*}
where $\mathrm{ReLU}$ denotes the Rectified Linear Unit activation function. 
This considerably deeper and wider architecture was chosen after observing that simpler networks failed to accurately reconstruct the force fields across the full parametric space. 
The increased complexity is therefore intended to mitigate possible underfitting and increase the model's representational capability.

\paragraph{Results.} The unsupervised and supervised training progress for both networks in the force-augmented setting is depicted in Figure~\ref{fig:UC_loss} (right). 
Compared to the model trained solely on solution field data, the force-augmented variant exhibits slightly higher loss values after 5000 epochs; nevertheless, the loss is effectively minimized to the order of $10^{-3}$.

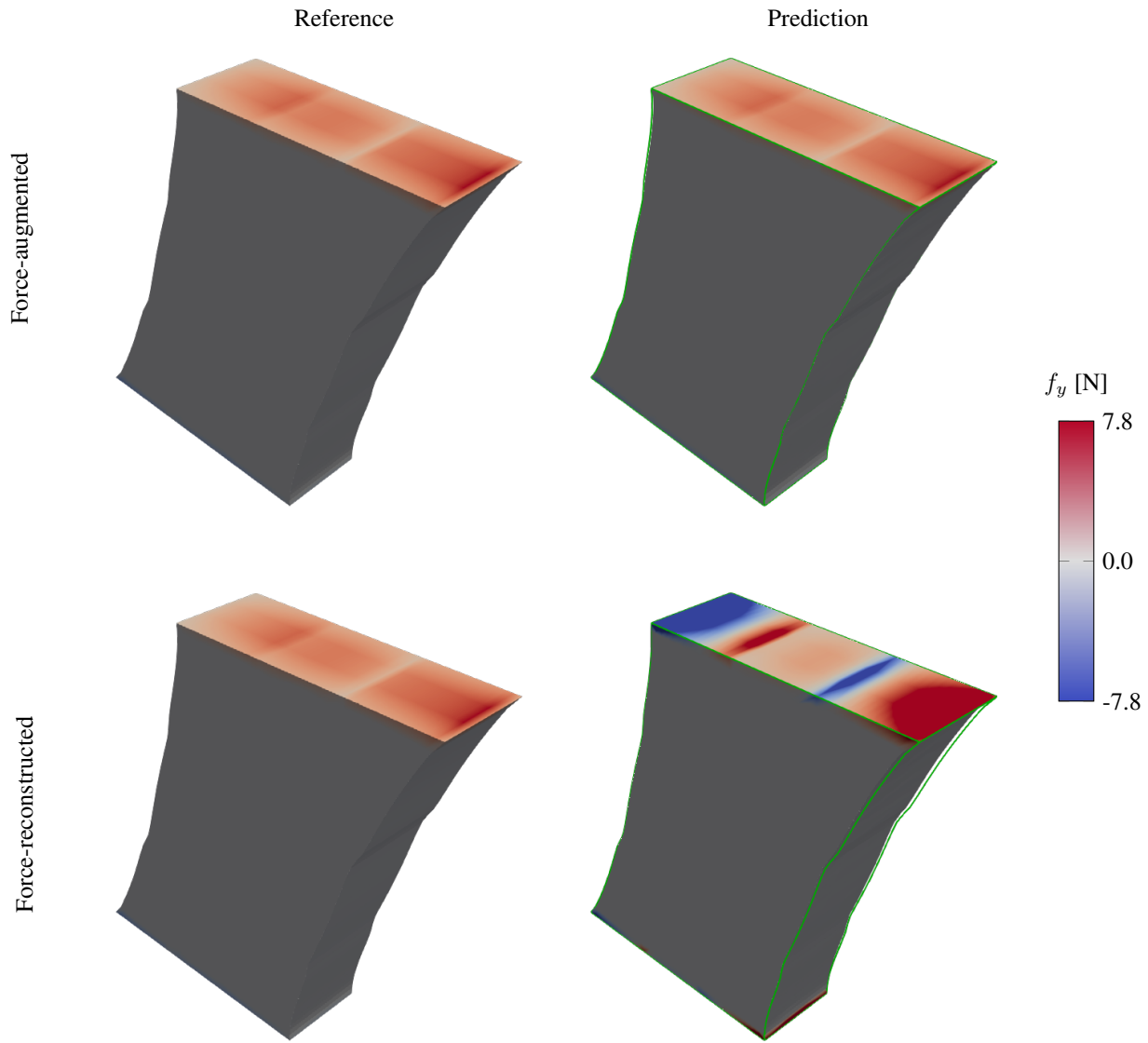
\begin{figure}
    \hspace*{-1cm}
    \centering
    \input{figures/UC/force/UC_force}
\caption{Force components at the Dirichlet boundary in the $y$-direction, evaluated at the natural coordinates $\xi = 0.55$ and $\eta = 0.35$. 
The reference is obtained from a high-fidelity full-order finite element simulation. 
Top: Predicted forces computed using the end-to-end force-augmented neural model. 
Bottom: Reconstructed forces obtained from the latent space trained solely on displacement field data, without explicit force supervision.}
    \label{fig:UC_prediction_force}
\end{figure}

Figure~\ref{fig:UC_prediction_force} presents a comparison between the reference solution and the predictions from both the force-augmented and force-reconstructed models. 
The force-augmented network demonstrates accurate prediction of the force terms, both qualitatively and quantitatively. 
Conversely, the force-reconstructed model performs significantly worse: it fails to replicate the qualitative contour shapes and exhibits predicted force magnitudes roughly an order of magnitude larger than the reference values. 
Given the complexity and depth of the force-reconstruction network architecture, this discrepancy is unlikely to be caused by insufficient model capacity.

\begin{figure}
    \hspace*{-1cm}
    \centering
    \input{figures/UC/force/UC_error_disp}
\caption{Prediction error between the full-order reference solution and the non-intrusive prediction for networks trained solely on solution field data (left) and the force-augmented version of the end-to-end model (right).
The contours are plotted on the undeformed configuration.
The operator $\mathrm{Err}(\hat{\bm{\phi}}, \bm{\phi})$ refers to the Euclidean norm, $\lVert \bullet \rVert$, of the difference $\hat{\bm{\phi}}-\bm{\phi}$ between the full-order reference solution $\bm{\phi}$ and its predicted counterpart $\hat{\bm{\phi}}$.
Notably, the errors at the top and bottom surfaces are zero because the solution field is entirely prescribed at these surfaces and thus a priori known.}
    \label{fig:UC_error_disp}
\end{figure}
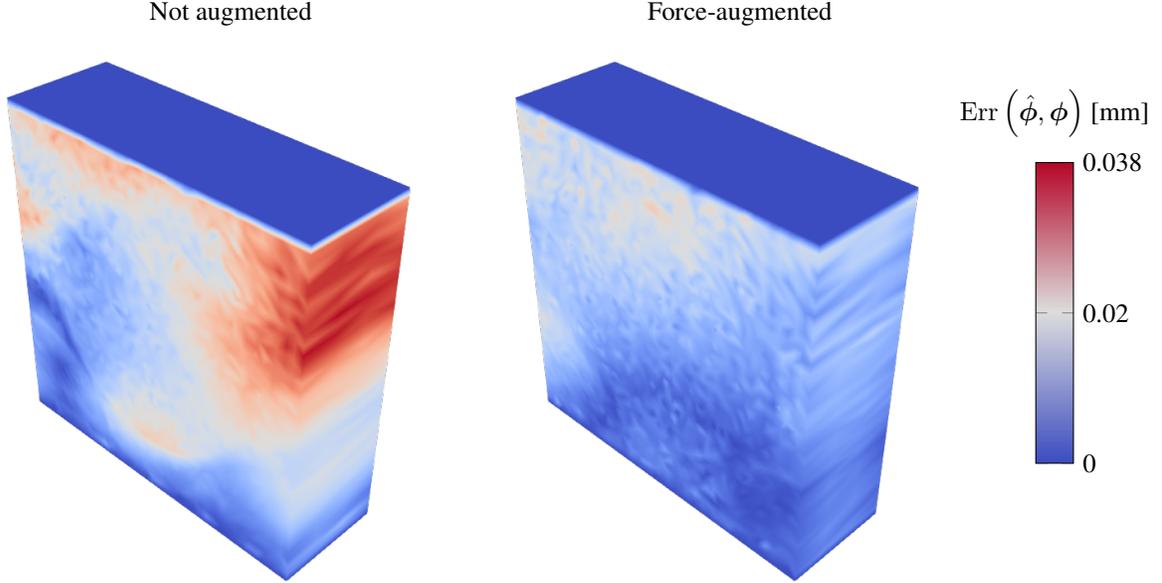

Moreover, as indicated by the green contours, the force-augmented model achieves marginally better performance than the model trained exclusively on solution field data (cf.~Section~\ref{sec3:2:1}). 
This improvement is also quantitatively confirmed in Figure~\ref{fig:UC_error_disp}. 
As noted previously, the model of the previous section exhibits reduced accuracy near the right surface of the specimen; however, the force-augmented approach enhances the predictive accuracy across the solution field. 
This improvement aligns with the inherently strong coupling between solution fields and boundary forces, underscoring the benefits of incorporating force information during training.
%
\subsection{Plate with elliptic hole}
\label{sec3:3}
\begin{figure}
    \centering
    \input{figures/PWEH_sketch}
    \caption{Sketch of the three-dimensional plate with an elliptic hole. The plate dimensions are \(10 \times 10 \times 1~\mathrm{mm}^3\). The hole is defined by its semi-axes \(a\) and \(b\), and oriented along the horizontal and vertical directions, respectively. The material orientation is characterized by the angle \(\alpha\), measured counterclockwise from the horizontal axis. The bottom surface is fully clamped, while a vertical displacement is prescribed on the top surface.}
    \label{fig:PWEH}
\end{figure}
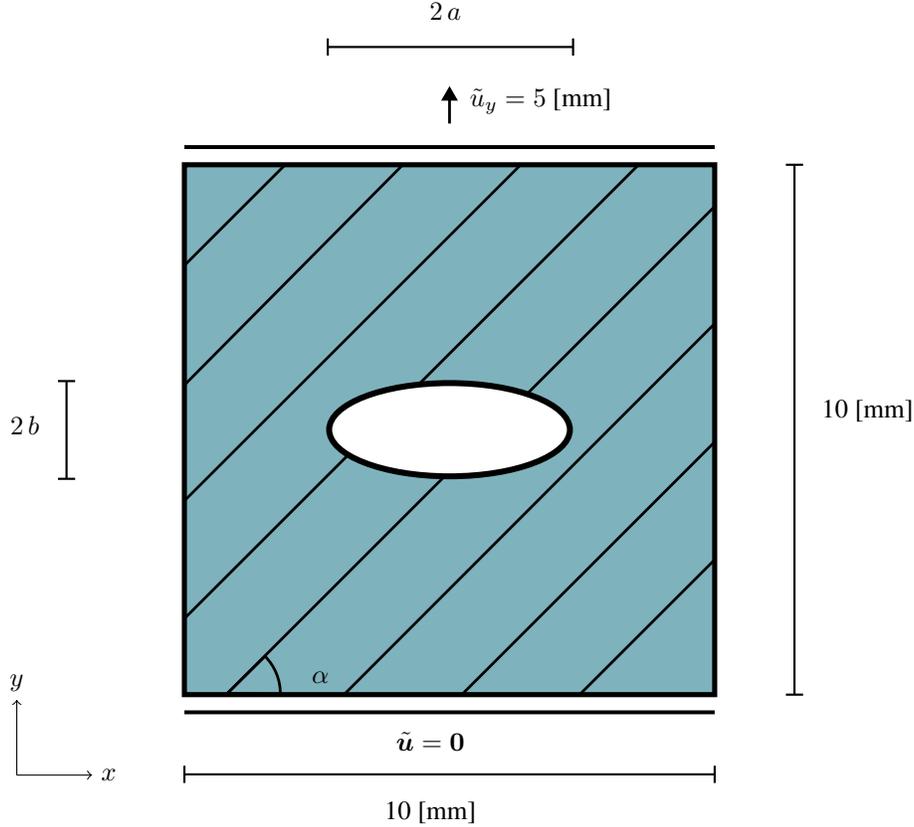

\paragraph{Boundary value problem.} Next, we consider a benchmark problem involving a plate with an elliptic hole, as illustrated in Figure~\ref{fig:PWEH}. The lower boundary of the specimen is fully clamped, while a prescribed vertical displacement is applied on the upper surface. The geometry of the boundary value problem is parameterized by the major and minor axes of the elliptic hole, denoted by \( 2a \) and \( 2b \), respectively.

The material behavior is characterized by hyperlelastic, transverse isotropy and modeled via a polyconvex Helmholtz free energy function \cite{HartmannNeff2003,SchroederNeff2003} of the form
\begin{equation}
    \psi = \frac{\mu}{2}\left( I_1' - 3\right) 
         + \frac{\kappa}{4} \left( I_3 - 1 - \ln(I_3) \right) 
         + \frac{K_1}{2} \left(I_4 - 1 \right)^2 
         + \frac{K_2}{2} \left(I_5 - 1 \right)^2,
\label{eq:PWEH_material}
\end{equation}
where the material parameters are given as \( \mu = 100~\mathrm{MPa} \), \( \kappa = 100~\mathrm{MPa} \), \( K_1 = 200~\mathrm{MPa} \), and \( K_2 = 10~\mathrm{MPa} \). The additional invariants are defined as follows: \( I_4 = \mathrm{tr}(\bm{C}\bm{M}) \), and \( I_5 = \mathrm{tr}(\mathrm{cof}\,\bm{C}\bm{M}) \),  with \( \bm{M} = \bm{m} \otimes \bm{m} \) being the structural tensor. The structural vector \( \bm{m} \) is defined by
\begin{equation}
    \bm{m} = 
    \begin{pmatrix}
        \cos{\alpha} \\
        \sin{\alpha} \\
        0
    \end{pmatrix},
\end{equation}
with \( \alpha \) denoting the material orientation angle that defines the preferred fiber direction. The operator \( \mathrm{cof} \) refers to the cofactor of the tensor \( \bm{C} \).

\paragraph{Parametrization and discretization.} In this example, we investigate the predictive performance of our method with respect to variations in both geometry and material anisotropy. Specifically, we consider changes in the ellipticity of the central hole and in the material orientation angle \( \alpha \). To this end, the solution space is parametrized by two natural coordinates \( \bm{\theta} = (\xi, \eta) \in [0,1]^2 \), according to
\begin{gather}
    a = 3 - 1.5\, \xi, \quad b = 1.5 + 1.5\, \xi, \\
    \alpha = 30^\circ + 120^\circ \, \eta,
\end{gather}
where \( a \) and \( b \) denote the semi-axes of the elliptic hole in millimeters and \( \alpha \) (in degrees) defines the orientation of the structural vector.

The parametric space is discretized using a structured sampling strategy, with \( \xi = \{0, \tfrac{1}{3}, \tfrac{2}{3}, 1\} \) and \( \eta = \{0, \frac{1}{4}, \frac{1}{2}, \frac{3}{4}, 1\} \), resulting in a total of 20 unique parameter combinations, i.e., snapshots.

Each snapshot is computed using a standard finite element formulation with low-order, trilinear hexahedral elements. 
The finite element mesh consists of 1647 nodes and 952 three-dimensional elements, ensuring a sufficiently accurate spatial resolution for capturing the mechanical response across the parameter space.
This results in a total of 4941 degrees of freedom, of which 4641 are \textit{active}.

In this example, we evaluate the plate with an elliptic hole at $\xi=0.75$ and $\eta=0.3$.
\subsubsection{Prediction of mesh morphing}
\label{sec3:3:1}
To account for changes in the geometry of the elliptical hole, the positions of all mesh nodes must be recomputed. 
Since this transformation is inherently nonlinear, we employ \emph{elastic mesh morphing}, wherein a full-order finite element simulation is used to obtain the updated nodal positions.

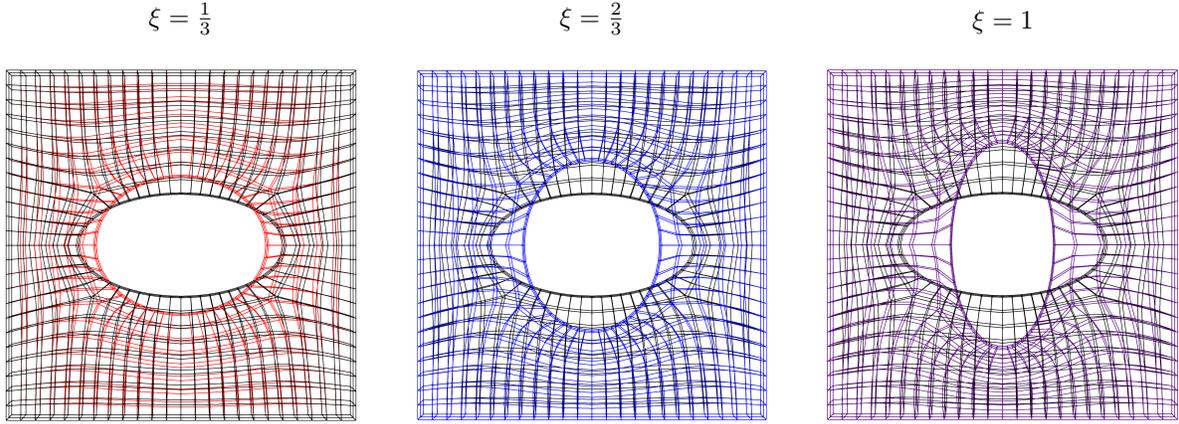
\begin{figure}
    \centering
    \input{figures/PWEH/morphing/morphings}
    \vspace*{-1cm}
    \caption{Snapshots illustrating the mesh morphing of the inner elliptical hole. 
    The black lines represent the original mesh corresponding to the parameter value $\xi = 0$. 
    The other $\xi$-configurations are obtained using an elastic mesh morphing strategy. 
    To preserve the geometric integrity, the $z$-components of all nodes located on the bounding box are constrained to zero. 
    In addition, the $x$- and $y$-displacement components of the nodes situated on the four lateral surfaces — corresponding to the extrema in $x$- and $y$-directions — are fixed.}
    \label{fig:PWEH_morphing}
\end{figure}

For this purpose, a pseudo-isotropic elastic material model is adopted by setting $K_1 = K_2 = 0$, and choosing $\mu = \kappa = 1$; see Equation~\eqref{eq:PWEH_material}. 
The associated boundary value problem is formulated such that the bounding box of the plate with the elliptical hole remains fixed, while the shape of the hole varies parametrically. 
The resulting geometries used for the snapshot dataset are illustrated in Figure~\ref{fig:PWEH_morphing}.

As the geometric transformation is nonlinear, the nodal displacements are also inherently nonlinear. 
Importantly, the element connectivity remains unchanged throughout the deformation process, as shown in Figure~\ref{fig:PWEH_morphing}.

However, if we aim to learn the nodal displacements using an Autoencoder, we face the challenge that the current positions of the nodes cannot be inferred without knowledge of the reference configuration. 
To address this, we first train a separate Autoencoder to predict the reference mesh resulting from the elastic morphing. 
This Autoencoder shares the same architecture as described previously (unsupervised Autoencoder with supervised latent space) and learns to predict the reference mesh in the parametric domain.

\paragraph{Network topology.} To encode and decode the nodal positions resulting from elastic mesh morphing, we employ an Autoencoder architecture consisting of four hidden layers in both the encoder and decoder, with a latent space of dimension two
\begin{equation*}
\begin{split}
    2871 \xrightarrow{\mathrm{GELU}} 256 \xrightarrow{\mathrm{GELU}} 128 \xrightarrow{\mathrm{GELU}} 64 \xrightarrow{\mathrm{GELU}} 32 \xrightarrow{\phantom{\mathrm{GELU}}} 2,
\end{split}
\end{equation*}
where the decoder is the mirrored counterpart of the encoder.  
The reduced number of active nodes (2871 compared to the original 4641) arises from the modified boundary value problem used during the mesh morphing procedure, which alters the domain of interest.

In addition, the latent space is predicted via a separate feed-forward network, which maps the parametric input to the latent code
\begin{equation*}
    1 \xrightarrow{\mathrm{GELU}} 32 \xrightarrow{\mathrm{GELU}} 32 \xrightarrow{\mathrm{GELU}} 32 \xrightarrow{\phantom{\mathrm{GELU}}} 2.
\end{equation*}
It is important to emphasize that only the natural coordinate $\xi$ affects the geometry of the elliptical hole. 
The coordinate $\eta$ has no influence on the deformation and is therefore excluded from the latent space input.

\paragraph{Results.} The training loss of the Autoencoder for elastic mesh morphing is shown in Figure~\ref{fig:PWEH_loss} (left).  
During the unsupervised training phase, the loss decreases to below the order of $10^{-2}$, while the subsequent supervised training phase reduces the loss to below $10^{-4}$.

A comparison between the high-fidelity reference solution for $\xi = 0.75$ and the prediction produced by the neural network is presented in Figure~\ref{fig:PWEH_morphing_comparison}.  
As observed, the discrepancy between the predicted and reference node positions is marginal.  
This high accuracy may, in part, be attributed to the boundary conditions: nodes on the bounding box remain fixed, and the nodes located on the surface of the hole are explicitly constrained.  
Despite these constraints, the remaining (unconstrained) nodal positions are predicted with high fidelity.
\begin{figure}
    \centering
    \input{figures/PWEH/PWEH_loss}
    \caption{\textbf{Plate with elliptic hole.} Training loss over the course of training for both neural networks, namely the Autoencoder and the Latent Space Prediction network, each trained for a maximum of 5000 epochs. 
Left: The end-to-end model is trained to predict the reference node positions (mesh morphing); see Section~\ref{sec3:3:1}. 
Right: The end-to-end model is trained to predict the nodal displacements; see Section~\ref{sec3:3:2}.}
    \label{fig:PWEH_loss}
\end{figure}
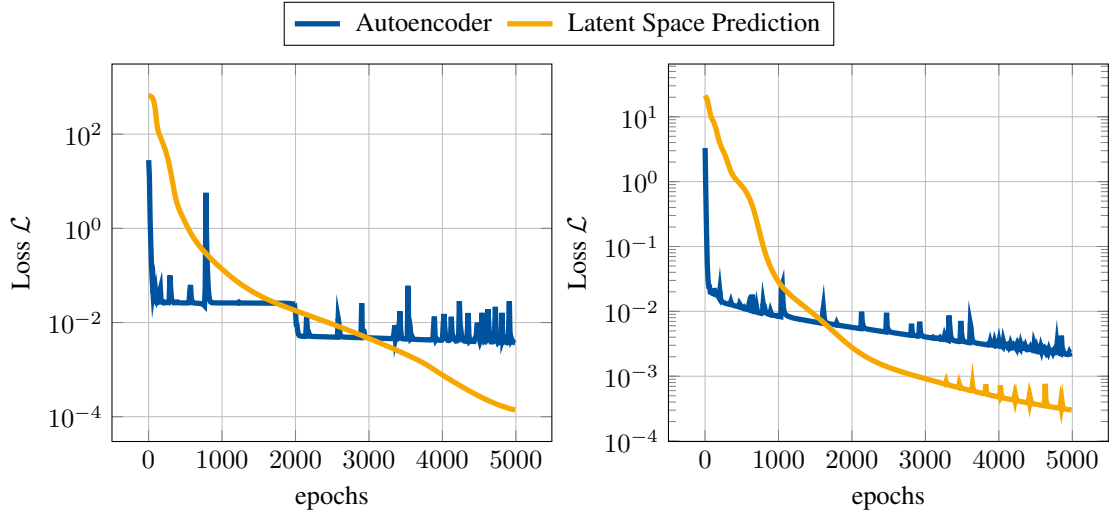
\begin{figure}
    \centering
    \input{figures/PWEH/morphing/morphing_comparison}
    \caption{Reference node positions resulting from elastic mesh morphing.  
The green lines indicate the nodal positions obtained from a high-fidelity full-order finite element simulation, while the orange lines represent the predictions of the end-to-end neural network model.  
The inset provides a zoomed perspective view of the elliptical hole and the surrounding mesh.}
    \label{fig:PWEH_morphing_comparison}
\end{figure}
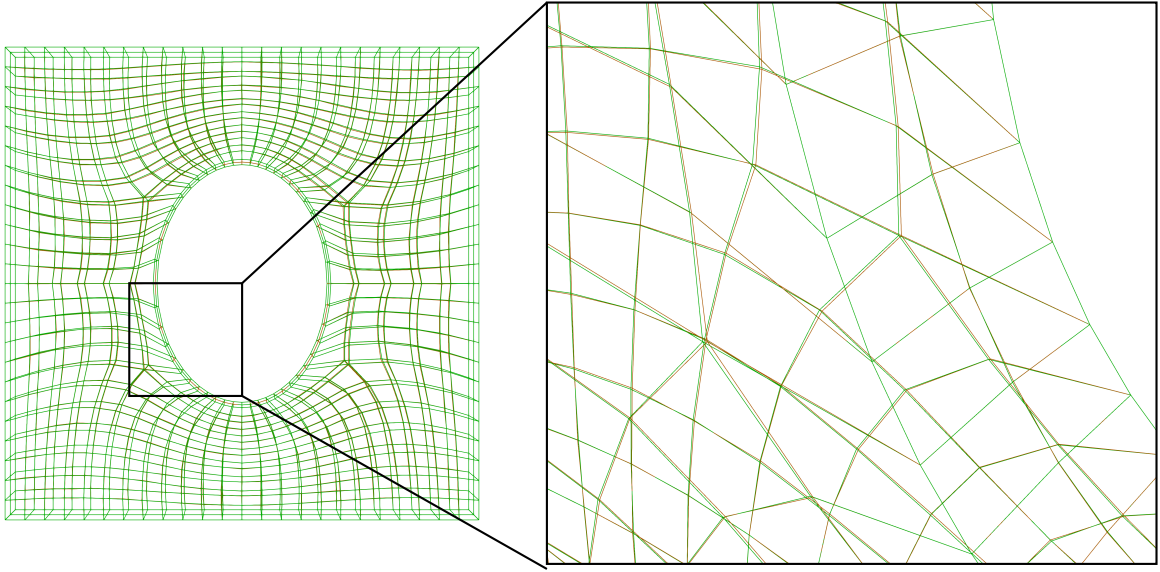
%
\subsubsection{Prediction of displacement field}
\label{sec3:3:2}
After training an Autoencoder to learn the referential positions of nodes, we now investigate its ability to predict displacement fields. The boundary value problem, including all relevant parameters, was previously detailed in Section~\ref{sec3:3}.

\paragraph{Network topology.} In this example, the number of neurons in the latent space is set to eight, while we employ four hidden layers to encode the data
\begin{equation*}
    4641 \xrightarrow{\mathrm{GELU}} 256 \xrightarrow{\mathrm{GELU}} 128 \xrightarrow{\mathrm{GELU}} 64 \xrightarrow{\mathrm{GELU}} 32 \xrightarrow{\phantom{\mathrm{GELU}}} 8
\end{equation*}
as well as five hidden layers for decoding the reconstructed solution field
\begin{equation*}
    8 \xrightarrow{\mathrm{GELU}} 32 \xrightarrow{\mathrm{GELU}} 64 \xrightarrow{\mathrm{GELU}} 128 \xrightarrow{\mathrm{GELU}} 256 \xrightarrow{\mathrm{GELU}} 512 \xrightarrow{\phantom{\mathrm{GELU}}} 4641.
\end{equation*}
The larger number of neurons in the latent space was chosen because we anticipate a greater amount of information to be compressed, owing to the parametrization of the fiber angle and the associated anisotropic nature of the underlying material model.

For latent space prediction, we maintain the same architecture as used for mesh morphing
\begin{equation*}
    2 \xrightarrow{\mathrm{GELU}} 32 \xrightarrow{\mathrm{GELU}} 32 \xrightarrow{\mathrm{GELU}} 32 \xrightarrow{\phantom{\mathrm{GELU}}} 8,
\end{equation*}
where this time both natural coordinates within the parametric space influence the solution field.

\paragraph{Results.} The training losses of both the Autoencoder and the latent space prediction network for reconstructing the displacement field are shown in Figure~\ref{fig:PWEH_loss} (right).

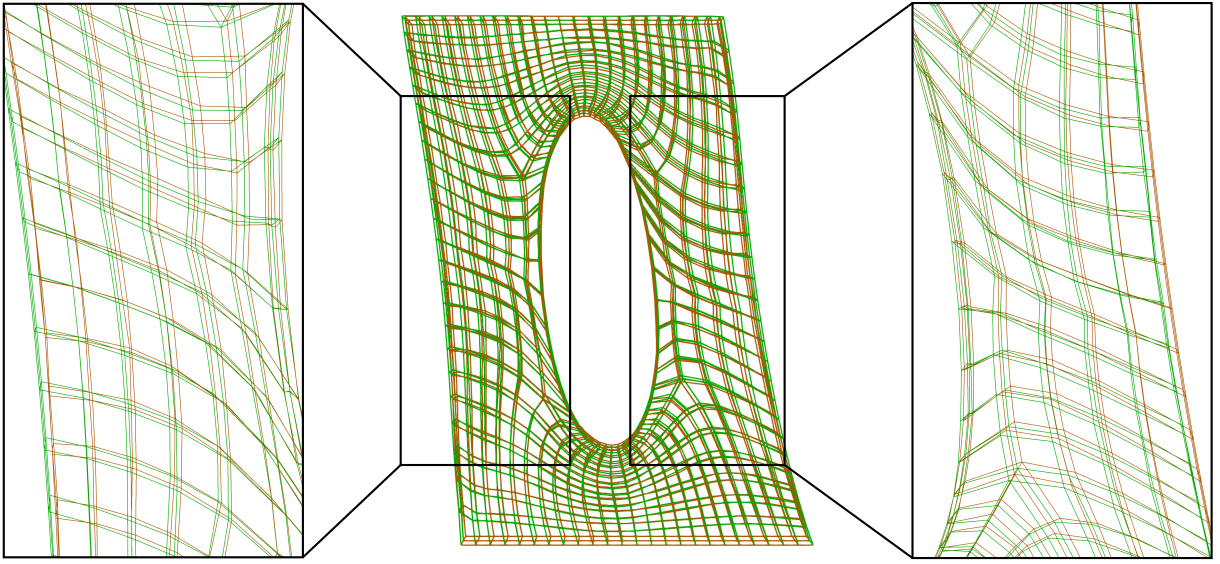
\begin{figure}
    \centering
    \input{figures/PWEH/displacement/displacement_comparison}
    \caption{Deformed mesh of the plate with an elliptic hole.
Green lines illustrate the high-fidelity reference solution, while orange lines correspond to the prediction of the end-to-end model.
To visualize the predicted deformed mesh, nodal positions were first computed using the end-to-end model trained in Section~\ref{sec3:3:1}; see Figure~\ref{fig:PWEH_morphing_comparison}.
The displacements were then predicted by the end-to-end model of Section~\ref{sec3:2:2} and added to the referential nodal positions to obtain the current positions.
The insets zoom in on the area of the hole.
This time, all nodes except for the lower and upper surfaces are \textit{active}.}
    \label{fig:PWEH_displacement_comparison}
\end{figure}

A comparison between the displacement field obtained via full-order finite element simulation and the prediction generated by the end-to-end model is presented in Figure~\ref{fig:PWEH_displacement_comparison}.  
The model is evaluated at the parametric values $\xi = 0.75$ and $\eta = 0.3$.  
As evident from the figure, the predicted deformation closely matches the reference solution.  
While minor discrepancies appear along the left and right outer surfaces, the shape of the deformed hole is accurately captured.  
Interestingly, within the interior of the specimen, the nodal positions are not perfectly matched with the reference finite element simulation.  
Nonetheless, the overall deformation trend is well reproduced.  
Incorporating convolutional layers or graph-based architectures could potentially reduce this residual error within the specimen's domain.

What is particularly noteworthy is that the end-to-end model achieves this performance despite never being explicitly informed that the reference nodal positions differ between snapshots used for training — this holds true for both the unsupervised Autoencoder and the supervised latent space predictor.  
Although this was already present in the mesh morphing task, the degree of variation between the snapshots is even greater in this case.  
This observation, coupled with the fact that the current nodal positions are reconstructed through the composition of two independently trained models (for reference position and displacement), highlights both the expressive power and the data-driven structure-learning capability of the proposed architecture.
%
\subsection{Multi-field problem of thermo-mechanics}
\label{sec3:4}
\paragraph{Boundary value problem.} Lastly, we investigate the multi-field extension of the non-intrusive Autoencoder approach, with a particular focus on transient thermo-elasticity at finite strains.
To this end, we again consider a plate featuring an elliptic hole; however, in contrast to previous studies, the semi-axes of the ellipse are kept fixed and the material is assumed to be isotropic.

The mechanical boundary conditions are defined such that the lower surface of the plate is fully clamped, while the displacement in the $y$-direction is constrained to zero at the top surface. Thus, no external mechanical loads are applied in this scenario.

On the thermal side, a temperature of zero is prescribed along the bottom surface, while the temperature at the top surface is increased linearly over time.
Although no displacements are explicitly prescribed, the thermal expansion induced by the temperature gradient leads to internal stresses due to the thermo-mechanical coupling.

Because the displacement in the $y$-direction is restricted, the plate experiences an effective compressive load in this direction, resulting in a bifurcation phenomenon. 
As a consequence, the system exhibits instability in the form of buckling, which highlights the strong coupling within the multi-physical context.
The onset of buckling varies with the material parameters and geometry, which may be critical for structural safety evaluations.

Here, we consider an isotropic material exhibiting isotropic thermal expansion. 
The governing equations of finite strain thermo-elasticity are summarized in~\ref{app:thermo}. 
Accordingly, we present here only the essential constitutive relations in terms of the Helmholtz free energy function and the referential heat flux
\begin{align}
    \psi &= \frac{\mu}{2}\left( I_{1M}' - 3 \right) 
         + \frac{\kappa}{4} \left( I_{3M} - 1 - \ln(I_{3M}) \right), \\
    \bm{q}_0 &= - \Lambda\, \sqrt{I_3}\, \bm{C}^{-1} \, \mathrm{Grad}(T),
\end{align}
where $\mu = 100~\mathrm{MPa}$ and $\kappa = 100~\mathrm{MPa}$ denote the shear and bulk modulus, respectively, while the heat conductivity $\Lambda$ is parameterized as described below. 
The term $\mathrm{Grad}(T)$ represents the referential temperature gradient.

The invariants $I_{1M}$ and $I_{3M}$ refer to the mechanical part of the right Cauchy-Green tensor, $\bm{C}_M$, and are defined as $I_{1M} = \mathrm{tr}\,\bm{C}_M$, $I_{3M} = \det \bm{C}_M$, and $I_{1M}' = I_{1M} / I_{3M}^{1/3}$.
In addition, the coefficient of thermal expansion is given by $\alpha_T = 0.005~\mathrm{K}^{-1}$, the heat capacity is set to $c=0.1~\mathrm{MPa~mm^3~K^{-1}}$, and the reference temperature is set to $T_0 = 273.15~\mathrm{K}$; see also \ref{app:thermo}.
\begin{figure}
    \centering
    \input{figures/PWEH_Thermo_sketch}
    \caption{Sketch of the three-dimensional plate with an elliptic hole. The plate dimensions are \(10 \times 10 \times d~\mathrm{mm}^3\). The thickness $d$ is parameterized. To promote a unique buckling deformation and break symmetry, an initial geometric imperfection is introduced by linearly inclining the top edge of the plate by \(0.01~\mathrm{mm}\) in $z-$direction, corresponding to an approximate angle of \(0.06^\circ\). The bottom surface is fully clamped and the temperature is prescribed to be zero, while the vertical displacement is constrained and the temperature is linearly increased over time $t$ on the top surface.}
    \label{fig:PWEH_thermo}
\end{figure}
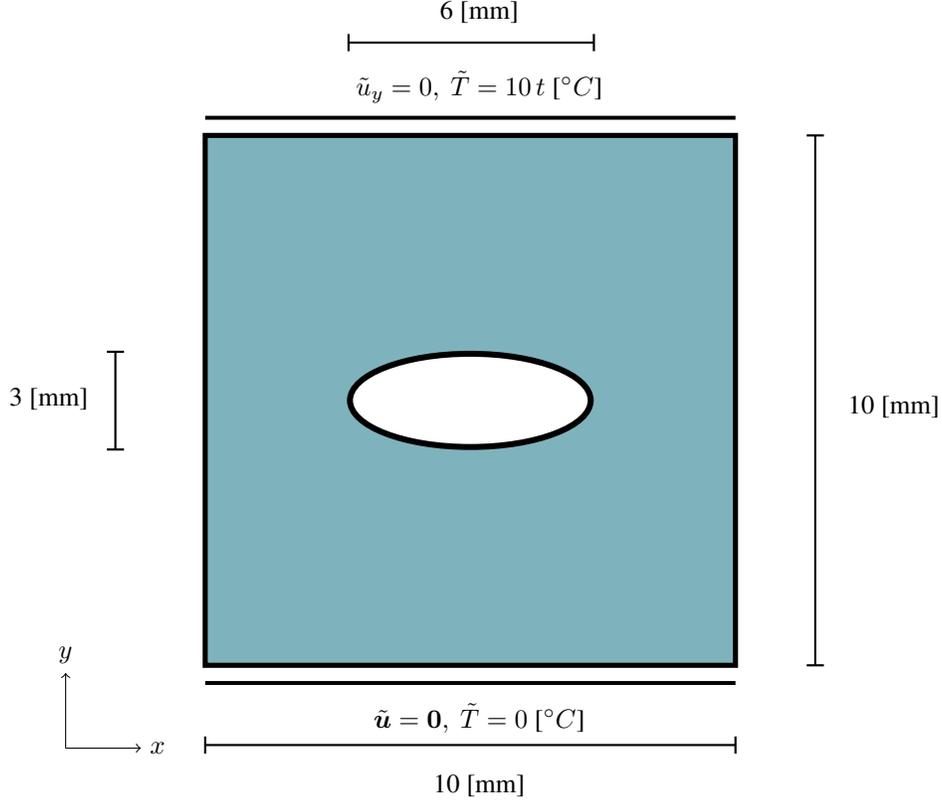

\paragraph{Parametrization and discretization.} In contrast to the previous examples, we now consider a transient problem, i.e., the underlying boundary value problem depends explicitly on time. 
Accordingly, temporal discretization is required in addition to the spatial discretization. 
As before, we parameterize the nonlinear solution space by two natural coordinates, $\xi$ and $\eta$, and additionally incorporate time as a third input parameter. 
This results in a three-dimensional input parameter space, $\bm{\theta} = (\xi, \eta, t) \in [0,1]^3$, where $t$ denotes the normalized time.

In this example, we vary the heat conductivity and the plate thickness according to
\begin{equation}
    \Lambda = 20 + 10\,\xi, \quad d = 0.2 + 0.8\,\eta.
\end{equation}
Unlike the previous example, the geometric variation due to the changing plate thickness scales linearly with $\eta$. 
Therefore, no elastic morphing or dedicated Autoencoder is required to compute the nodal coordinates.

We employ the same finite element mesh as in Section~\ref{sec3:3}. 
However, due to the additional degree of freedom introduced by the temperature field, the total number of degrees of freedom increases to 6588, of which 6138 are \textit{active}, i.e., not constrained by boundary conditions.

To solve the transient problem, we set the time step size to $\Delta t = 0.01$ and simulate the system over 100 time steps. 
The parameter space is discretized as $\xi, \eta \in \{0, \tfrac{1}{3}, \tfrac{2}{3}, 1\}$, which yields a total of 1600 snapshots.

We evaluate the model at $\xi=0.5$, $\eta=0.5$, and $t=0.8$.

\paragraph{Network topology.} In this example, we employ the proposed multi-field extension of the end-to-end model.  
Specifically, two distinct encoder networks are utilized to map each solution field — namely, the displacement field $\bm{u}$ and the temperature field $T$ — into their respective latent space representations.  
For simplicity, both encoder networks share the same architecture and consist of four latent neurons each. These latent vectors are subsequently concatenated to form a combined latent representation of eight neurons.

The architecture of the encoders is given by
\begin{equation*}
    \mathcal{X} \xrightarrow{\mathrm{GELU}} 1024 \xrightarrow{\mathrm{GELU}} 512 \xrightarrow{\mathrm{GELU}} 128 \xrightarrow{\mathrm{GELU}} 32 \xrightarrow{\mathrm{GELU}} 16 \xrightarrow{\phantom{\mathrm{GELU}}} 4,
\end{equation*}
where $\mathcal{X}$ denotes the input dimension for each solution field.  
For the displacement field, we have $\mathcal{X} = 4641$, and for the temperature field, $\mathcal{X} = 1497$.

From the concatenated latent space of dimension eight, a shared decoder network is used to reconstruct both solution fields
\begin{equation*}
    8 \xrightarrow{\mathrm{SiLU}} 16 \xrightarrow{\mathrm{SiLU}} 32 \xrightarrow{\mathrm{SiLU}} 64 \xrightarrow{\mathrm{SiLU}} 128 \xrightarrow{\mathrm{GELU}} 1024 \xrightarrow{\phantom{\mathrm{GELU}}} 6138,
\end{equation*}
from which the predicted values for each individual solution field can be extracted.

Finally, two independent latent space predictors are introduced — one for each solution field.  
Again, identical architectures are used for simplicity
\begin{equation*}
    3 \xrightarrow{\mathrm{GELU}} 32 \xrightarrow{\mathrm{GELU}} 32 \xrightarrow{\mathrm{GELU}} 32 \xrightarrow{\phantom{\mathrm{GELU}}} 4,
\end{equation*}
which predict the corresponding four-dimensional latent space.  
The input to each predictor consists of the natural coordinates $\xi$ and $\eta$, as well as the time variable $t$.

\paragraph{Results.} Although the geometry is once again parameterized, no elastic mesh morphing is required in this case.  
This is due to the fact that the geometry is modified solely by linear scaling in the thickness direction, which allows for straightforward analytical computation of the referential nodal positions.  
Nevertheless, it remains the case that during training, the end-to-end model is not explicitly informed about variations in referential nodal positions.

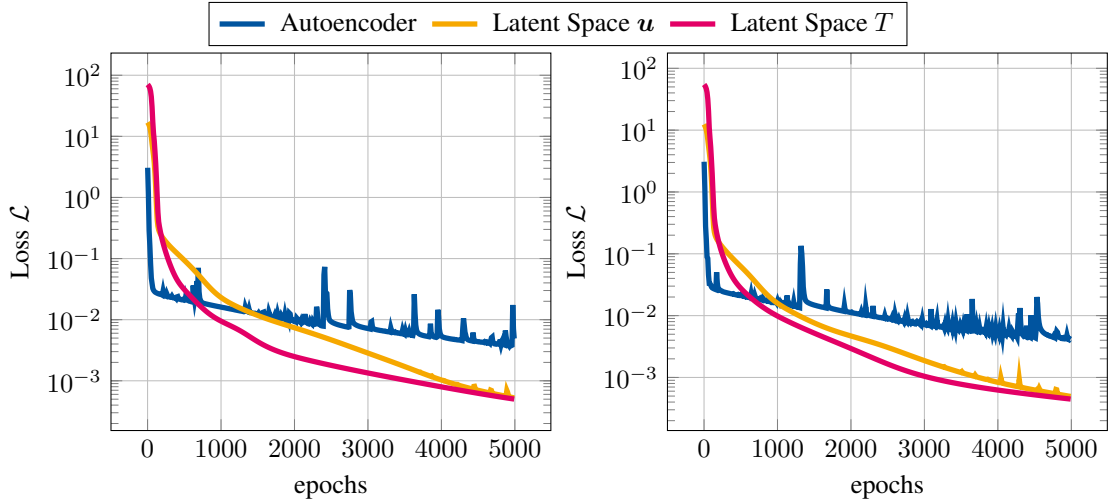
\begin{figure}
    \centering
    \input{figures/Thermo_imperfection/Thermo_loss}
    \caption{\textbf{Thermo-mechanics.} Training loss during the minimization procedure for all three networks: the Autoencoder, the latent space predictor for the displacement field $\bm{u}$, and the latent space predictor for the temperature field $T$.  
In all training sessions, the number of epochs is fixed at 5000.  
Left: End-to-end model trained on 1600 snapshots.  
Right: End-to-end model trained on 528 snapshots, with reduced resolution in the temporal dimension.}
    \label{fig:Thermo_loss}
\end{figure}

The training loss is shown in Figure~\ref{fig:Thermo_loss}.  
Two training scenarios are considered: one using the complete dataset consisting of 1600 snapshots, and a second using only 33 equally spaced time steps, resulting in a total of 528 snapshots.  
Since the latent space is shared between fields but comprises separate contributions, two distinct latent space predictors are trained — one for the displacement field and one for the temperature field.  
Regardless of the number of snapshots used, all neural networks achieve loss magnitudes comparable to those observed in the previous examples.

In Figure~\ref{fig:Thermo_comparison_thermo}, the predicted temperature field across the specimen is shown.  
The predictions are in good agreement with the high-fidelity reference solution — both qualitatively and quantitatively.  
Notably, even the model trained on the reduced snapshot set yields satisfactory results.

As can clearly be seen, the specimen exhibits pronounced buckling behavior under thermal loading.  
This response is driven by the strong thermo-mechanical coupling, caused by thermal expansion and deformation-dependent heat conduction, as governed by Fourier’s law (see \ref{app:thermo}).  
Although no mechanical loads or prescribed displacements are applied, the system experiences an effective thermal pressure, which ultimately induces buckling.

The predictive performance of the end-to-end model with respect to the displacement field is investigated in Figure~\ref{fig:Thermo_comparison_disp}.  
Across both training setups, the prediction error in the displacement field remains marginal.  
Interestingly, the model trained on only 528 snapshots slightly outperforms the one trained on the full dataset at the specific evaluation point.  
However, we attribute this to a coincidental alignment of the parametric configuration rather than to any systematic advantage.

Lastly, we investigate the performance of the end-to-end model in the absence of any geometric imperfection; see Figure~\ref{fig:Thermo_wo_imperfection}.  
The original purpose of introducing an inclined imperfection was to avoid the bifurcation issue, where it is inherently ambiguous whether the plate buckles in the positive or negative $z$-direction.  
However, when allowing for bifurcation without an imperfection, we observe a significant deterioration in the predictive accuracy.  
The predicted deformed configuration deviates substantially from the full-order finite element solution, and the resulting surface is notably less smooth.

This behavior is likely related to the nature of the training snapshots.  
Unlike in the previous study, the snapshots here are generated from specimens without geometric imperfections, leading to a mixture of buckling modes — some in the positive and others in the negative $z$-direction.  
Such variability hinders the end-to-end model’s ability to extract coherent structural patterns from the data.

A potential remedy could be to pre-process the snapshots by reflecting those buckling in one direction to align all buckling modes consistently.  
Whether such a strategy improves surface smoothness remains an open question beyond the scope of this contribution.

It is further worth noting that only 1400 snapshots were used for training the geometrically perfect model because the finite element simulations at parametric points $(\xi,\eta) = \left(\frac{2}{3}, 0\right)$ and $(1, 0)$ could not be solved for the entire time series using a constant time increment.
\begin{figure}
\hspace*{-1cm}
    \centering
    \input{figures/Thermo_imperfection/comparison_themo}
    \caption{Temperature field across the specimen evaluated at $\xi = \eta = 0.5$ and $t = 0.8$.  
The left panel shows the reference solution obtained from a high-fidelity finite element simulation.  
The middle panel displays the prediction generated by the end-to-end model trained on 1600 snapshots.  
The right panel shows the prediction obtained from a model trained with fewer temporal snapshots.  
Green contours indicate the deformed configuration of the reference simulation.}
    \label{fig:Thermo_comparison_thermo}
\end{figure}
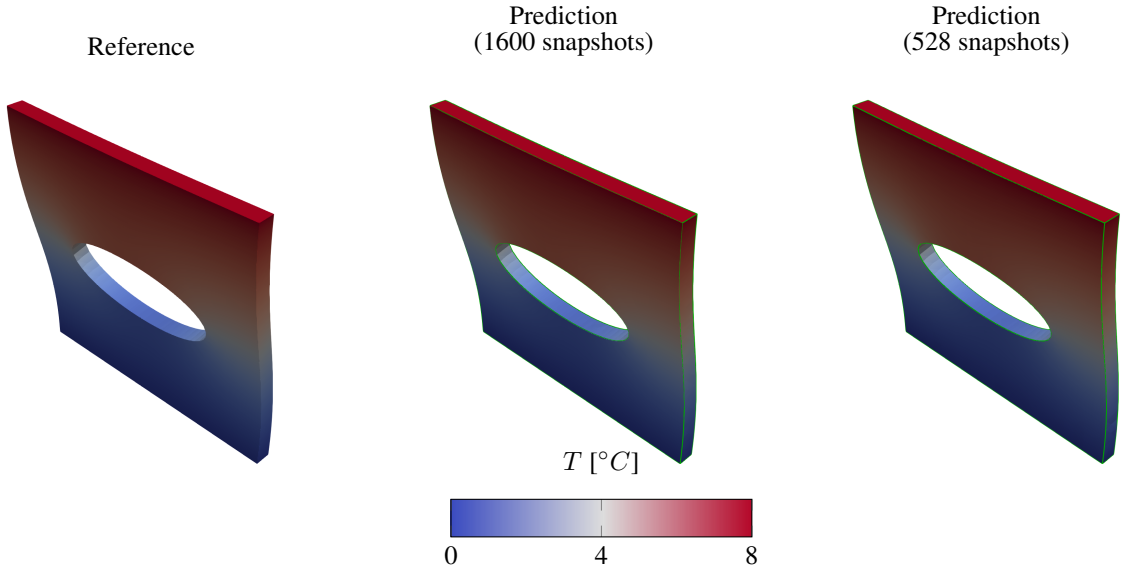
\begin{figure}
\hspace*{-2cm}
    \centering
    \input{figures/Thermo_imperfection/comparison_disp}
    \caption{Comparison between the predicted displacement fields and the high-fidelity reference solution obtained from a finite element simulation.  
Top: Euclidean norm $\lVert \bullet \rVert$ of the displacement field $\bm{u}$ for the deformed reference configuration.  
Bottom: Error contours showing the Euclidean norm of the difference $\lVert \hat{\bm{u}} - \bm{u} \rVert$ between the predicted displacement field $\hat{\bm{u}}$ and the reference field $\bm{u}$. 
The end-to-end model is evaluated once after being trained on 1600 snapshots and once on a reduced dataset with 528 snapshots (lower temporal resolution).  
Green contours indicate the deformed configuration of the reference simulation.}
    \label{fig:Thermo_comparison_disp}
\end{figure}
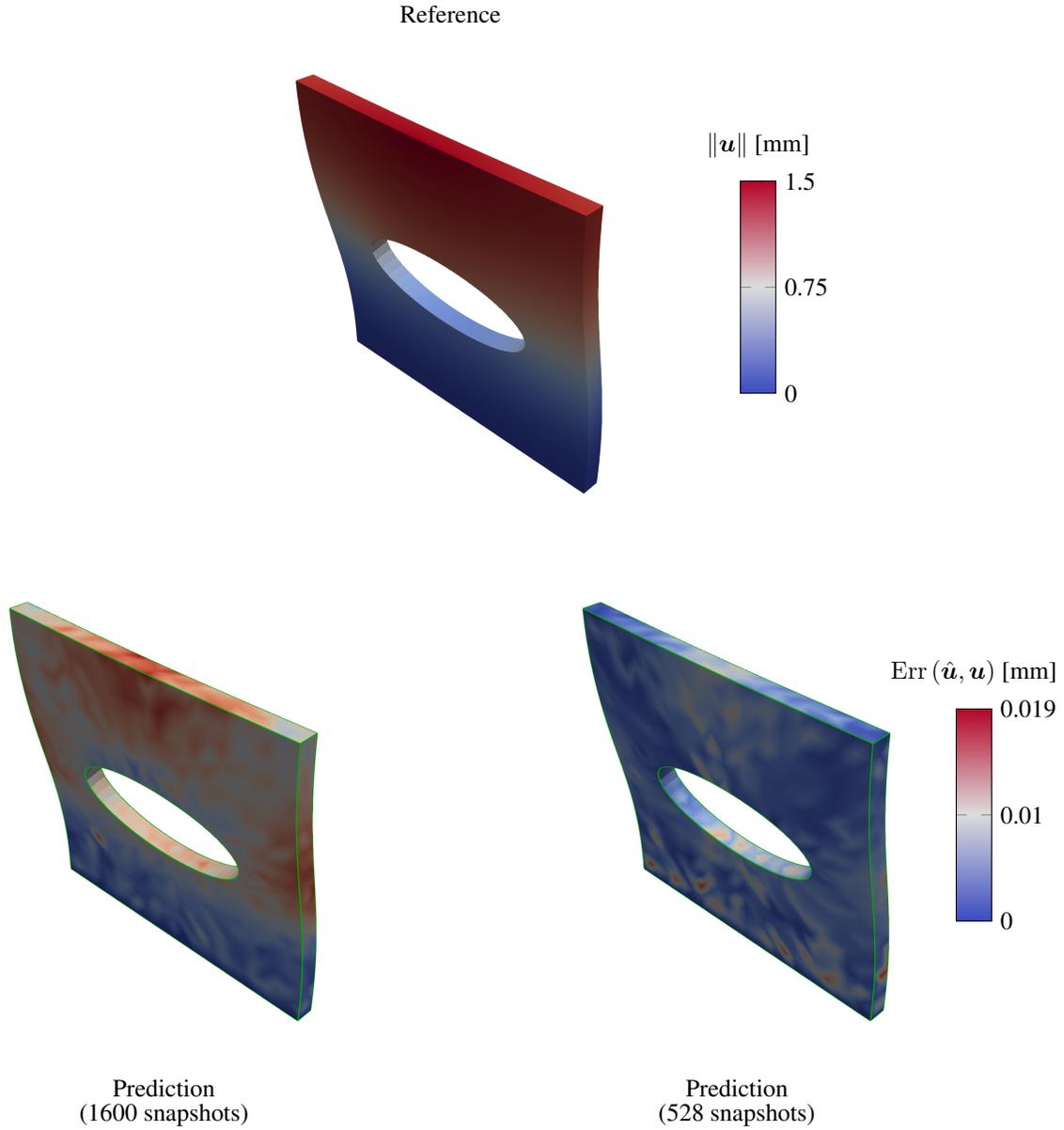
\begin{figure}
    \centering
    \includegraphics[width=\textwidth]{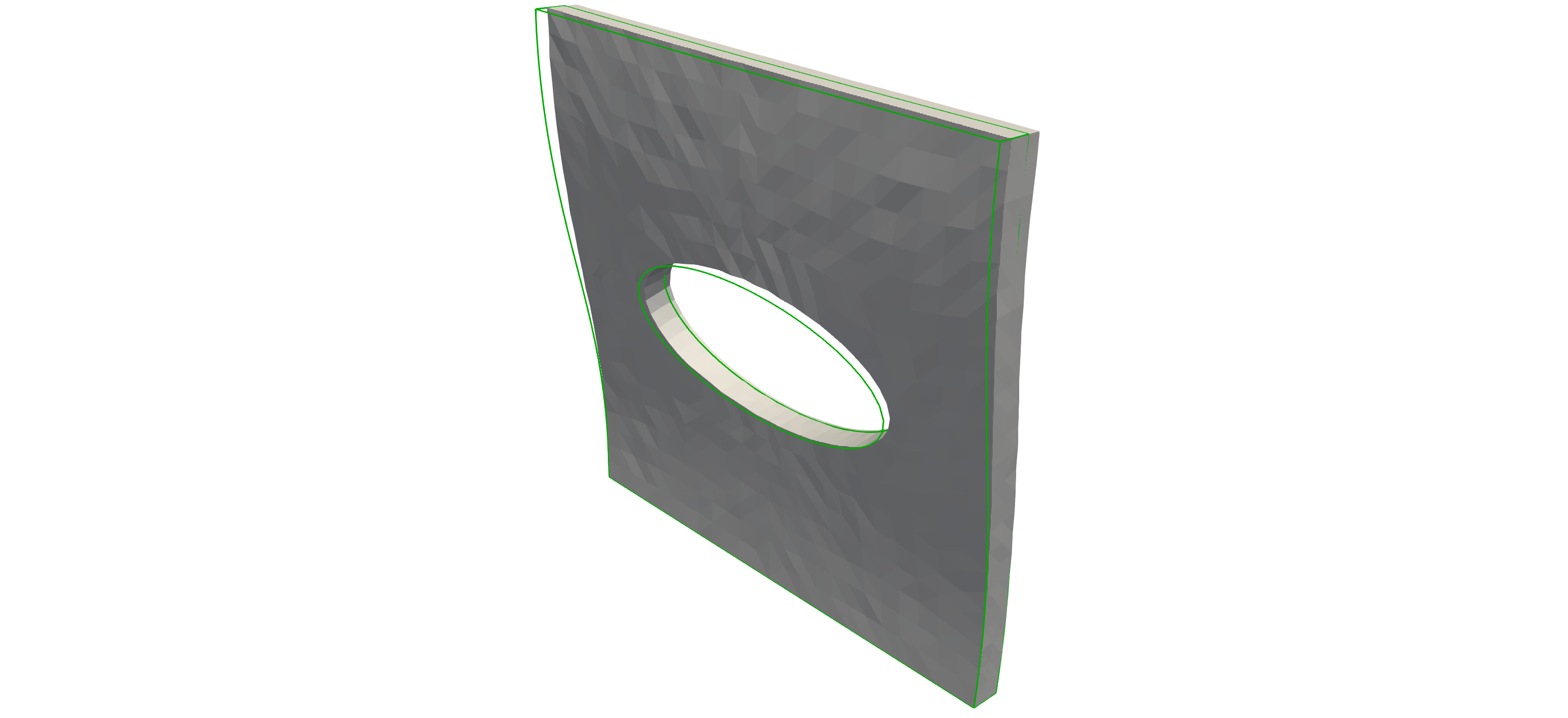}
    \caption{Computation and prediction of the deformation of a plate with an elliptic hole subjected to thermal loading, without any geometric imperfections; see Figure~\ref{fig:PWEH_thermo}.  
The solid body represents the predicted deformed configuration obtained by the end-to-end model, while green contours indicate the reference deformation computed via a full-order finite element simulation.  
The model is trained on 1400 snapshots and evaluated at the parametric values $\xi = \eta = 0.5$ and time $t = 0.8$.}
    \label{fig:Thermo_wo_imperfection}
\end{figure}

%% file: figures/UC_sketch.tex
\begin{tikzpicture}
    \node (pic) at (0,0) {\includegraphics[width=0.6\textwidth]{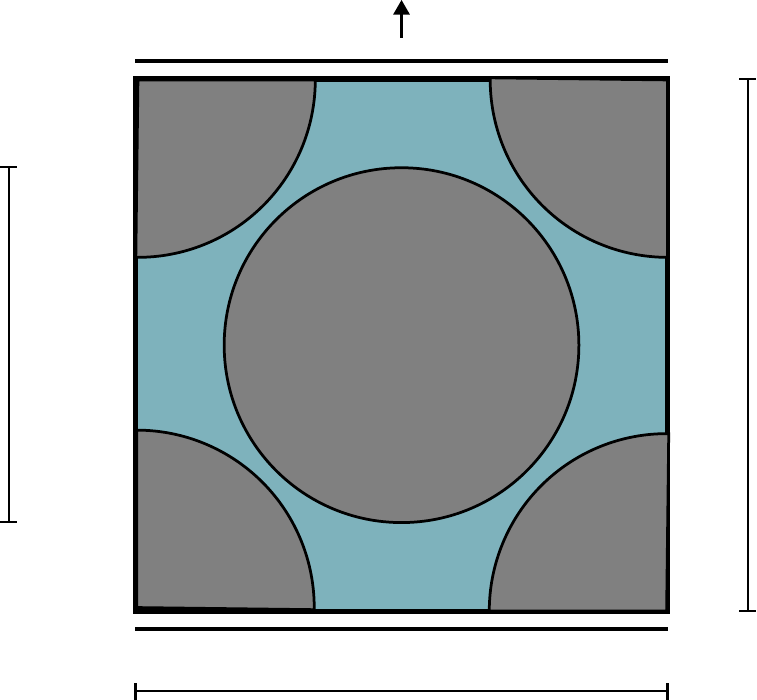}};
    \node[anchor=west] (height) at (pic.east) {3 [mm]};
    \node[anchor=north] (length) at (pic.south) {3 [mm]};
    \node[anchor=east] (radius) at (pic.west) {2 [mm]};
    \node[anchor=west] (utop) at ($(pic.north)+(.4,-0.3)$) {$\tilde{u}_x$, $\tilde{u}_y$};
    \node[] (ubottom) at ($(length)+(+0,+.95)$) {$\tilde{\bm{u}}=\bm{0}$};
    \coordinate (origin) at ($(length)+(-5.5,+0.5)$);
    \draw[->] (origin) -- ($(origin)+(1,0)$) node[right] {$x$};
    \draw[->] (origin) -- ($(origin)+(0,1)$) node[above] {$y$};
\end{tikzpicture}

%% file: figures/UC/UC_loss.tex
\begin{tikzpicture}[]
    \begin{axis}[xlabel= Nice $x$ label, ylabel= Nice $y$ label, width=55mm, height=35mm,
				hide axis,
				xmin=1,
				xmax=50,
				ymin=0,
				ymax=0.4,
				legend columns=-1,
				legend style={column sep=1mm}
				]
\addlegendimage{rwth1, line width=2pt}
\addlegendentry{Autoencoder}
\addlegendimage{rwth8, line width=2pt}
\addlegendentry{Latent Space Prediction} 
    \end{axis}
\end{tikzpicture}

\begin{tikzpicture}
\begin{semilogyaxis}[
    grid = major,
    xlabel = {epochs},
    ylabel = {Loss $\mathcal{L}$},
    width=0.45\textwidth,
    height=0.4\textwidth,
    ymin = 0.001,
    ymax = 100,
    /pgf/number format/1000 sep={},
    legend style={
        at={(0.5,1.05)},
        anchor=south,
        legend columns=-1, 
        /tikz/every even column/.append style={column sep=0.5em}
    }
]
    \addplot[rwth1, line width=2pt] table[x expr=\coordindex*10+1, y index=0] {figures/UC/lossAE_sampled.txt};
    \addplot[rwth8, line width=2pt] table[x expr=\coordindex*10+1, y index=0] {figures/UC/lossFFNtoLS_sampled.txt};
\end{semilogyaxis}
\end{tikzpicture}
\begin{tikzpicture}
\begin{semilogyaxis}[
    grid = major,
    xlabel = {epochs},
    ylabel = {Loss $\mathcal{L}$},
    ymin = 0.001,
    ymax = 100,
    width=0.45\textwidth,
    height=0.4\textwidth,
    /pgf/number format/1000 sep={},
]
    \addplot[rwth1, line width=2pt] table[x expr=\coordindex*10+1, y index=0] {figures/UC/force/lossAE_sampled.txt};
    \addplot[rwth8, line width=2pt] table[x expr=\coordindex*10+1, y index=0] {figures/UC/force/lossFFNtoLS_sampled.txt};
\end{semilogyaxis}
\end{tikzpicture}

%% file: figures/UC/pure_solution.tex
\begin{tikzpicture}
\node[anchor=west, inner sep=0] (image1) at (0,0) {\includegraphics[width=0.5\textwidth]{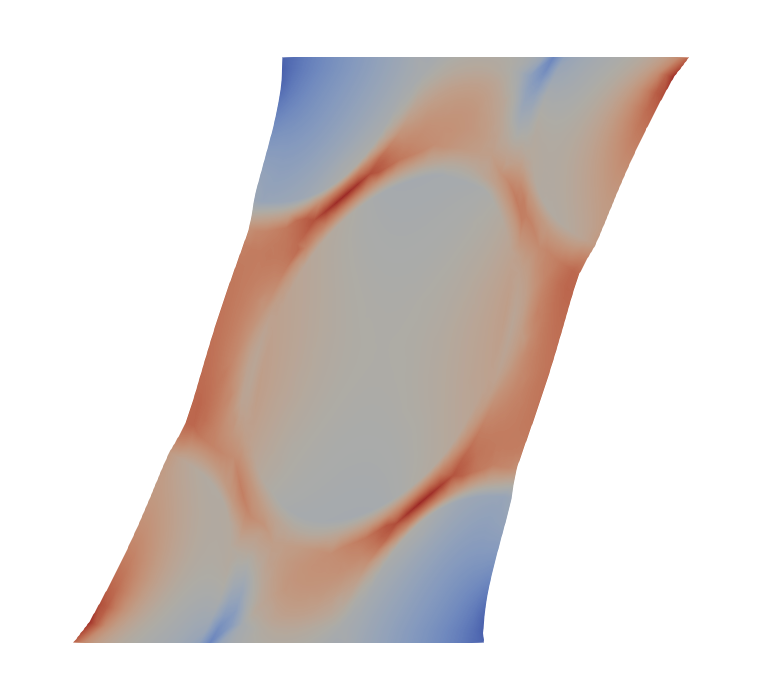}};
\node[anchor=west, inner sep=0] (image2) at ($(image1.east) + (-1.5cm,0)$) {\includegraphics[width=0.5\textwidth]{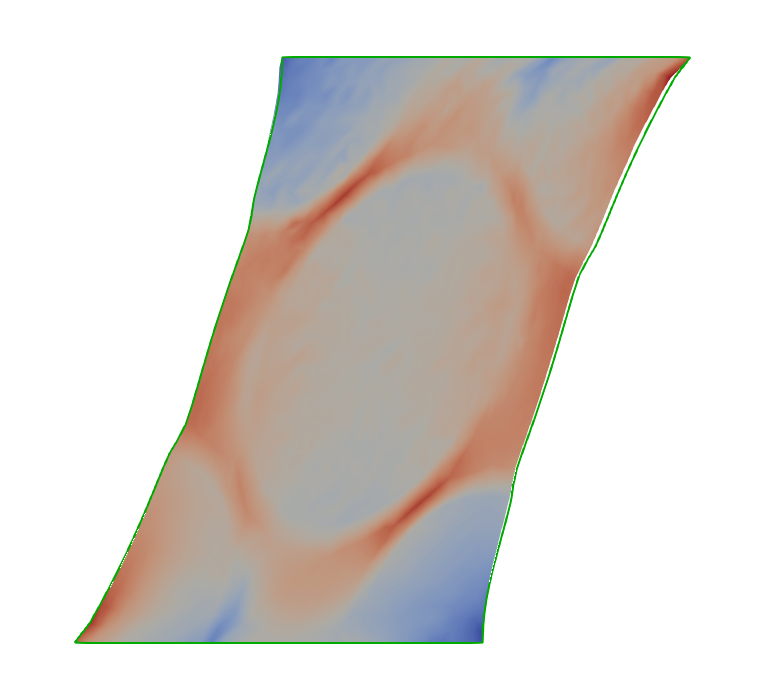}};
\node[anchor=south, inner sep=0] (name1) at ($(image1.north) + (+1cm,0)$) {Reference};
\node[anchor=south, inner sep=0] (name2) at ($(image2.north) + (+1cm,0)$) {Prediction};
\begin{axis}[
    at={($(image2.east) + (-1cm,0)$)}, anchor=west,
    xshift=0.5cm,
    height=4cm,
    width=1cm,
    scale only axis,
    hide axis,
    colorbar,
    point meta min=0,
    point meta max=1,
    colormap={cooltowarm}{
        rgb(0cm)=(0.23137,0.29804,0.75294);
        rgb(0.5cm)=(0.865,0.865,0.865);
        rgb(1cm)=(0.70588,0.01569,0.14902)
    },
    colorbar style={
        ytick={0, 0.5, 1},
        yticklabels={0, 0.3, 0.6},
        title=$\displaystyle \frac{\partial\phi_y}{\partial y}$ [-]
    }
]
\addplot [draw=none] coordinates {(0,0)};
\end{axis}
\end{tikzpicture}

%% file: figures/UC/force/UC_force.tex
\begin{tikzpicture}
\node[anchor=west, inner sep=0] (image1) at (0,0) {\includegraphics[width=0.5\textwidth]{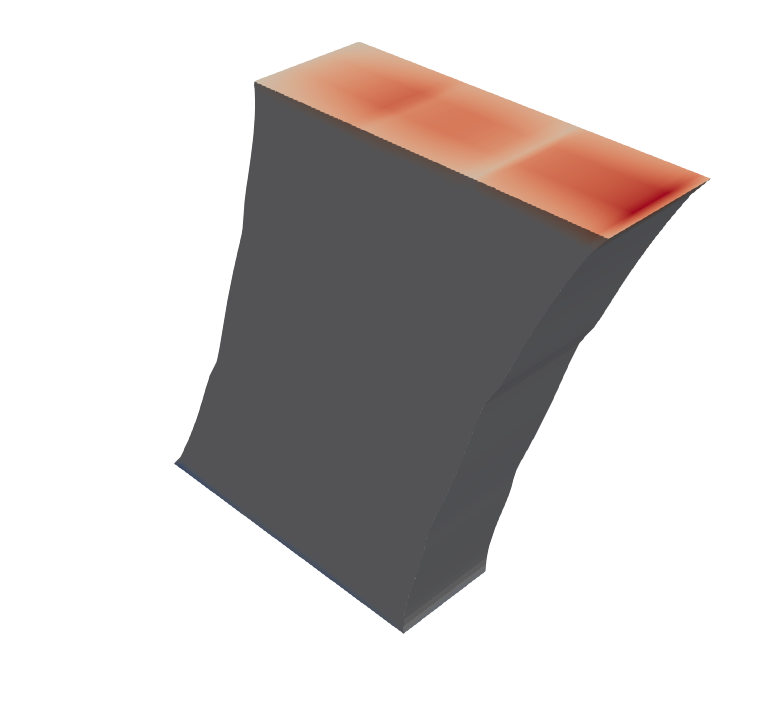}};
\node[anchor=west, inner sep=0] (image2) at ($(image1.east) + (-1.5cm,0)$) {\includegraphics[width=0.5\textwidth]{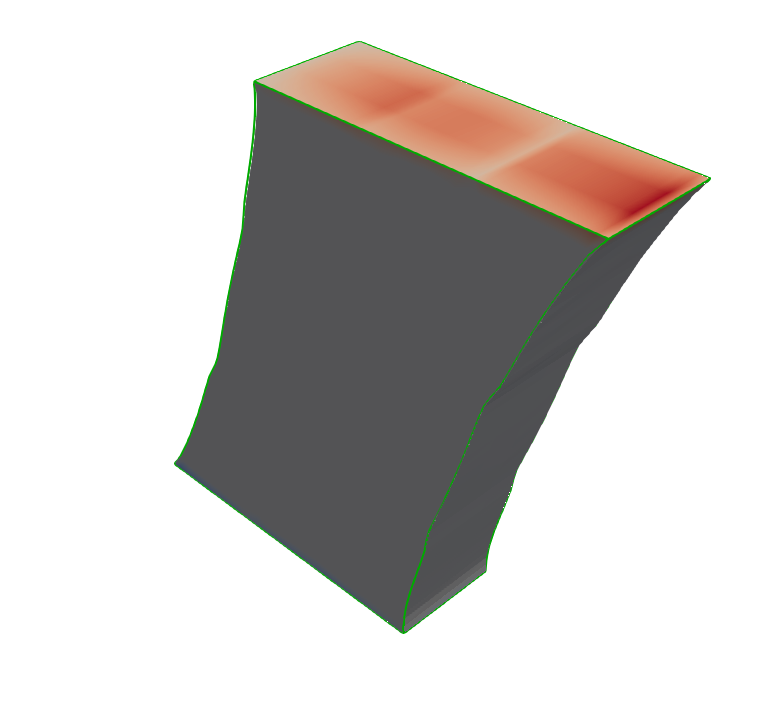}};
\node[anchor=north, inner sep=0] (image3) at ($(image1.south) + (0,0)$) {\includegraphics[width=0.5\textwidth]{figures/UC/force/reference_force.png}};
\node[anchor=north, inner sep=0] (image4) at ($(image2.south) + (0,0)$) {\includegraphics[width=0.5\textwidth]{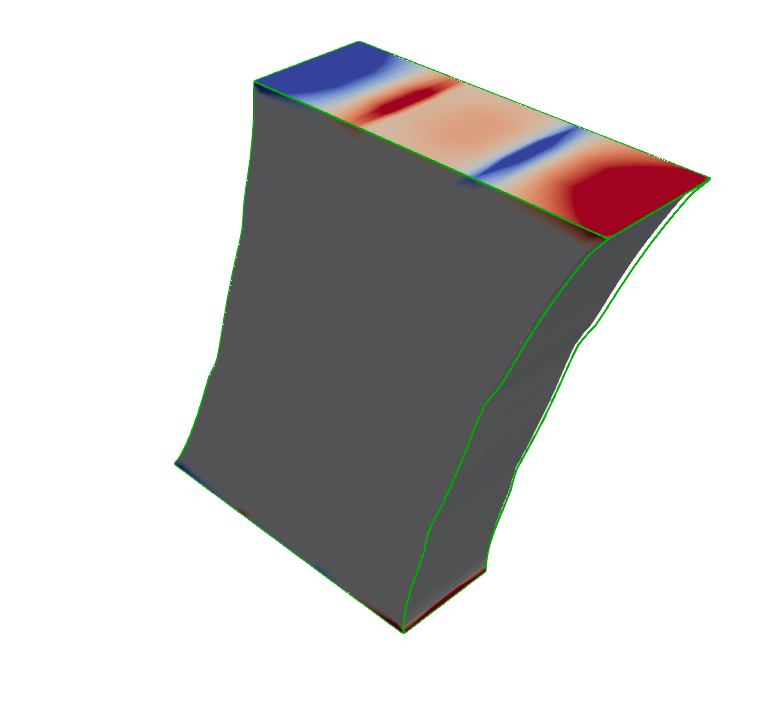}};
\node[anchor=south, inner sep=0] (name1) at ($(image1.north) + (+1cm,0)$) {Reference};
\node[anchor=south, inner sep=0] (name2) at ($(image2.north) + (+1cm,0)$) {Prediction};
\node[anchor=south, inner sep=0, rotate=90] (name3) at ($(image1.west) + (+0.7cm,+0.8cm)$) {Force-augmented};
\node[anchor=south, inner sep=0, rotate=90] (name4) at ($(image3.west) + (+0.7cm,+0.2cm)$) {Force-reconstructed};
\begin{axis}[
    at={($(image2.south east)!0.5!(image4.north east) + (-1cm,0)$)}, anchor=west,
    xshift=0.5cm,
    height=4cm,
    width=1cm,
    scale only axis,
    hide axis,
    colorbar,
    point meta min=0,
    point meta max=1,
    colormap={cooltowarm}{
        rgb(0cm)=(0.23137,0.29804,0.75294);
        rgb(0.5cm)=(0.865,0.865,0.865);
        rgb(1cm)=(0.70588,0.01569,0.14902)
    },
    colorbar style={
        ytick={0, 0.5, 1},
        yticklabels={-7.8, 0.0, 7.8},
        title=$f_y$ [N]
    }
]
\addplot [draw=none] coordinates {(0,0)};
\end{axis}
\end{tikzpicture}

%% file: figures/UC/force/UC_error_disp.tex
\begin{tikzpicture}
\node[anchor=west, inner sep=0] (image1) at (0,0) {\includegraphics[width=0.5\textwidth]{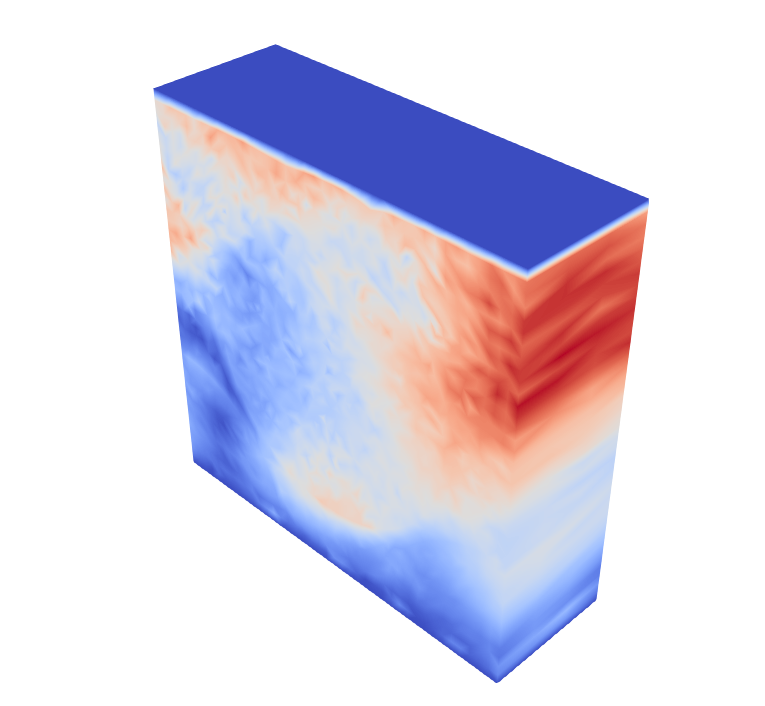}};
\node[anchor=west, inner sep=0] (image2) at ($(image1.east) + (-1.5cm,0)$) {\includegraphics[width=0.5\textwidth]{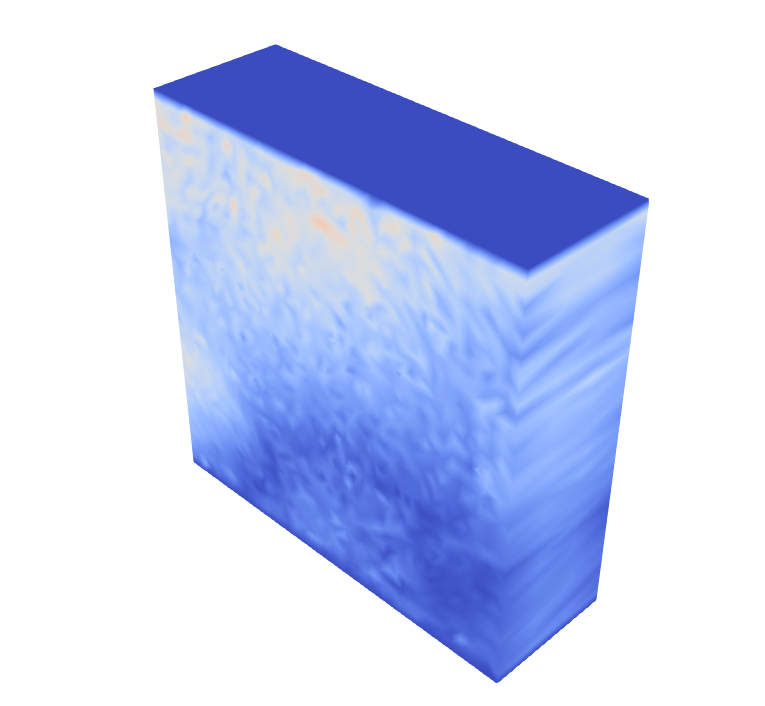}};
\node[anchor=south, inner sep=0] (name1) at ($(image1.north) + (+0.5cm,0)$) {Not augmented};
\node[anchor=south, inner sep=0] (name2) at ($(image2.north) + (+0.5cm,0)$) {Force-augmented};
\begin{axis}[
    at={($(image2.east) + (-1cm,0)$)}, anchor=west,
    xshift=0.5cm,
    height=4cm,
    width=1cm,
    scale only axis,
    hide axis,
    colorbar,
    point meta min=0,
    point meta max=1,
    colormap={cooltowarm}{
        rgb(0cm)=(0.23137,0.29804,0.75294);
        rgb(0.5cm)=(0.865,0.865,0.865);
        rgb(1cm)=(0.70588,0.01569,0.14902)
    },
    colorbar style={
        ytick={0, 0.5, 1},
        yticklabels={0, 0.02, 0.038},
        title={$\mathrm{Err}\left(\hat{\bm{\phi}}, \bm{\phi}\right)$ [mm]}
    }
]
\addplot [draw=none] coordinates {(0,0)};
\end{axis}
\end{tikzpicture}

%% file: figures/PWEH_sketch.tex
\begin{tikzpicture}
    \node (pic) at (0,0) {\includegraphics[width=0.6\textwidth]{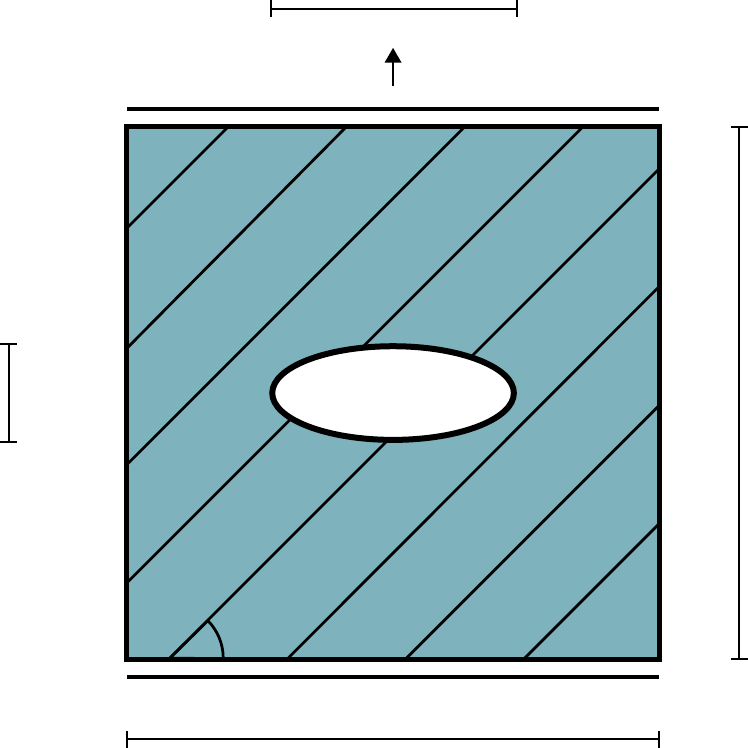}};
    \node[anchor=west] (height) at (pic.east) {10 [mm]};
    \node[anchor=north] (length) at (pic.south) {10 [mm]};
    \node[anchor=south] (a) at ($(pic.north)+(.2,0)$) {$2\, a$};
    \node[anchor=east] (b) at ($(pic.west)+(0,-.2)$) {$2\, b$};
    \node[anchor=west] (utop) at ($(a)+(.2,-1.2)$) {$\tilde{u}_y=5$ [mm]};
    \node[] (ubottom) at ($(length)+(+0,+.95)$) {$\tilde{\bm{u}}=\bm{0}$};
    \coordinate (origin) at ($(length)+(-5.5,+0.5)$);
    \draw[->] (origin) -- ($(origin)+(1,0)$) node[right] {$x$};
    \draw[->] (origin) -- ($(origin)+(0,1)$) node[above] {$y$};
    \node[anchor=west] (angle) at ($(length)+(-1.7,+1.8)$) {$\alpha$};
\end{tikzpicture}

%% file: figures/PWEH/morphing/morphings.tex
\begin{tikzpicture}
\node[anchor=west, inner sep=0] (image1) at (0,0) {\includegraphics[width=0.33\textwidth]{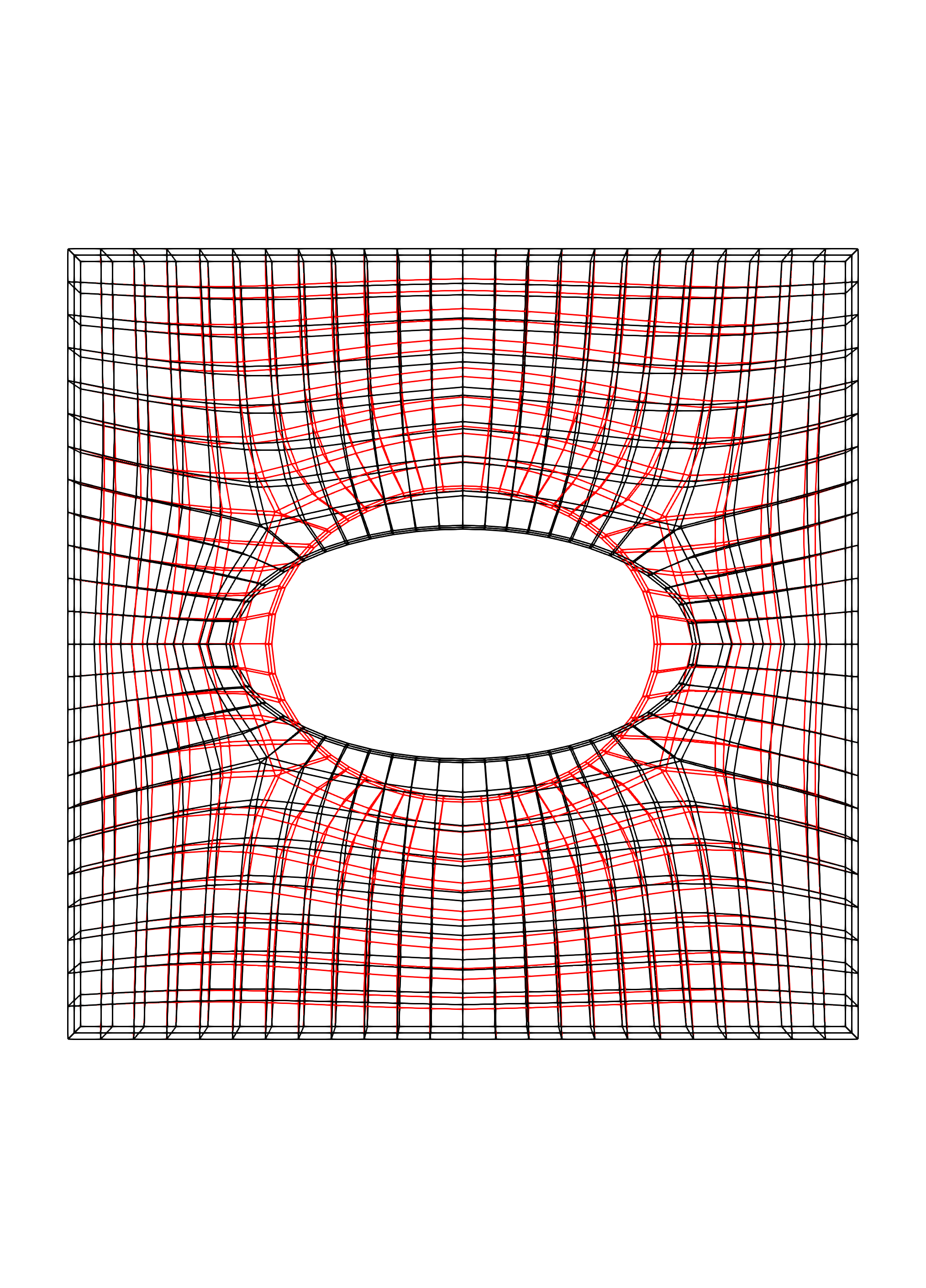}};
\node[anchor=west, inner sep=0] (image2) at ($(image1.east) + (0,0)$) {\includegraphics[width=0.33\textwidth]{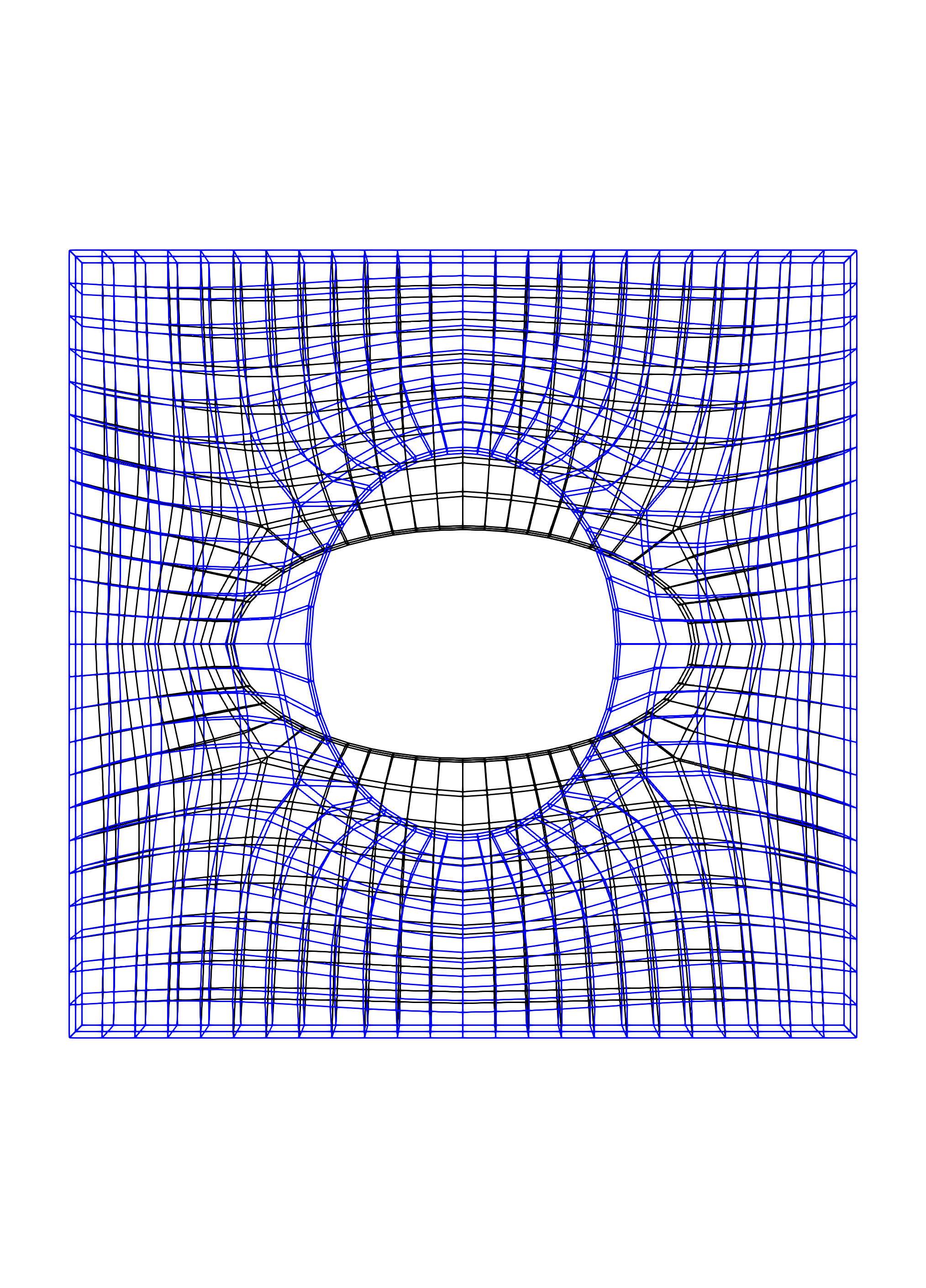}};
\node[anchor=west, inner sep=0] (image3) at ($(image2.east) + (0,0)$) {\includegraphics[width=0.33\textwidth]{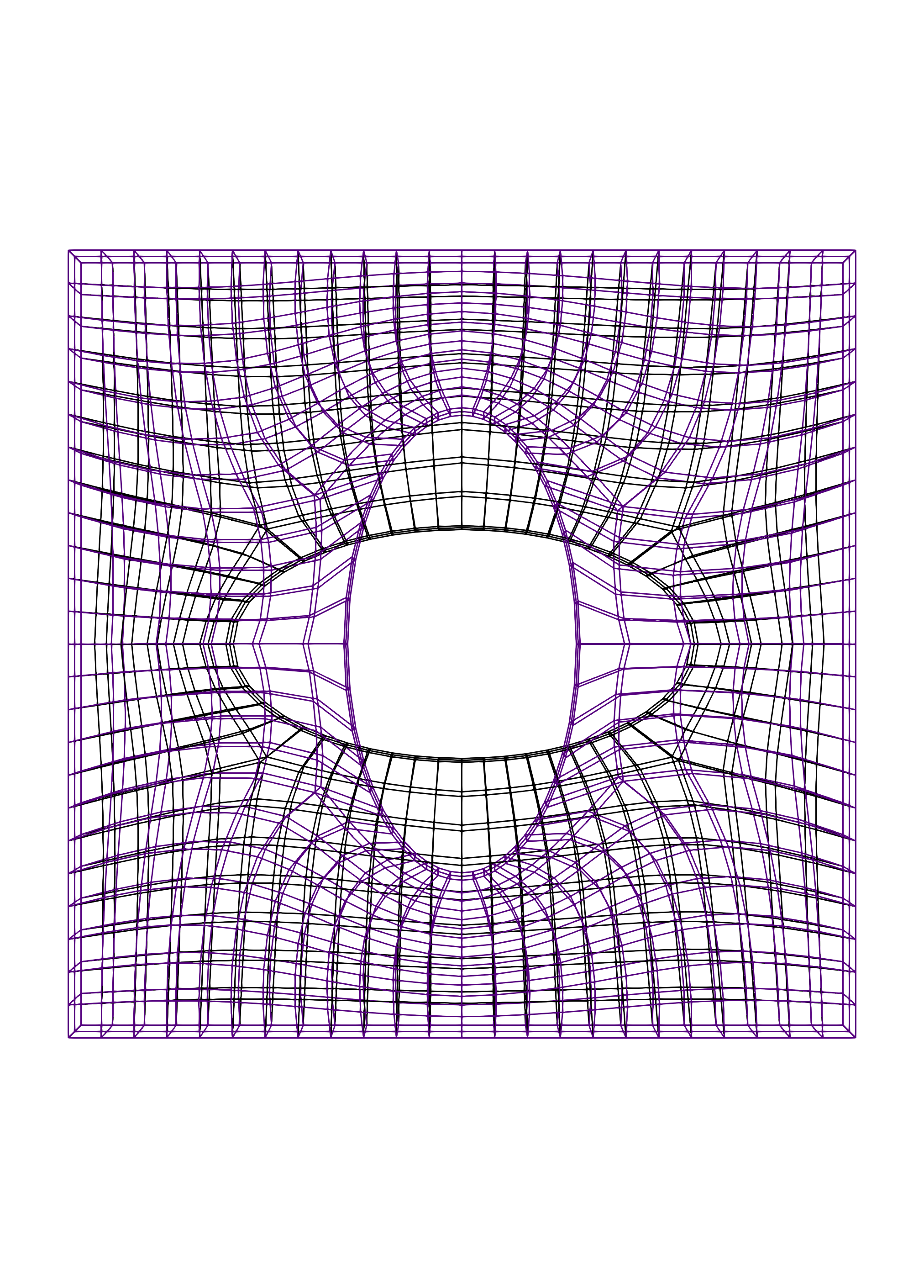}};
\node[anchor=south, inner sep=0] (name1) at ($(image1.north) + (0,-1cm)$) {$\xi=\frac{1}{3}$};
\node[anchor=south, inner sep=0] (name2) at ($(image2.north) + (0,-1cm)$) {$\xi=\frac{2}{3}$};
\node[anchor=south, inner sep=0] (name2) at ($(image3.north) + (0,-1cm)$) {$\xi=1$};
\end{tikzpicture}

%% file: figures/PWEH/PWEH_loss.tex
\begin{tikzpicture}[]
    \begin{axis}[xlabel= Nice $x$ label, ylabel= Nice $y$ label, width=55mm, height=35mm,
				hide axis,
				xmin=1,
				xmax=50,
				ymin=0,
				ymax=0.4,
				legend columns=-1,
				legend style={column sep=1mm}
				]
\addlegendimage{rwth1, line width=2pt}
\addlegendentry{Autoencoder}
\addlegendimage{rwth8, line width=2pt}
\addlegendentry{Latent Space Prediction} 
    \end{axis}
\end{tikzpicture}

\vspace*{.2cm}
\begin{tikzpicture}
\begin{semilogyaxis}[
    grid = major,
    xlabel = {epochs},
    ylabel = {Loss $\mathcal{L}$},
    width=0.45\textwidth,
    height=0.4\textwidth,
    /pgf/number format/1000 sep={},
    legend style={
        at={(0.5,1.05)},
        anchor=south,
        legend columns=-1, 
        /tikz/every even column/.append style={column sep=0.5em}
    }
]
    \addplot[rwth1, line width=2pt] table[x expr=\coordindex*10+1, y index=0] {figures/PWEH/morphing/lossAE_sampled.txt};
    \addplot[rwth8, line width=2pt] table[x expr=\coordindex*10+1, y index=0] {figures/PWEH/morphing/lossFFNtoLS_sampled.txt};
\end{semilogyaxis}
\end{tikzpicture}
\begin{tikzpicture}
\begin{semilogyaxis}[
    grid = major,
    xlabel = {epochs},
    ylabel = {Loss $\mathcal{L}$},
    width=0.45\textwidth,
    height=0.4\textwidth,
    /pgf/number format/1000 sep={},
]
    \addplot[rwth1, line width=2pt] table[x expr=\coordindex*10+1, y index=0] {figures/PWEH/displacement/lossAE_sampled.txt};
    \addplot[rwth8, line width=2pt] table[x expr=\coordindex*10+1, y index=0] {figures/PWEH/displacement/lossFFNtoLS_sampled.txt};
\end{semilogyaxis}
\end{tikzpicture}

%% file: figures/PWEH/morphing/morphing_comparison.tex
\begin{tikzpicture}
\node[anchor=west, inner sep=0] (image1) at (0,0) {\includegraphics[width=0.49\textwidth]{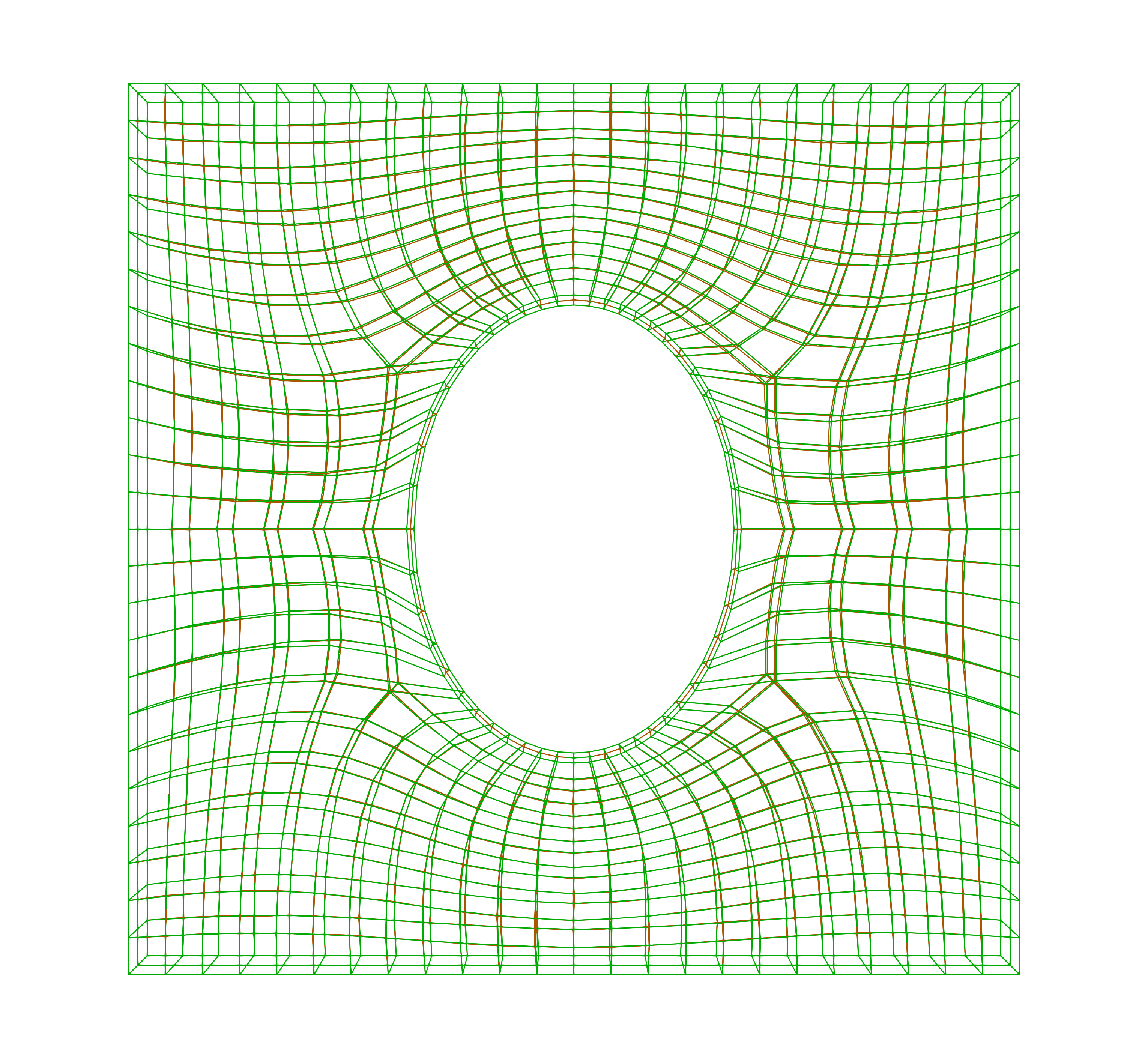}};
\node[anchor=west, inner sep=0] (image2) at ($(image1.east) + (0,0)$) {\includegraphics[width=0.49\textwidth]{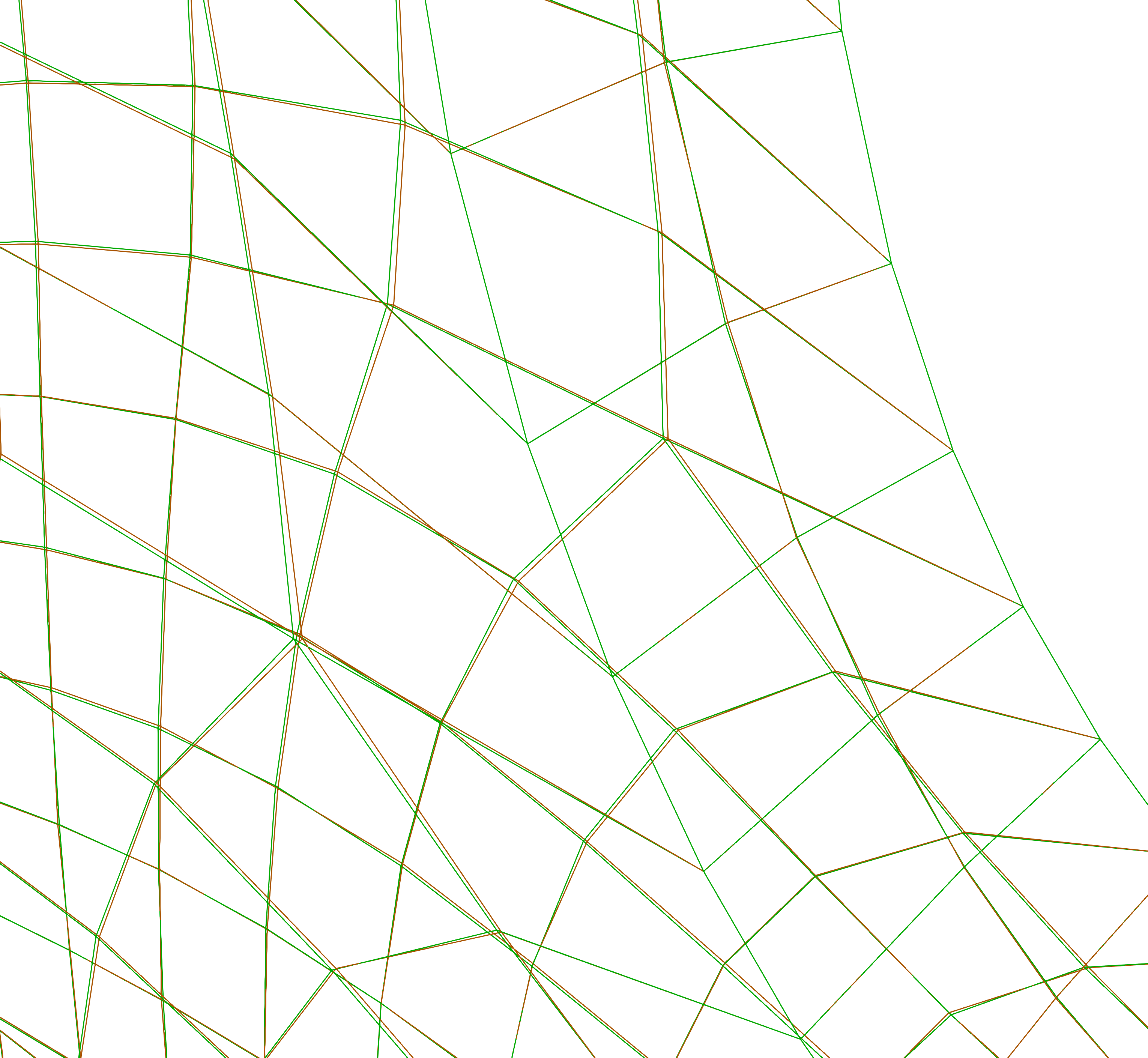}};
\draw[thick] (image2.south west) rectangle (image2.north east);
\draw[thick] (image1.center) rectangle ($(image1.center) + (-1.5,-1.5)$);
\coordinate (temp) at ($(image1.center) + (-1.5,-1.5)$);
\coordinate (pt1) at (image1.center |- temp);
\draw[thick] (pt1) -- (image2.west |- image3.south);
\draw[thick] (image1.center) -- (image2.west |- image2.north);
\end{tikzpicture}

%% file: figures/PWEH/displacement/displacement_comparison.tex
\begin{tikzpicture}
\node[anchor=west, inner sep=0] (image1) at (0,0) {\includegraphics[width=0.24\textwidth]{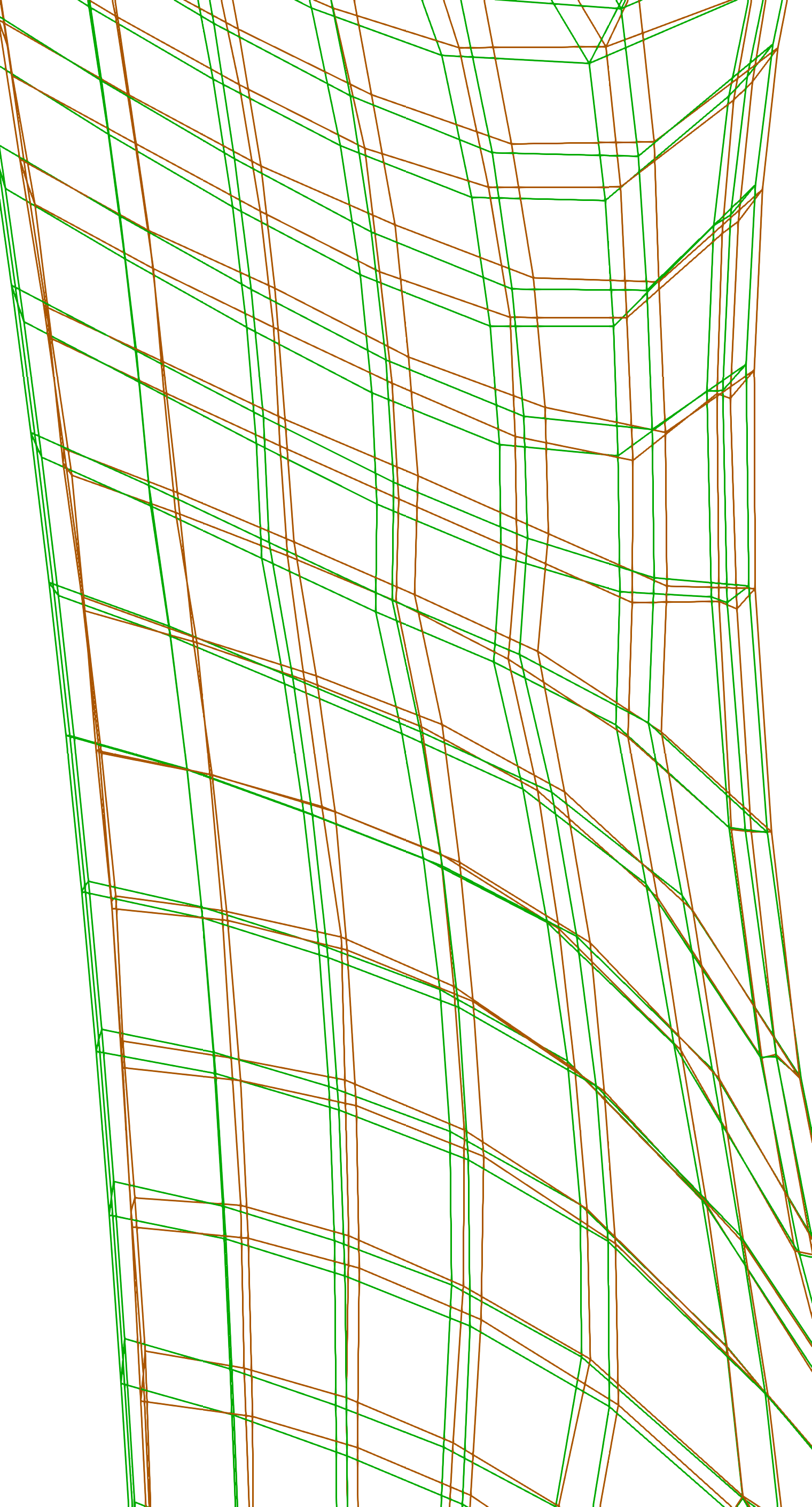}};
\node[anchor=west, inner sep=0] (image2) at ($(image1.east) + (0,0)$) {\includegraphics[width=0.49\textwidth]{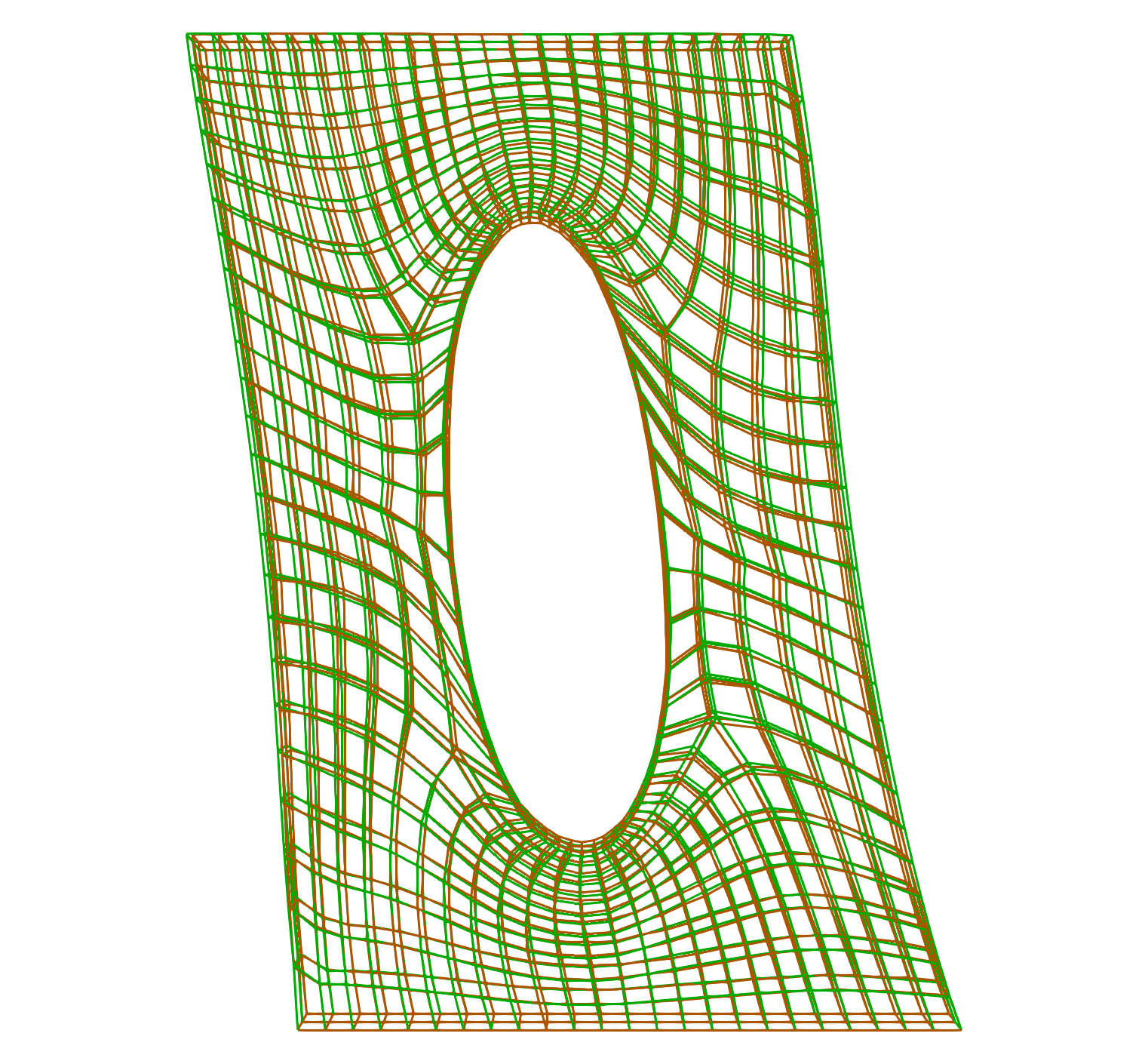}};
\node[anchor=west, inner sep=0] (image3) at ($(image2.east) + (0,0)$) {\includegraphics[width=0.24\textwidth]{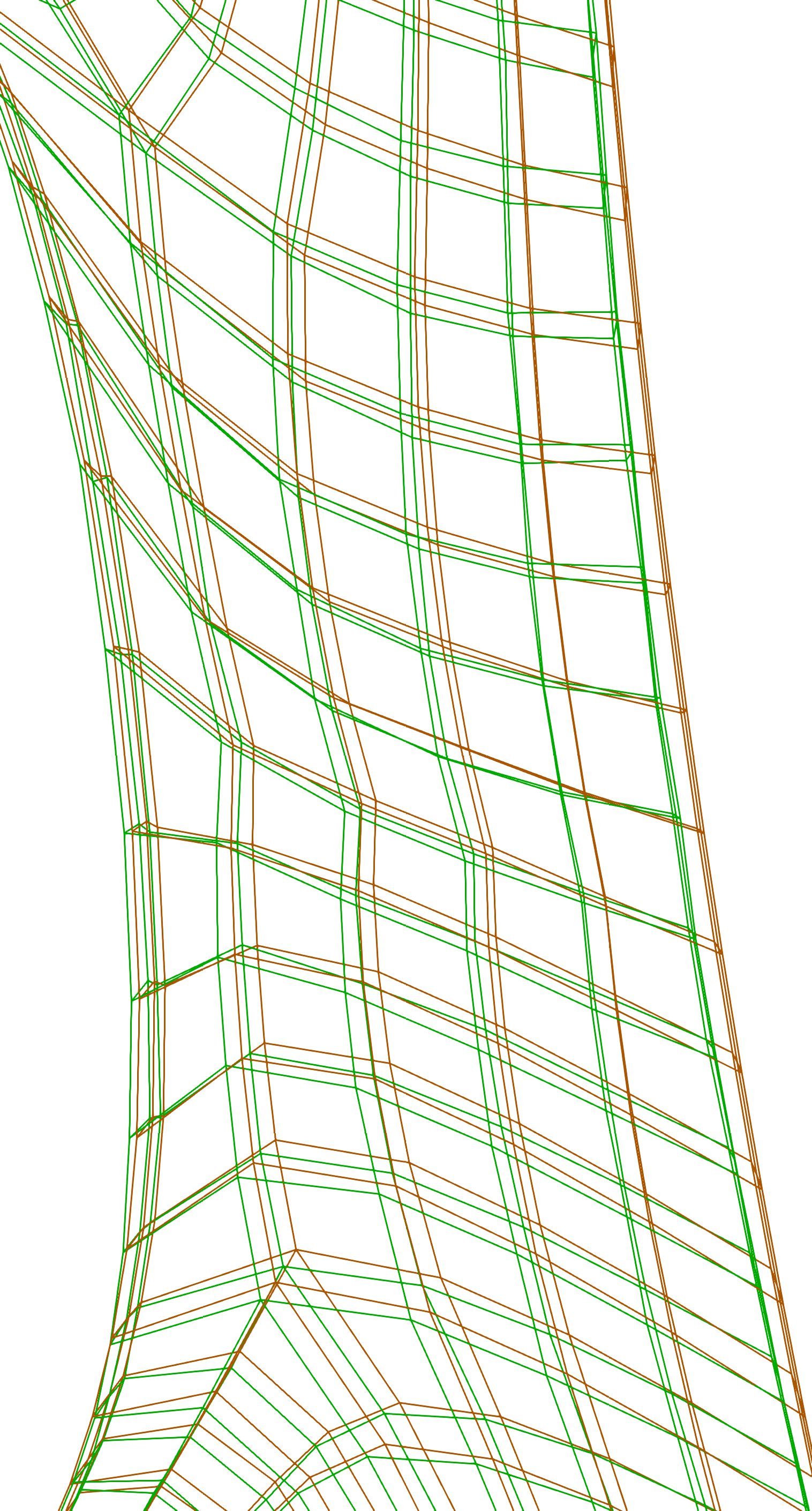}};
\draw[thick] (image1.south west) rectangle (image1.north east);
\draw[thick] (image3.south west) rectangle (image3.north east);
\draw[thick] ($(image2.south) + (-0.5,1.3)$) rectangle ($(image2.north west) + (1.3,-1.3)$);
\draw[thick] ($(image2.south) + (0.3,1.3)$) rectangle ($(image2.north east) + (-1.7,-1.3)$);
\draw[thick] ($(image2.north west) + (1.3,-1.3)$) -- (image1.south east |- image1.north west);
\coordinate (pt1) at ($(image2.south) + (-0.5,1.3)$);
\coordinate (pt2) at ($(image2.north west) + (1.3,-1.3)$);
\draw[thick] (pt2 |- pt1) -- (image1.south east |- image1.south west);
\coordinate (pt3) at ($(image2.south) + (0.3,1.3)$);
\coordinate (pt4) at ($(image2.north east) + (-1.7,-1.3)$);
\draw[thick] (pt4) -- (image3.south west |- image3.north);
\draw[thick] (pt4 |- pt3) -- (image3.south west |- image3.south);
\end{tikzpicture}

%% file: figures/PWEH_Thermo_sketch.tex
\begin{tikzpicture}
    \node (pic) at (0,0) {\includegraphics[width=0.6\textwidth]{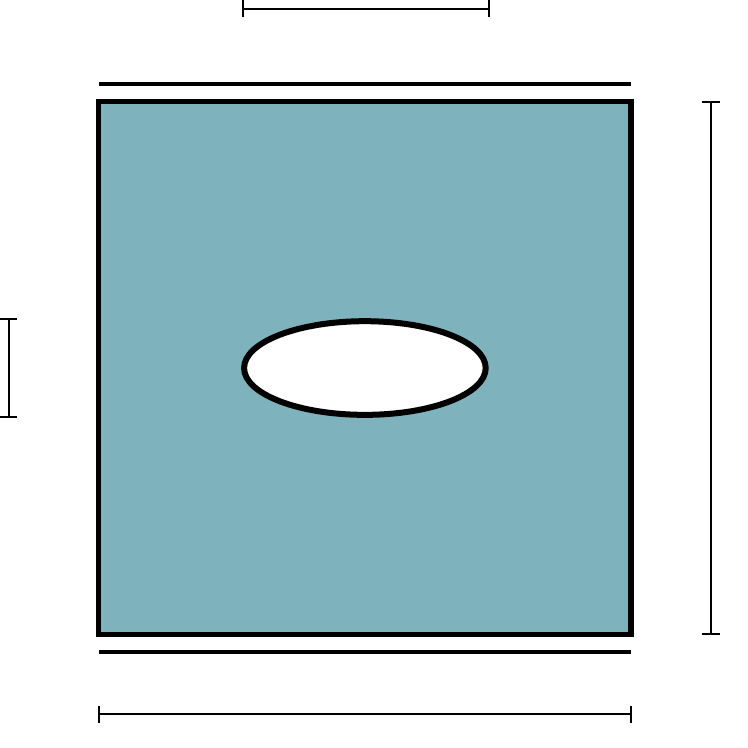}};
    \node[anchor=west] (height) at ($(pic.east)+(-0.3,0.0)$) {10 [mm]};
    \node[anchor=north] (length) at ($(pic.south)+(0.0,0.3)$) {10 [mm]};
    \node[anchor=south] (a) at ($(pic.north)+(0.0,-0.1)$) {6 [mm]};
    \node[anchor=east] (b) at ($(pic.west)+(0,0.1)$) {3 [mm]};
    \node[] (utop) at ($(a)+(0.0,-1.0)$) {$\tilde{u}_y=0,\ \tilde{T} = 10\, t$ [$^\circ C$]};
    \node[] (ubottom) at ($(length)+(+0,0.9)$) {$\tilde{\bm{u}}=\bm{0},\ \tilde{T} = 0$ [$^\circ C$]};
    \coordinate (origin) at ($(length)+(-5.5,+0.5)$);
    \draw[->] (origin) -- ($(origin)+(1,0)$) node[right] {$x$};
    \draw[->] (origin) -- ($(origin)+(0,1)$) node[above] {$y$};
\end{tikzpicture}

%% file: figures/Thermo_imperfection/Thermo_loss.tex
\begin{tikzpicture}[]
    \begin{axis}[xlabel= Nice $x$ label, ylabel= Nice $y$ label, width=55mm, height=35mm,
				hide axis,
				xmin=1,
				xmax=50,
				ymin=0,
				ymax=0.4,
				legend columns=-1,
				legend style={column sep=1mm}
				]
\addlegendimage{rwth1, line width=2pt}
\addlegendentry{Autoencoder}
\addlegendimage{rwth8, line width=2pt}
\addlegendentry{Latent Space $\bm{u}$} 
\addlegendimage{rwth9, line width=2pt}
\addlegendentry{Latent Space $T$} 
    \end{axis}
\end{tikzpicture}

\begin{tikzpicture}
\begin{semilogyaxis}[
    grid = major,
    xlabel = {epochs},
    ylabel = {Loss $\mathcal{L}$},
    width=0.45\textwidth,
    height=0.4\textwidth,
    /pgf/number format/1000 sep={},
    legend style={
        at={(0.5,1.05)},
        anchor=south,
        legend columns=-1, 
        /tikz/every even column/.append style={column sep=0.5em}
    }
]
    \addplot[rwth1, line width=2pt] table[x expr=\coordindex*10+1, y index=0] {figures/Thermo_imperfection/all_snapshots/lossAE_sampled.txt};
    \addplot[rwth8, line width=2pt] table[x expr=\coordindex*10+1, y index=0] {figures/Thermo_imperfection/all_snapshots/lossFFNtoLS_disp_sampled.txt};
    \addplot[rwth9, line width=2pt] table[x expr=\coordindex*10+1, y index=0] {figures/Thermo_imperfection/all_snapshots/lossFFNtoLS_Temp_sampled.txt};
\end{semilogyaxis}
\end{tikzpicture}
\begin{tikzpicture}
\begin{semilogyaxis}[
    grid = major,
    xlabel = {epochs},
    ylabel = {Loss $\mathcal{L}$},
    width=0.45\textwidth,
    height=0.4\textwidth,
    /pgf/number format/1000 sep={},
]
    \addplot[rwth1, line width=2pt] table[x expr=\coordindex*10+1, y index=0] {figures/Thermo_imperfection/less_snapshots/lossAE_sampled.txt};
    \addplot[rwth8, line width=2pt] table[x expr=\coordindex*10+1, y index=0] {figures/Thermo_imperfection/less_snapshots/lossFFNtoLS_disp_sampled.txt};
    \addplot[rwth9, line width=2pt] table[x expr=\coordindex*10+1, y index=0] {figures/Thermo_imperfection/less_snapshots/lossFFNtoLS_Temp_sampled.txt};
\end{semilogyaxis}
\end{tikzpicture}

%% file: figures/Thermo_imperfection/comparison_themo.tex
\begin{tikzpicture}
\node[anchor=west, inner sep=0] (image1) at (0,0) {\includegraphics[width=0.4\textwidth]{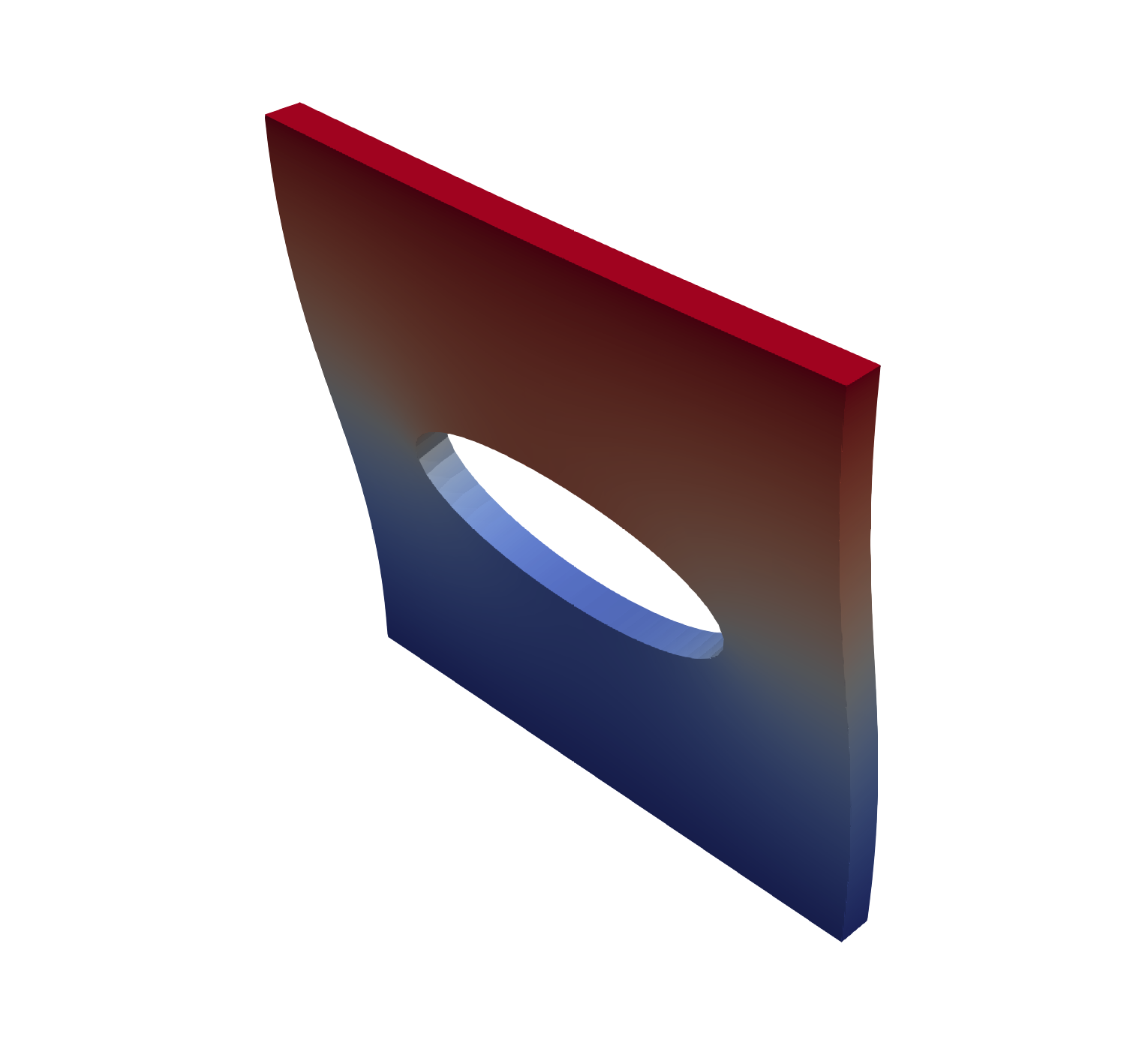}};
\node[anchor=west, inner sep=0] (image2) at ($(image1.east) + (-1cm,0)$) {\includegraphics[width=0.4\textwidth]{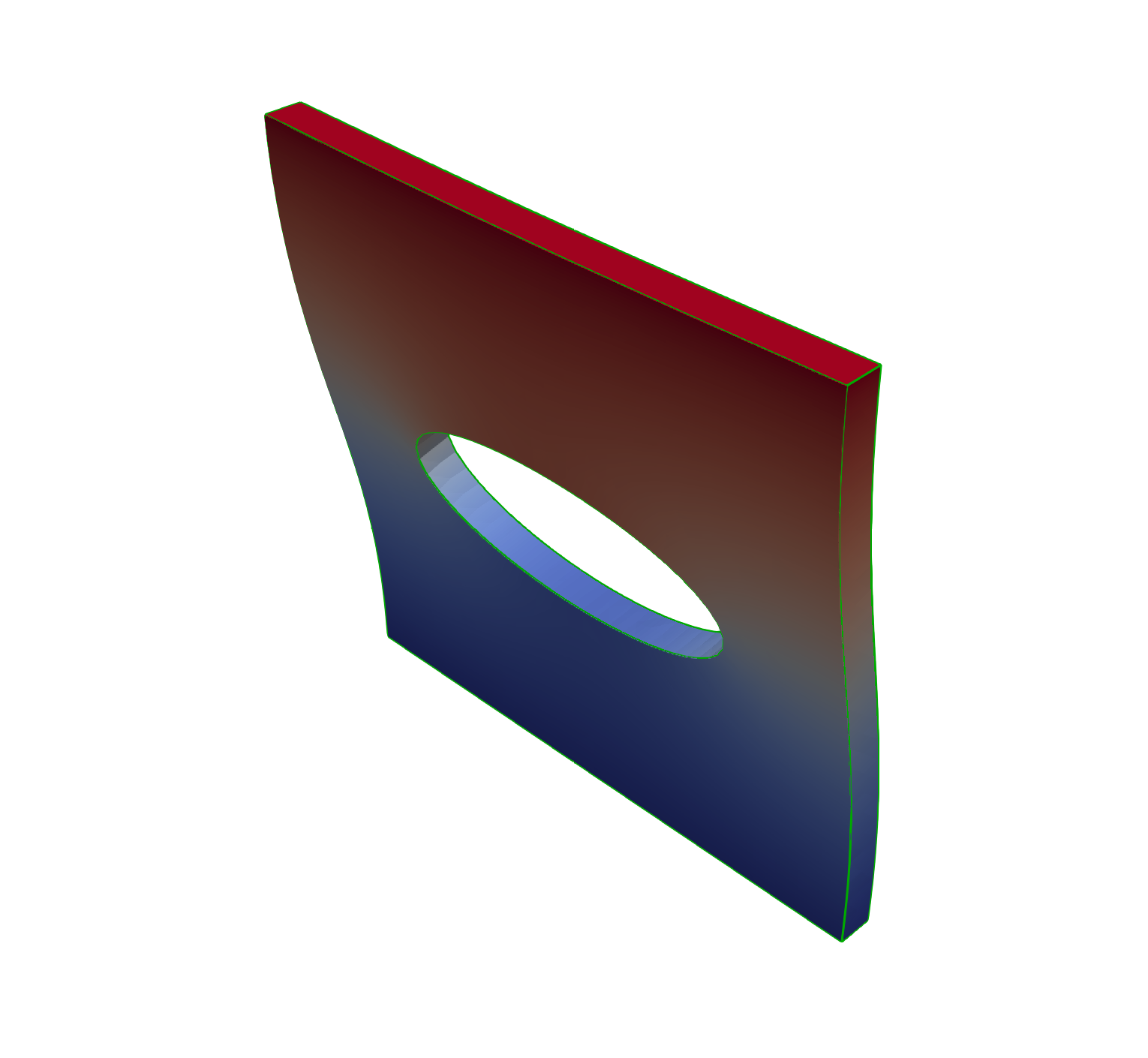}};
\node[anchor=west, inner sep=0] (image3) at ($(image2.east) + (-1cm,0)$) {\includegraphics[width=0.4\textwidth]{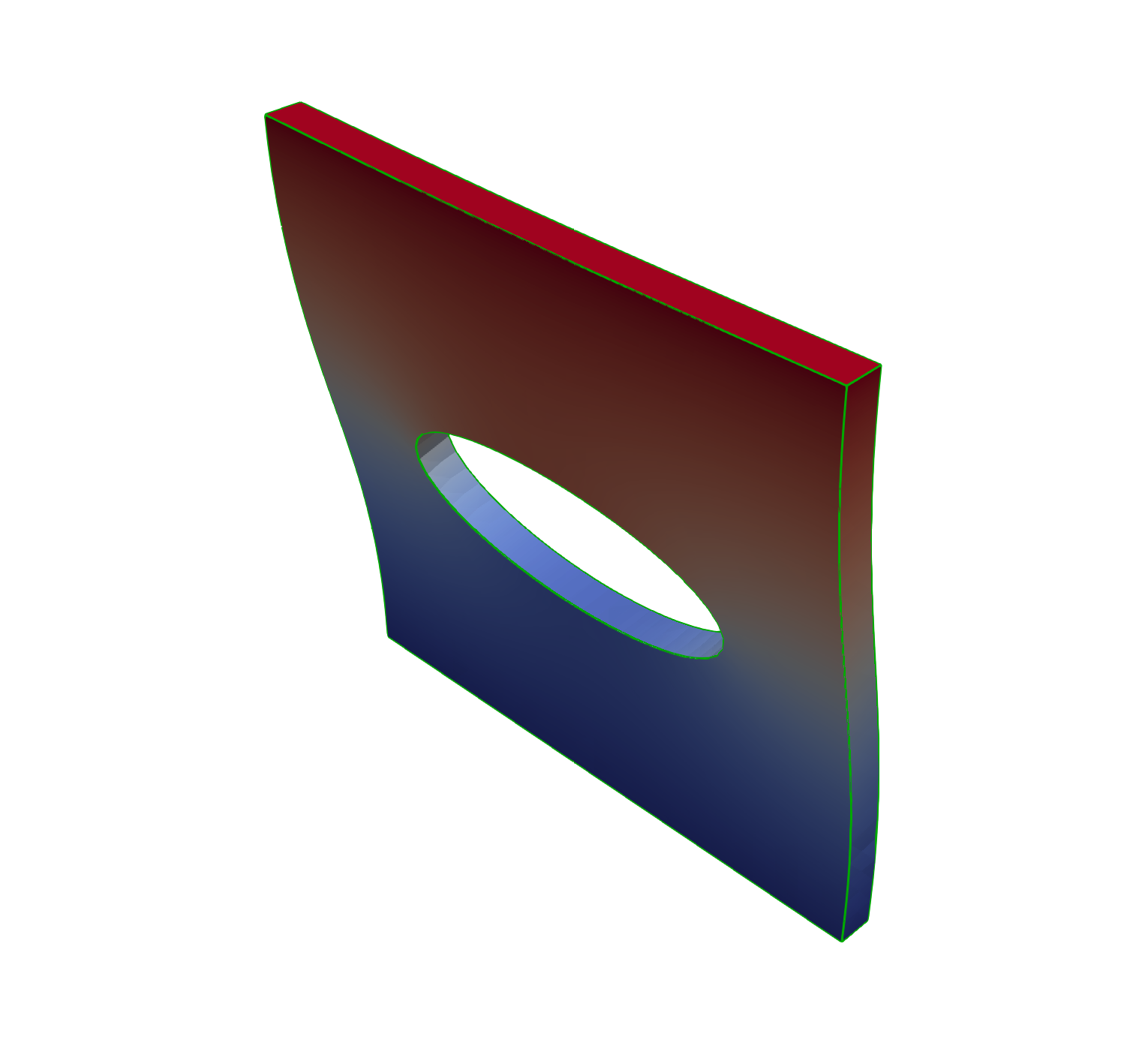}};
\node[anchor=south, inner sep=0] (name1) at ($(image1.north) + (0,0)$) {Reference};
\node[anchor=south, inner sep=0] (name2) at ($(image2.north) + (0,0)$) {\shortstack{Prediction \\ (1600 snapshots)}};
\node[anchor=south, inner sep=0] (name3) at ($(image3.north) + (0,0)$) {\shortstack{Prediction \\ (528 snapshots)}};
\begin{axis}[
    at={($(image2.south) + (0,+1.5cm)$)}, anchor=north,
    xshift=0.5cm,
    height=2cm,  
    width=4cm,
    scale only axis,
    hide axis,
    colorbar horizontal,  
    point meta min=0,
    point meta max=1,
    colormap={cooltowarm}{
        rgb(0cm)=(0.23137,0.29804,0.75294);
        rgb(0.5cm)=(0.865,0.865,0.865);
        rgb(1cm)=(0.70588,0.01569,0.14902)
    },
    colorbar style={
        xtick={0, 0.5, 1},
        xticklabels={0, 4, 8},
        title={$T$ [$^\circ C$]}
    }
]
\addplot [draw=none] coordinates {(0,0)};
\end{axis}
\end{tikzpicture}

%% file: figures/Thermo_imperfection/comparison_disp.tex
\begin{tikzpicture}
\node[anchor=west, inner sep=0] (image1) at (0,0) {\includegraphics[width=0.49\textwidth]{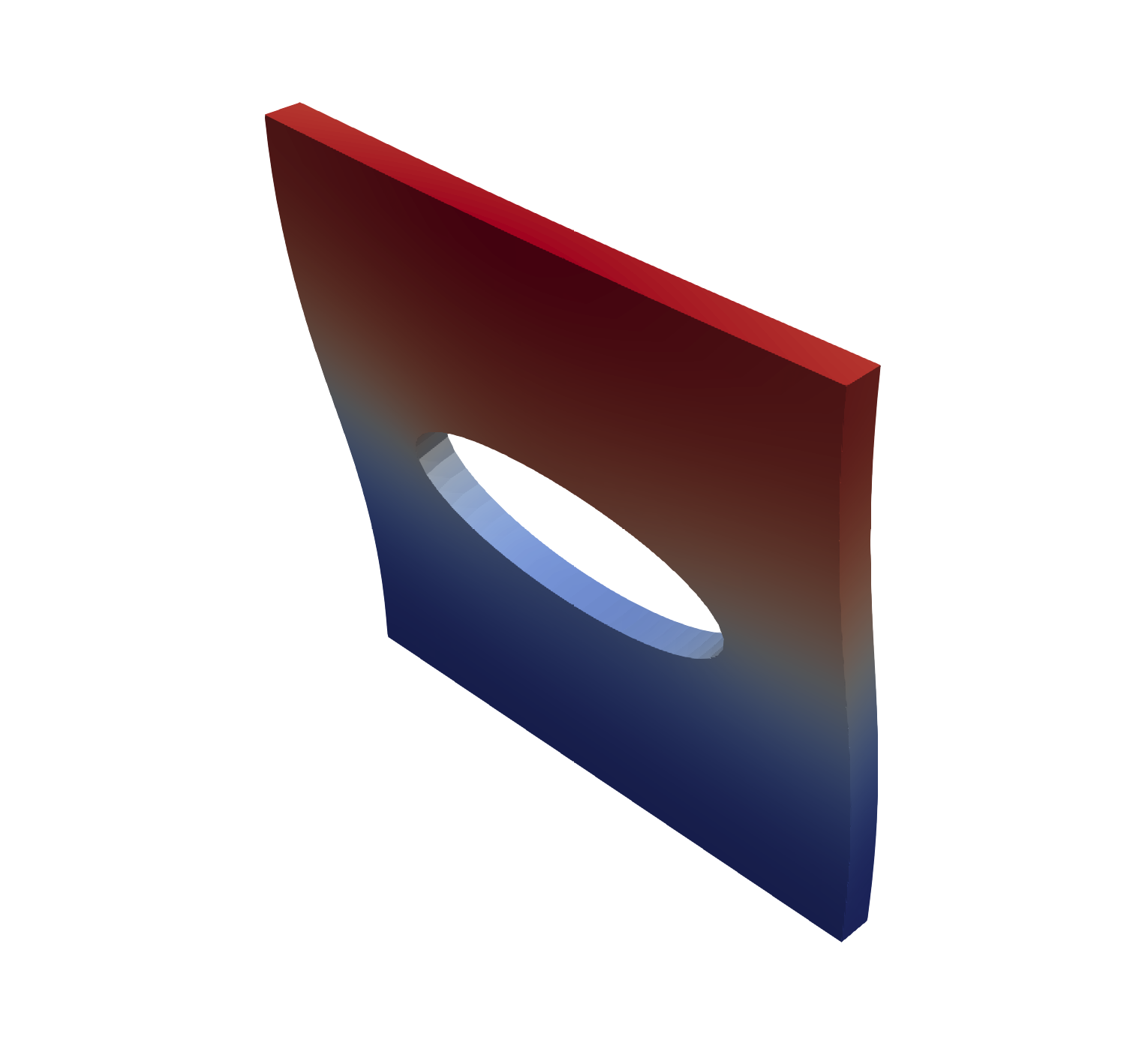}};
\node[anchor=north east, inner sep=0] (image2) at ($(image1.south) + (0,0)$) {\includegraphics[width=0.49\textwidth]{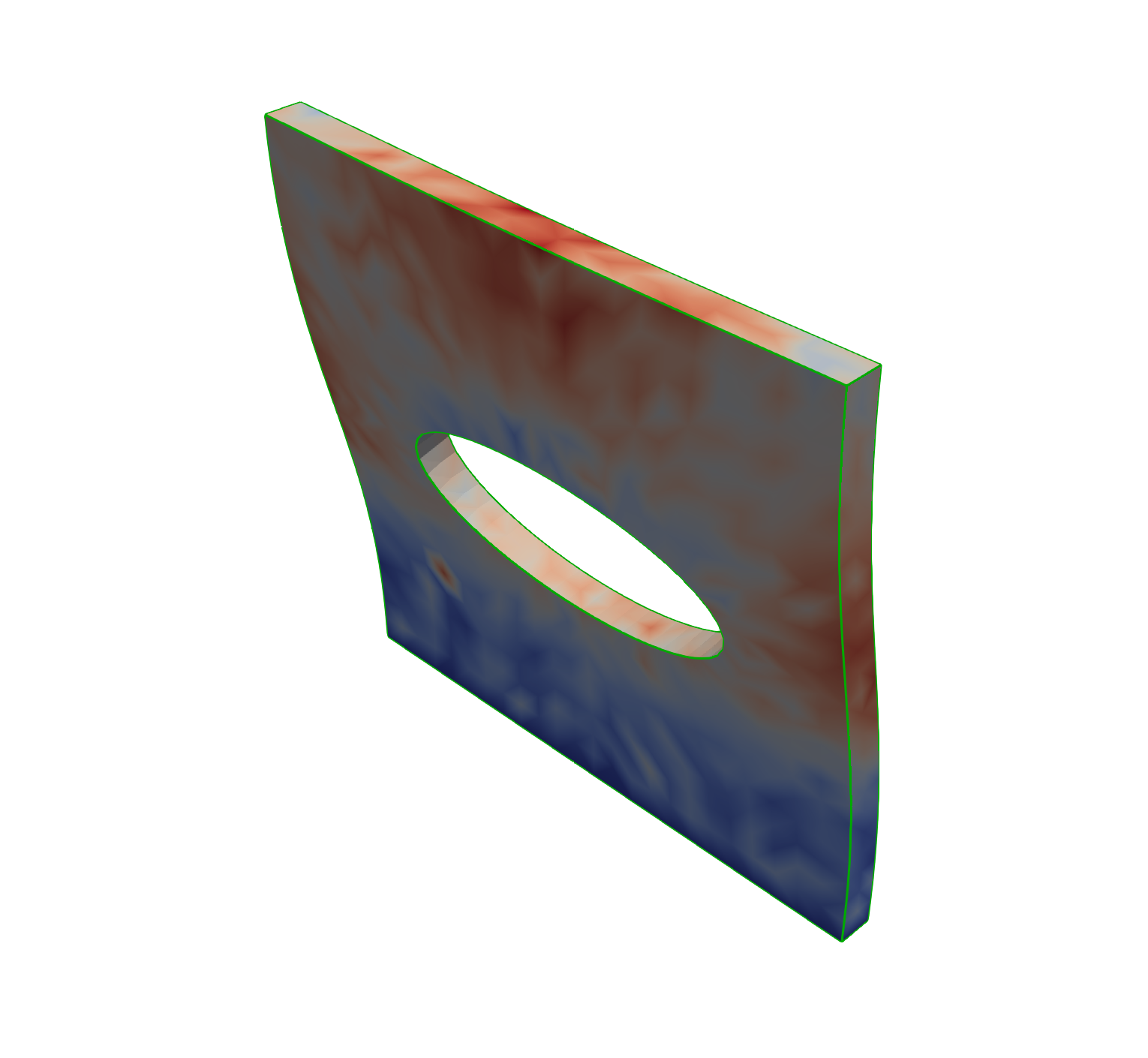}};
\node[anchor=north west, inner sep=0] (image3) at ($(image1.south) + (0,0)$) {\includegraphics[width=0.49\textwidth]{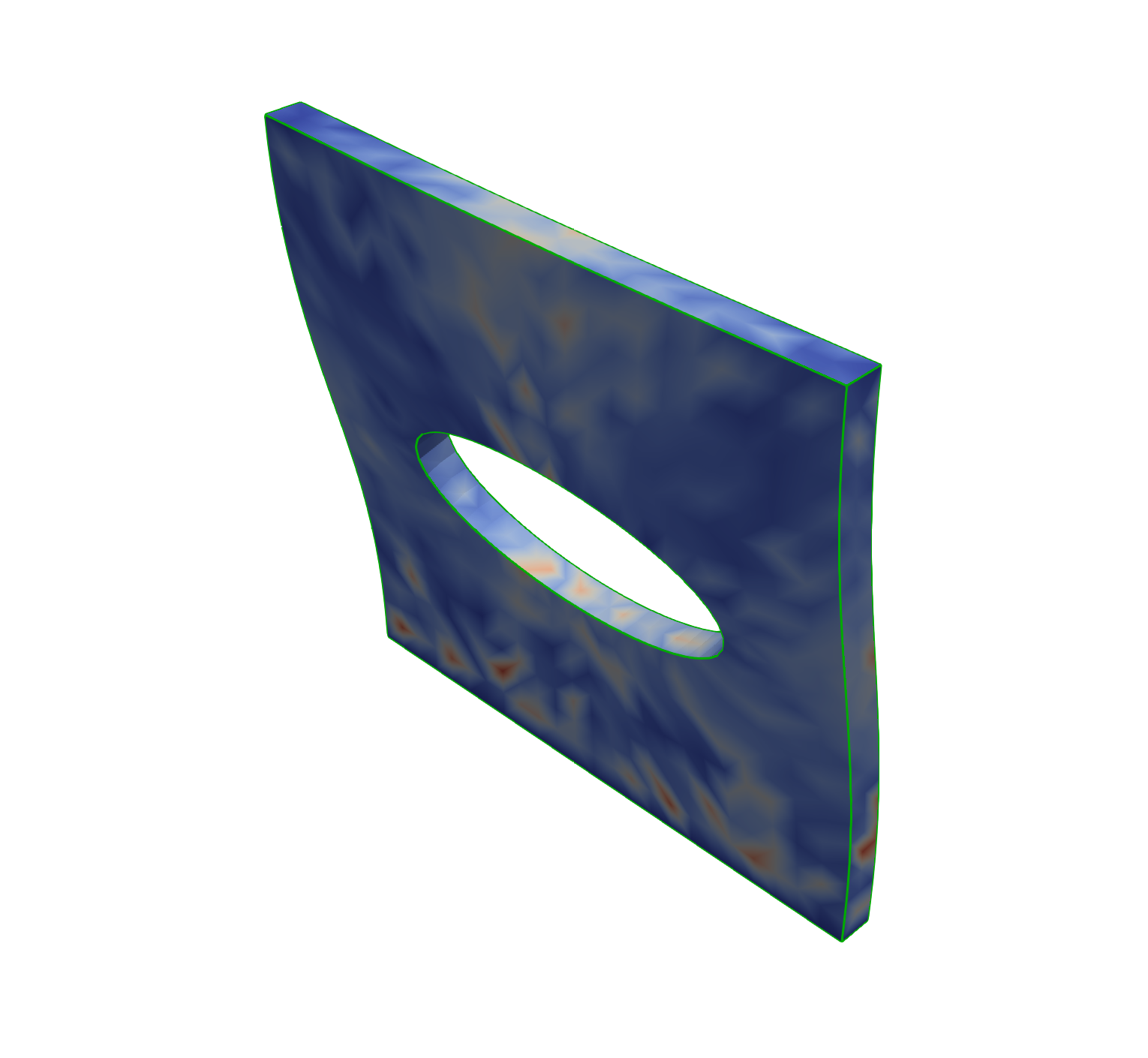}};
\node[anchor=south, inner sep=0] (name1) at ($(image1.north) + (0,0)$) {Reference};
\node[anchor=north, inner sep=0] (name2) at ($(image2.south) + (0,0)$) {\shortstack{Prediction \\ (1600 snapshots)}};
\node[anchor=north, inner sep=0] (name3) at ($(image3.south) + (0,0)$) {\shortstack{Prediction \\ (528 snapshots)}};
\begin{axis}[
    at={($(image1.east) + (-1.5cm,0)$)}, anchor=west,
    xshift=0.5cm,
    height=3cm,
    width=1.5cm,
    scale only axis,
    hide axis,
    colorbar,
    point meta min=0,
    point meta max=1,
    colormap={cooltowarm}{
        rgb(0cm)=(0.23137,0.29804,0.75294);
        rgb(0.5cm)=(0.865,0.865,0.865);
        rgb(1cm)=(0.70588,0.01569,0.14902)
    },
    colorbar style={
        ytick={0, 0.5, 1},
        yticklabels={0, 0.75, 1.5},
        title={$\lVert \bm{u} \rVert$ [mm]}
    }
]
\addplot [draw=none] coordinates {(0,0)};
\end{axis}
\begin{axis}[
    at={($(image3.east) + (-2.5cm,0)$)}, anchor=west,
    xshift=0.5cm,
    height=3cm,
    width=1.5cm,
    scale only axis,
    hide axis,
    colorbar,
    point meta min=0,
    point meta max=1,
    colormap={cooltowarm}{
        rgb(0cm)=(0.23137,0.29804,0.75294);
        rgb(0.5cm)=(0.865,0.865,0.865);
        rgb(1cm)=(0.70588,0.01569,0.14902)
    },
    colorbar style={
        ytick={0, 0.5, 1},
        yticklabels={0, 0.01, 0.019},
        title={$\mathrm{Err}\left(\hat{\bm{u}},\bm{u}\right)$ [mm]}
    }
]
\addplot [draw=none] coordinates {(0,0)};
\end{axis}
\end{tikzpicture}

%% file: sections/section4.tex
\section{Discussion}
\label{sec4:1}

We presented an Autoencoder-based framework for non-intrusive model order reduction (MOR) in parameterized continuum mechanics. The method combines (i) an unsupervised Autoencoder to compress high-dimensional finite element (FE) snapshots into a low-dimensional latent space, (ii) a supervised regression network to map input parameters to latent codes, and (iii) an end-to-end surrogate model to reconstruct full-field solutions. Two key extensions were introduced: a force-augmented variant, which jointly predicts reaction forces and displacements using a shared latent representation, and a multi-physics extension, which employs dedicated encoder–decoder pairs for each physical field and concatenates their latent representations.

The framework was validated on three benchmark problems: (i) a heterogeneous unit cell problem with simultaneous prediction of displacement fields and reaction forces, (ii) a fiber-reinforced plate with an elliptic hole involving parameterized geometry and anisotropy, and (iii) a transient thermo-mechanical problem, demonstrating the multi-field capability by predicting both temperature and displacement fields. Across all examples, the approach achieved accurate reconstructions of high-fidelity FE solutions while being entirely non-intrusive and straightforward to extend to new fields or quantities of interest.

Compared to classical projection-based techniques such as Proper Orthogonal Decomposition or reduced-basis methods, our framework bypasses the need for intrusive code modifications and expert-driven basis construction. It aligns with the broader trend of end-to-end learning in scientific computing, where modern generative models have shown remarkable ability to extract and compress physical patterns. These findings highlight the potential of data-driven approaches to learn low-dimensional manifolds of physical systems directly from solution data, thus providing an accessible and flexible alternative to traditional MOR.

Recent studies have explored Autoencoder-based model order reduction (MOR), yet key differences distinguish our framework. Simpson et al.~\cite{SimpsonEtAl2021} addressed nonlinear dynamics using LSTMs but did not model parameter dependence or multi-field outputs. Deshpande et al.~\cite{DeshpandeRappelEtAl2025} combined Autoencoders with Gaussian processes for probabilistic surrogates, focusing on uncertainty rather than architectural modularity. Graph-based methods like Pichi et al.~\cite{PichiMoyaEtAl2024} emphasize spatial expressiveness but remain limited to single-field settings. Finally, Kneifl et al.~\cite{KneiflRosinEtAl2023} targeted a specific biomechanical system without addressing force prediction or mesh morphing. In contrast, our work proposes a general, extensible architecture for multi-physics, reaction force augmentation, and geometry-aware surrogate modeling.

Nonetheless, the current approach has limitations. It lacks extrapolation capability beyond the training domain, exhibits residual inconsistencies between predicted fields (e.g., forces and displacements), and relies on a fixed computational mesh—geometry variations require an additional morphing step. Moreover, the method compresses solution information but does not reduce the number of degrees of freedom in the FE system itself.

Overall, this work constitutes a step towards flexible, data-driven surrogate modeling for complex continuum mechanical systems. Future efforts will focus on improving physical consistency via physics-informed loss functions, enhancing generalization, and exploring more expressive architectures such as convolutional or graph-based networks. These developments may ultimately enable machine learning–based MOR to complement or replace traditional intrusive techniques in selected applications.

%% file: sections/section5.tex
\section{Conclusion}
\label{sec5:1}

In summary, the proposed framework demonstrates that nonlinear Autoencoder-based MOR can accurately and efficiently approximate complex finite element solutions while remaining non-intrusive and easily extensible to additional physical fields. These findings highlight the potential of data-driven approaches to make reduced-order modeling more accessible and versatile, with further developments — such as physics-informed constraints and expressive network architectures — promising to enhance extrapolation capabilities and support efficient, general-purpose surrogate models that may eventually be integrated into hybrid finite element codes.

%% file: sections/appendixA.tex
\section{Additions}
\subsection{Analogy to intrusive Model Order Reduction}
\label{app:analogy}
Let us briefly discuss the relationship between our approach and classical intrusive model order reduction techniques. 
For clarity and without loss of generality, we focus on projection-based model order reduction, specifically Proper Orthogonal Decomposition.
In the numerical solution of partial differential equations, the solution field is typically obtained by (iteratively) solving a linear system of the form
\begin{equation}
    \bm{K} \bm{\phi} = \bm{r},
    \label{eq:MOR_Kuf}
\end{equation}
where $\bm{K}$ denotes the stiffness matrix of the system and $\bm{r}$ the corresponding right-hand side vector. 
Regardless of the specific numerical method employed, solving this system is computationally expensive. 
To mitigate this cost, the system is projected onto a reduced-order subspace.

To construct such a subspace, a collection of solution snapshots $\{\bm{\phi}^s\}_{s=1}^{n_S} \in \mathbb{R}^{n_\text{dim}\times 1}$ is assembled into the snapshot matrix $\bm{\Phi} = \left[\bm{\phi}^1, \hdots, \bm{\phi}^{n_S}\right]$. 
A singular value decomposition (SVD) is then performed
\begin{equation}
    \bm{\Phi} = \bm{\Omega}\bm{\Sigma}\bm{V}^T,
    \label{eq:MOR_SVD}
\end{equation}
from which a reduced basis can be extracted by the columns of $\bm{\Omega}\in\mathbb{R}^{n_\text{dim}\times n_\text{dim}}$, i.e., $\bm{\Xi} = \left[\bm{\Omega}^1, \hdots, \bm{\Omega}^r\right]$, where $r \ll n_\text{dim}$ is the reduced dimension. 
This choice of basis dimensionality is analogous to the selection of the number of latent neurons in the Autoencoder. 
Similarly, the use of SVD as a basis generation mechanism loosely corresponds to the architectural choices (e.g., feedforward, convolutional) in encoder-decoder networks.

The projection matrix is chosen in a way such that the high-dimensional solution space can be approximated by
\begin{equation}
    \bm{\phi} \approx \bm{\Xi}\,\bm{a},
    \label{eq:MOR_reduce}
\end{equation}
where $\bm{a}\in\mathbb{R}^{r\times 1}$ represents the reduced solution. 
Substituting this relation into the governing system~\eqref{eq:MOR_Kuf} and applying a Galerkin projection yields the reduced system
\begin{equation}
    \tilde{\bm{K}} \bm{a} = \tilde{\bm{r}}, \quad \text{with} \quad \tilde{\bm{K}} = \bm{\Xi}^T \bm{K} \bm{\Xi}, \quad \tilde{\bm{r}} = \bm{\Xi}^T \bm{r}.
    \label{eq:MOR_reduced_system}
\end{equation}
The solution $\bm{a}$ in the reduced space can then be mapped back to the high-dimensional space via Equation~\eqref{eq:MOR_reduce}.

However, in many practical scenarios, the operators $\bm{K}$ and $\bm{r}$ depend nonlinearly on the state variable $\bm{\phi}$. 
In such cases, the linear subspace projection alone may fail to accurately capture the system dynamics. To address this limitation, nonlinear projection techniques such as quadratic manifold or locally linear embedding approaches are employed.
The overall process of projection, reduced system solution, and back-projection can be interpreted as an analogue to the encoder-decoder mechanism in the Autoencoder, with the training phase corresponding to the offline construction of the projection matrix.

Furthermore, when system parameters vary, one must interpolate not only $\bm{K}$ and $\bm{r}$, but also the reduced basis $\bm{\Xi}$. 
While standard interpolation methods may suffice for the former, interpolating the basis matrix $\bm{\Xi}$ is generally more complex. 
One established technique in this context is interpolation on the Grassmann manifold (see, e.g., \cite{FriderikosEtAl2022}). 
This operation is conceptually similar to the supervised learning of latent representations in our framework.
It is important to emphasize that this comparison is necessarily simplified and is intended solely to highlight conceptual parallels. 
Nonetheless, it provides a useful perspective on how machine learning-based model reduction relates to classical approaches.

\subsection{Staggered Surrogate Force Model (Force-reconstructed)}
\label{app:force}
In contrast to the End-to-End Force-Augmented Model, one may try to reconstruct the force terms by the learned latent space representation of the corresponding solution space.

\begin{figure}[b]
    \centering
    \input{figures/net_ReacForce.tex}
    \caption{\textbf{Staggered Surrogate Force Model.} The decoder network for the force terms is trained on latent representations generated by the regression network, which itself was trained solely on solution field data. 
    Hence, the regression network has never been exposed to force information. 
    The force decoder is trained in a supervised manner using the true force values as reference.}
    \label{app_fig:force}
\end{figure}
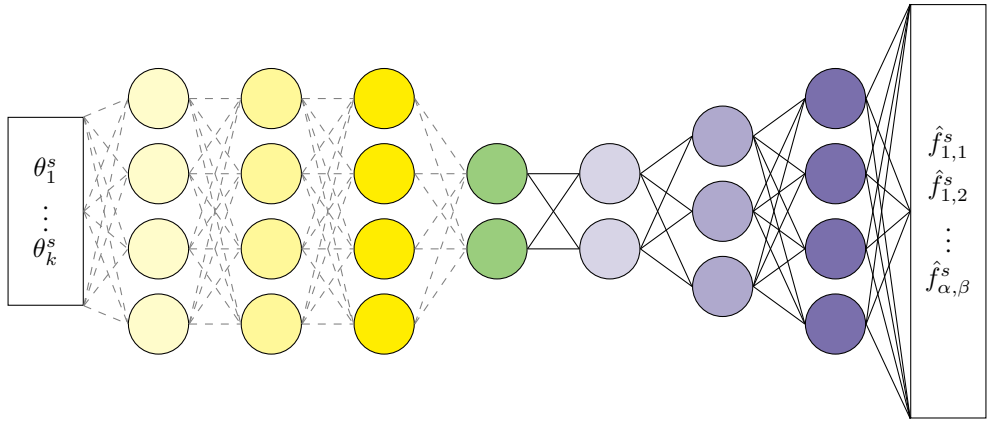

Therefore, we replace the decoder \( \mathcal{D} \) in the original End-to-End architecture with a dedicated supervised regression network \( \mathcal{T} \) designed to infer these force terms from the latent space representation. Formally, given the latent embedding \(\hat{\bm{z}}\), the network \( \mathcal{T} \) parameterized by \( \bm{w}_\mathcal{T} \) yields a prediction of the force terms \(\hat{\bm{f}}\), i.e.,
\begin{equation}
    \hat{\bm{f}} = \mathcal{T}(\hat{\bm{z}}; \bm{w}_\mathcal{T}) = \mathcal{T}\big( \mathcal{P}(\bm{\theta}; \bm{w}_\mathcal{P}); \bm{w}_\mathcal{T} \big),
\end{equation}
which is illustrated in Figure~\ref{app_fig:force}.

The training objective of the force prediction network is to minimize the error between the predicted force components \(\hat{f}^s_{\alpha,\beta}\) and the corresponding ground-truth force values \(f^s_{\alpha,\beta}\), aggregated over all snapshots \(s\), all \textit{non-active} degrees of freedom \(\alpha\), and all force components or spatial indices \(\beta\). Optionally, a regularization term \(\mathcal{F}(\bm{w}_\mathcal{T})\) can be included to impose constraints or promote generalization. Thus, the loss function is defined as
\begin{equation}
    \bm{w}_\mathcal{T}^* = \arg\min_{\bm{w}_\mathcal{T}} \frac{1}{n_s n_A n_B} \sum_{s=1}^{n_s} \sum_{\alpha=1}^{n_A} \sum_{\beta=1}^{n_B} \left( \hat{f}^s_{\alpha,\beta} - f^s_{\alpha,\beta} \right)^2 + \mathcal{F}(\bm{w}_\mathcal{T}),
\end{equation}
where \(n_s\) denotes the total number of snapshots, \(n_A\) is the number of \textit{non-active} degrees of freedom for which force values are predicted, and \(n_B\) represents the number of components per force vector.
\subsection{Governing equations of thermo-mechanics at finite strains}
\label{app:thermo}
The weak forms of the displacement, $\bm{u}$, and temperature, $T$, fields with respect to the reference configuration are given by
\begin{align}
    g_u\left(\bm{u},T,\delta\bm{u}\right) &:= \int_\mathcal{B} \bm{S} : \delta\bm{E}\,\mathrm{d}V - \int_\mathcal{B} \bm{f} \cdot \delta\bm{u}\,\mathrm{d}V - \int_{\partial_t\mathcal{B}} \bm{t} \cdot \delta\bm{u}\,\mathrm{d}A = 0 \\
    \begin{split}
        g_T\left(\bm{u},T,\dot{T},\delta T\right) &:= \int_\mathcal{B} c\,\dot{T}\,\delta T\,\mathrm{d}V - \int_\mathcal{B} \bm{q}_0 \cdot \mathrm{Grad}(\delta T)\,\mathrm{d}V \\
        &- \int_\mathcal{B} r\,\delta T \,\mathrm{d}V - \int_{\partial_q\mathcal{B}} q\,\delta T\, \mathrm{d}A = 0
    \end{split}
\end{align}
where $\delta\bm{u}$ and $\delta T$ denote the test functions, respectively.
Furthermore, $\bm{f}$ and $\bm{t}$ refer to the volume and traction source terms, respectively.
Analogously, $r$ and $q$ can be interpreted for the temperature field as well as $c$ being the heat capacity.
Lastly, $\bm{S} = 2\,\partial\psi/\partial\bm{C}$ denotes the second Piola-Kirchhoff stress, $\delta\bm{E}$ is referred to as the variation of the Green-Lagrange strain tensor, and the referential heat flux according to Fourier's law is given as $\bm{q}_0 = - \Lambda\,\det\bm{F}\,\bm{C}^{-1}\mathrm{Grad}(T)$ with the heat conductivity $\Lambda$ and the deformation gradient $\bm{F}$.

To take thermal expansion into account, we multiplicatively decompose the deformation gradient into a thermal ($T$) and mechanical ($M$) part \cite{StojanovicEtAl1964}
\begin{equation}
    \bm{F} = \bm{F}_M\,\bm{F}_T, \quad \bm{C}_M = \bm{F}_M^T\bm{F}_M.
\end{equation}
For isotropic thermal expansion, the thermal part can uniquely be given as
\begin{equation}
    \bm{F}_T = \vartheta(T)\,\bm{I}, \quad \text{with}\ \vartheta(T) = \exp(\alpha_T\,(T-T_0)),
\end{equation}
where $\alpha_T$ is the constant thermal expansion coefficient and $T_0$ denotes a reference temperature \cite{VujovsevicLubarda2002}.
With this at hand, the mechanical part of the right Cauchy-Green tensor reduces to $\bm{C}_M = 1/\vartheta^2\,\bm{C}$.

For a more in depth discussion including the derivations based on the strong forms, the time discretization scheme, and the linearization, the interested reader is kindly referred to \cite{FelderEtAl2022}.

%% file: figures/net_ReacForce.tex
\begin{center}
\begin{tikzpicture}[
    >=stealth,
    neuronReacOne/.style={circle, draw, fill=rwth13!30, minimum size=8mm},
    neuronReacTwo/.style={circle, draw, fill=rwth13!60, minimum size=8mm},
    neuronReacThree/.style={circle, draw, fill=rwth13, minimum size=8mm},
    neuronLatentFFN/.style={circle, draw, fill=rwth5!60, minimum size=8mm},
    neuronFFNOne/.style={circle, draw, fill=rwthg1!30, minimum size=8mm},
    neuronFFNTwo/.style={circle, draw, fill=rwth7!40, minimum size=8mm},
    neuronFFNThree/.style={circle, draw, fill=rwth7, minimum size=8mm},
    neuron/.style={circle, draw, minimum size=8mm}
]

\def\hsep{1.5cm}

\def\nLatent{2}
\def\nReacOne{2}
\def\nReacTwo{3}
\def\nReacThree{4}

\def\nFFNOne{4}
\def\nFFNTwo{4}
\def\nFFNThree{4}

\newcommand{\drawlayer}[5]{ 
  \foreach \i in {1,...,#2} {
    \node[#5] (#1\i) at (#3,{-#2/2 + \i}) {};
  }
  \node[below=0.2cm of #1#2, yshift=-1mm] {#4};
}

\drawlayer{FFN1}{\nFFNOne}{1*\hsep}{}{neuronFFNOne}
\drawlayer{FFN2}{\nFFNTwo}{2*\hsep}{}{neuronFFNTwo}
\drawlayer{FFN3}{\nFFNThree}{3*\hsep}{}{neuronFFNThree}

\drawlayer{LatentFFN}{\nLatent}{4*\hsep}{}{neuronLatentFFN}

\drawlayer{reac1}{\nReacOne}{5*\hsep}{}{neuronReacOne}
\drawlayer{reac2}{\nReacTwo}{6*\hsep}{}{neuronReacTwo}
\drawlayer{reac3}{\nReacThree}{7*\hsep}{}{neuronReacThree}

\node[draw, fill=white, minimum width=1cm, minimum height=2.5cm, anchor=center] (inputBoxFFN) at (0,0.5) {\shortstack{$\theta_{1}^s$ \\ $\vdots$ \\ $\theta_{k}^s$}};
\node[below=2mm of inputBoxFFN] {};

\node[draw, fill=white, minimum width=1cm, minimum height=5.5cm, anchor=center] (outputBox) at (8*\hsep,0.5) {\shortstack{$\hat{f}_{1,1}^s$ \\ $\hat{f}_{1,2}^s$ \\ $\vdots$ \\ $\hat{f}_{\alpha,\beta}^s$}};
\node[below=2mm of outputBox] {};

\foreach \j in {1,...,\nFFNOne} {
  \draw[-,dashed,gray] (inputBoxFFN.north east) -- (FFN1\j.west);
  \draw[-,dashed,gray] (inputBoxFFN.east) -- (FFN1\j.west);
  \draw[-,dashed,gray] (inputBoxFFN.south east) -- (FFN1\j.west);
}
\foreach \i in {1,...,\nFFNOne} {
  \foreach \j in {1,...,\nFFNTwo} {
    \draw[-,dashed,gray] (FFN1\i.east) -- (FFN2\j.west);
  }
}
\foreach \i in {1,...,\nFFNTwo} {
  \foreach \j in {1,...,\nFFNThree} {
    \draw[-,dashed,gray] (FFN2\i.east) -- (FFN3\j.west);
  }
}
\foreach \i in {1,...,\nFFNThree} {
  \foreach \j in {1,...,\nLatent} {
    \draw[-,dashed,gray] (FFN3\i.east) -- (LatentFFN\j.west);
  }
}
\foreach \i in {1,...,\nLatent} {
  \foreach \j in {1,...,\nReacOne} {
    \draw[-] (LatentFFN\i.east) -- (reac1\j.west);
  }
}
\foreach \i in {1,...,\nReacOne} {
  \foreach \j in {1,...,\nReacTwo} {
    \draw[-] (reac1\i.east) -- (reac2\j.west);
  }
}
\foreach \i in {1,...,\nReacTwo} {
  \foreach \j in {1,...,\nReacThree} {
    \draw[-] (reac2\i.east) -- (reac3\j.west);
  }
}
\foreach \j in {1,...,\nReacThree} {
  \draw[-] (reac3\j.east) -- (outputBox.south west);
  \draw[-] (reac3\j.east) -- (outputBox.west);
  \draw[-] (reac3\j.east) -- (outputBox.north west);
}
\end{tikzpicture}
\end{center}

%% file: sections/appendixB.tex
\section{Declarations}
%
\subsection{Acknowledgements}
This work was supported by the NSF CMMI Award 2320933 Automated Model Discovery for Soft Matter and by the ERC Advanced Grant 101141626 DISCOVER to Ellen Kuhl.
In addition, Tim Brepols gratefully acknowledges financial support of the projects 453596084 and 561202254 by the Deutsche Forschungsgemeinschaft.
Further, Kevin Linka is supported by the Emmy Noether Grant 533187597 by the Deutsche Forschungsgemeinschaft.
Lastly, Hagen Holthusen and Tim Brepols gratefully acknowledge financial support of the project 417002380 by the Deutsche Forschungsgemeinschaft.
%
%
\subsection{Conflict of interest}
The authors of this work certify that they have no affiliations with or involvement in any organization or entity with any financial interest (such as honoraria; participation in speakers’ bureaus; membership, employment, consultancies, stock ownership, or other equity interest; and expert testimony or patent-licensing arrangements), or non-financial interest (such as personal or professional relationships, affiliations, knowledge or beliefs) in the subject matter or materials discussed in this manuscript.
%
\subsection{Availability of data, source code and material}
Our data, source code and examples of the \texttt{JAX}/\texttt{Flax} implementation are accessible to the public at \url{https://doi.org/10.5281/zenodo.16992619}.
%
\subsection{Contributions by the authors}
\textbf{Jannick Kehls:} Conceptualization, Methodology, Software, Validation, Investigation, Formal Analysis, Writing - Original Draft, Writing - Review \& Editing \\
\textbf{Ellen Kuhl:} Writing - Review \& Editing, Supervision, Funding acquisition \\
\textbf{Tim Brepols:} Conceptualization, Formal Analysis, Writing - Original Draft, Writing - Review \& Editing, Supervision, Funding acquisition \\
\textbf{Kevin Linka:} Conceptualization, Formal Analysis, Writing - Original Draft, Writing - Review \& Editing, Supervision, Funding acquisition\\
\textbf{Hagen Holthusen:} Conceptualization, Methodology, Software, Validation, Investigation, Formal Analysis, Data Curation, Writing - Original Draft, Writing - Review \& Editing, Funding acquisition
%
\subsection{Statement of AI-assisted tools usage}
This document was prepared with the assistance of OpenAI's ChatGPT, an AI language model. ChatGPT was used for language refinement. The authors reviewed, edited, and take full responsibility for the content and conclusions of this work.